\documentclass[journal]{IEEEtran}
\usepackage[latin9]{inputenc}
\usepackage{color}
\usepackage{amsmath}
\usepackage{graphicx}

\makeatletter

\DeclareTextSymbolDefault{\textquotedbl}{T1}

\usepackage{amssymb}
\usepackage{amsfonts}
\usepackage{algorithmic}
\usepackage{url}
\usepackage{hyperref}
\usepackage{wasysym}
\usepackage{xcolor}
\usepackage[ruled,vlined]{algorithm2e}

\makeatother

\begin{document}
\global\long\def\ve#1{\boldsymbol{#1}}%

\global\long\def\mat#1{\boldsymbol{#1}}%

\title{A Robust Stochastic Method of Estimating the Transmission Potential
of 2019-nCoV}
\author{Jun Li\\{\tt FirstName.LastName@uts.edu.au}
\\University of Technology Sydney, Broadway 123, NSW 2007}
\maketitle
\begin{abstract}
The recent outbreak of a novel coronavirus (2019-nCoV) has quickly
evolved into a global health crisis. The transmission potential of
2019-nCoV has been modelled and studied in several recent research
works. The key factors such as the basic reproductive number, $R_{0}$,
of the virus have been identified by fitting contagious disease spreading
models to aggregated data. The data include the reported cases both
within China and in closely connected cities over the world. 

In this paper, we study the transmission potential of 2019-nCoV from
the perspective of the robustness of the statistical estimation, in
light of varying data quality and timeliness in the initial stage
of the outbreak. Sample consensus algorithm has been adopted to improve
model fitting when outliers are present.

The robust estimation enables us to identify two \emph{clusters} of
transmission models, both are of substantial concern, one with $R_{0}:8\sim14$,
comparable to that of measles and the other dictates a large initial
infected group.

\end{abstract}

\subsection*{Highlights}
\begin{itemize}
\item We introduce robust transmission model fitting. We employed random
sample consensus algorithm for the fitting of a susceptible-exposed-infectious-recovered
(SEIR) infection model.
\item We identify data consistency issues and raise flags for i) a potentially
high-infectious epidemic and ii) further investigation of records
with unexplained statistical characteristics.
\item This analysis accounts for the spreading in 80+ China cities with
multi-million \emph{individual} populations, which are connected to
the original outbreak location (Wuhan) during the massive people transportation
period (\emph{chunyun})\footnote{\begin{itemize}
\item Traffic is considered in \cite{Wu2020}, but for the purpose of modelling
the population variation within Wuhan, the outbreak site.
\end{itemize}
}.
\item As the virus is active and the analytics and control of the epidemic
is an urgent endeavour, we choose to release all source code and implementation
details despite the research is on-going. The scientific ramification
is that conclusions may need further revision with richer and better
prepared data made available. We have published our implementation
on Github \texttt{\footnotesize{}https://www.github.com/junjy007/ransac\_seir}.
All procedures are included in a single Python notebook.
\item We have only used publicly available data in the research, which have
been also made available with the project. The quality and reliability
of estimation could be further improved by adopting richer data from
commercial sources or authorities. More discussion in this regard
can be found in the conclusion section.
\end{itemize}

\section{Introduction}

Since December 2019, a new strain of coronavirus (2019-nCoV) has started
spreading in Wuhan, Hubei Province, China \cite{Wu2020}. The initial
cases of infection have suspicious exposure to wild animals. However,
when cases are reported in globally in middle January 2020, including
Southeast and East Asia as well as the United States and Australia,
the virus shows sustained human-to-human transmission (On 21 January
2020, the WHO suggested there was possible sustained human-to-human
transmission). With the massive people transport prior to Chinese
New Year (Chunyun), the virus spreads to major cities in China and
densely populated cities within Hubei Province.

There are a number of epidemiological analysis on the transmission
potential of 2019-nCoV. Read et al. \cite{Read2020} fit a susceptible-exposed-infectious-recovered
(SEIR) metapopulation infection model to reported cases in Wuhan and
major cities connected by air traffic. In \cite{Wu2020}, an SEIR
model has been estimated by including surface traffic from location-based
services data of Tencent.  However, neither the air traffic to international
destinations nor the aggregated people throughput to Wuhan can help
establish the transmission model among populous China cities connected
to Wuhan mainly via surface traffic. Significantly, the reported cases
in those populations connected to Wuhan are important to help robust
estimation of the transmission potential of the virus. This is particularly
important in the initial stage of the outbreak, as the initial reports
can be prone to various disturbances, such as to delay or misdiagnosis,
which is identified in our robust analysis below.

In this work, we present a study on robust methods of fitting the
infection models to empirical data. We propose to employ the random
sample consensus (RANSAC) algorithm \cite{Fischler1981} to achieve
robust parameter estimation. SEIR and most infection models of contagious
diseases are designed for review analysis \cite{Chowell2009}. On
the other hand, to provide a useful forecast in the out-breaking stage
of a new disease, transmission models must be established using data
that are insufficient in terms of both quantity and quality. The maximum
likelihood model estimation used by most existing studies is sensitive
to outliers. Therefore, the estimated parameters can be unreliable
due to the quality of the data in the initial stage of an epidemic.
The issue is rooted in the combination of the quality of the data
and sensitivity of the fitting method, therefore it is not easily
addressed/captured by traditional sensitivity analysis techniques
such as bootstrapping.

Random sample consensus algorithm alleviates the predominant influence
on the model fitting of the records of infections in the original
place, Wuhan, and close-by cities. The selected model reveals different
statistical characteristics in the spreading of the virus in different
cities, according to the local records, which deserves further investigation.

By identifying and accounting for a large volume of records of uncertain
timeliness and accuracy, we have identified two candidate groups of
models that agree with empirical records. One with significantly higher
$R_{0}$, at the level of measles, and the other model cluster has
$R_{0}$ similar to previously reported values \cite{Wu2020,Read2020}
but suggests there were already a large number of infected individuals
on 1 January 2020.

\section{Method}

\subsection{Data Source}

This research follows a similar procedure of acquiring and processing
data of confirmed cases and public transportation as in \cite{Wu2020}.
The infection report is summarised daily by Pengpai News\cite{Pengpai},
who collects reports from the Health Commissions of local administrations
of different provinces and cities. We include the major populated
areas with strong connections with Wuhan in this study. We selected
the locations which i) have a population greater than 3 million ii)
are among the top-100 destinations for travellers departing from Wuhan
on 22 January (the day before the lockdown of the city for quarantine
purposes. We include 84 cities, including Wuhan, in this study.

We collect data of population from various sources on the World Wide
Web. The transportation data is from Baidu migaration index \cite{Qianxi},
based on their record of location-based services. We estimated the
absolute number of travellers by aligning the index of a reported
number of 4.09M during the period of 10-20 January 2020.

In the data collection, infections outside China are summarised at
the country level and the specific cities are missing. We exclude
this part of infection records since entire countries have a different
distribution of population than individual populated areas. Such evidence
can be considered in future research by employing more geographical/demographical
data as well as volumes of traffic connections.

\subsection{Transmission Model and Ftting to Data}

\subsubsection{SEIR metapopulation infection model}

In this research, we adopt the susceptible-exposed-infectious-recovered
(SEIR) model of the development and infection process of 2019-nCoV,
similar to that in \cite{Read2020}. The model includes a dynamic
component corresponding to people movement between populated areas.
The transmission model is defined as follows
\begin{align}
\frac{dS_{j}(t)}{dt} & =-\beta\left(\sum_{c}\frac{K_{c,j}(t)}{n_{c}}I_{c}+I_{j}\right)\cdot\frac{S_{j}(t)}{n_{j}}\label{eq:s}\\
\frac{dE_{j}(t)}{dt} & =\beta\left(\sum_{c}\frac{K_{c,j}(t)}{n_{c}}I_{c}+I_{j}\right)\cdot\frac{S_{j}(t)}{n_{j}}-\alpha E_{j}(t)\\
\frac{dI_{j}(t)}{dt} & =\alpha E_{j}(t)-\gamma I_{j}(t)\\
\frac{dR_{j}(t)}{dt} & =\gamma I_{j}(t)\label{eq:r}
\end{align}
where $S,E,I,R$ represent the number of susceptible, exposed, infected
and recovered (non-infectable) subjects. Equation set (\ref{eq:s}\nobreakdash-\ref{eq:r})
specify the dynamics of the disease spreading in a set of populated
areas connected by a traffic network. The subscript $j$ is over the
areas, e.g. cities.

\textbf{Spreading dynamics}: The model parameters $\alpha,\beta,\gamma$
control the dynamics of the disease spreading. In a unit of time,
exposed subjects become infected with a rate of $\alpha$. Thus the
mean latent (incubation) period is $1/\alpha$, which were ranging
from 3.8-9 in previous epidemiological studies of CoV's \cite{Virlogeux2016,Leung2004}.
We use $\alpha=1/7$ according to empirical observation as of Feb
2020. The model and the fitting process is not hypersensitive to this
parameter \cite{Read2020}. Parameter $\beta$ represents the rate
of conversion from the status of ``exposed'' to ``infected'' in
one time unit. Parameter $\gamma$ determines the rate of recovery,
while the recovered subjects are removed from the repository of susceptible
subjects. The parameters $\beta$ and $\gamma$ are estimated by fitting
the model to data using a stochastic searching strategy, as discussed
below.

\textbf{Transportation dynamics}: Between-area dynamics is specified
by a traffic model, which entails a set of connectivity matrices $\mat K(t)$,
where an entry $K_{i,j}(t)$ is the number of travellers from area-i
to area-j at time $t$. The transportation model dictates that at
time $t$, $\sum_{c}\frac{K_{c,j}(t)}{n_{c}}I_{c}$ infected subjects
arrive at area-j and start infecting susceptible subject in the destination
area-j. 

\textbf{Initial infections}: At $t=0$, which is set to 1 January
2020 in this study, the number of infected cases at Wuhan is set to
a seeding number $I_{W}(0)$. $I_{W}(0)$ is a parameter inferred
from data as in \cite{Read2020}. Alternatively, a zoonotic infection
model is used in \cite{Wu2020}, considering the evidence of an animal
origin of the2019-nCoV. 

\subsubsection{Model Fitting via Maximum Likelihood and Challenges}

There are three parameters to specify in the metapopulation SEIR model,
denoted by a vector $\ve{\theta}$: $(\beta,\gamma,I_{W}(0))$. Most
existing studies adopt the maximum likelihood method to infer model
parameters from empirical data. The inference is an optimisation process,
with the objective defined as the probability of observing the empirical
data given the model predictions, e.g. 
\begin{align}
\ve{\theta}^{*} & :=\arg\min_{\ve{\theta}}\sum_{t}-\log P\left(\ve x_{t}|SEIR(t;\ve{\theta})\right)\label{eq:nll}
\end{align}
where $P(\ve x|\ve{\mu})$ represents the probability density/mass
of observing $\ve x$ given model prediction $\ve{\mu}$. The probability
is accumulated over time $t$. Note that we use boldface symbols to
indicate that both observed data $\ve x$ and model prediction $\ve{\mu}$
can be vectors containing the information of the disease at multiple
locations. Theoretically, the inference optimisation in \eqref{eq:nll}
can be established by using any observation model. However, in practice,
to estimate the transmission characteristics of a contagious disease
during the out-breaking stage, the empirical observations are usually
limited to the sporadic report of confirmed infection cases, as the
exposed latent subjects are unable to identify and waiting for recovery
cases is not a viable option for nowcasting and forecasting study.

Relying on confirmed infections can make model parameter estimation
difficult. On one hand, the initial observations are often of suboptimal
quality in terms of both timeliness and accuracy. As a new disease
starts spreading, the first cases can be misdiagnosed, especially
when the symptoms are mild in a significant portion of infectious
subjects/period. On the other hand, the negative log-likelihood objective
function is usually dominated by the observations in the original
location, where the disease starts spreading. Therefore, it is possible
that significantly disturbed observations in the original location
lead to biased estimation of the model. The systematic bias is not
easily dealt with by traditionally statistical techniques such as
boot-strapping.

\subsubsection{RANSAC Algorithm of Robust Model Fitting}

The random sample consensus (RANSAC) method is designed for model
estimation with a significant amount of outliers in data. The essential
idea is to fit a simple model (3 adjustable parameters in the SEIR
model) using the minimum number of data points randomly drawn from
the dataset. Algorithm 1. The following Algorithm \ref{alg:ransac}
shows the steps of the algorithm.

\begin{algorithm}
\LinesNumbered
\SetAlgoLined
\DontPrintSemicolon
\KwIn{Rounds of random sampling, $n_{R}$ and number of random samples in each round of model fitting, $n_s$}
\KwIn{Daily records of infectons of $T$ days and $n_L$ locations, $X: [n_L \times T]$}
\KwIn{Model fitting function:  $f: \{x_1, \dots, x_{n_s}\} \mapsto (\beta, \gamma, I_W(0))$}
\KwIn{Inlier Counting: $g: (\beta, \gamma, I_W(0)), X \mapsto n_{In}$}
\KwResult{Optimal parameters: $\beta^*, \gamma^*, I_W^*(0)$}  
Initialise $n^*_{In} \leftarrow -1$\;
\For{$i\leftarrow 1$ \KwTo $n_R$}{
  Randomly draw $l_i$ from $\{1, \dots, L\}$\;
  Randomly draw $n_s$ samples from $X[l_i, \dots]$: $\{x_{i_1}, \dots, x_{i_{n_s}}\}$ \;
  $\beta, \gamma, I_W(0) \leftarrow f(x_{i_1}, \dots, x_{i_{n_s}})$\;
  $n_{In} \leftarrow g((\beta, \gamma, I_W(0)), X)$\;
  \If{$n_{In} > n^*_{In}$}{ \label{algln:r0}
	$n^*_{In} \leftarrow n_{In}$ \;
    $\beta^*, \gamma^*, I_W^*(0) \leftarrow \beta, \gamma, I_W(0)$ \label{algln:r1}
  }
}
\caption{RANSAC Algorithm of Fitting SEIR Model to Infection Data \label{alg:ransac}} 
\end{algorithm}

In the algorithm, the steps from line \ref{algln:r0} to line \ref{algln:r1}
choose the model achieving maximum consensus among the random samples.
The function $f$ executes the maximum likelihood model fitting. However,
the optimisation has been made straightforward, as there are only
$n_{s}$ daily infection data points from one location $l_{i}$ to
fit to. We choose $n_{s}=4$ in this study to determine the 3 parameters
of the SEIR model. So there are 4 constraints and 3 degrees of freedom,
where the one extra constraint helps stabilise the optimisation. The
function $g$ counts inliers in the whole data for a given SEIR model.
To be considered as an inlier, a recorded infection number at time
$t$ in place $l$ needs to fall within the 5\% to 95\% CI of the
model prediction at the time and location. Following \cite{Read2020},
we use the Poisson distribution to approximate the probability distribution
of the infection number within one day in a location. 

\section{Estimation and Prediction of Epidemic Size}

\subsection{Parameters of SEIR Transmission Model}

\begin{figure}
\begin{centering}
\includegraphics[width=0.9\columnwidth]{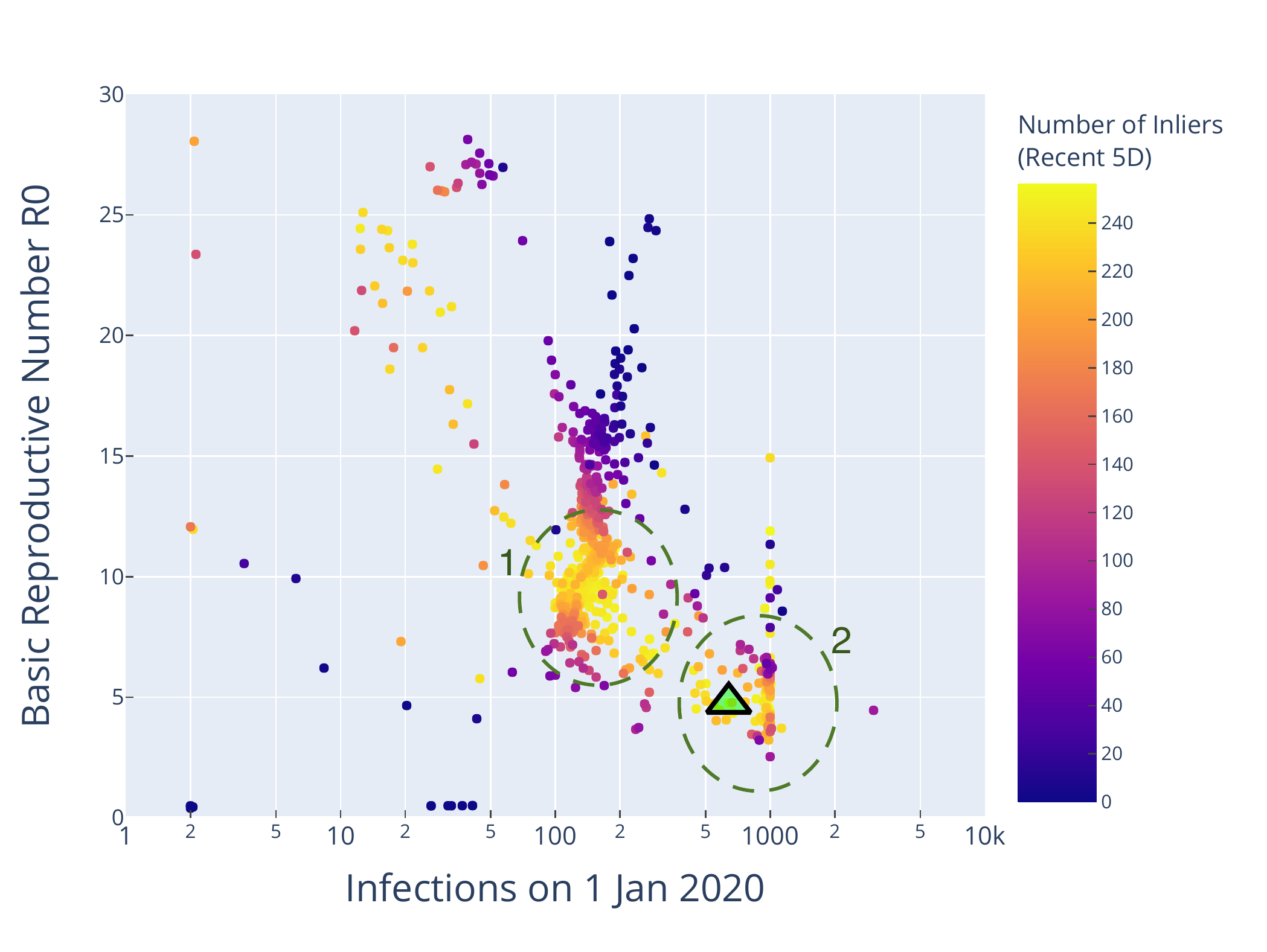}
\par\end{centering}
\begin{centering}
(a)
\par\end{centering}
\begin{centering}
\includegraphics[width=0.9\columnwidth]{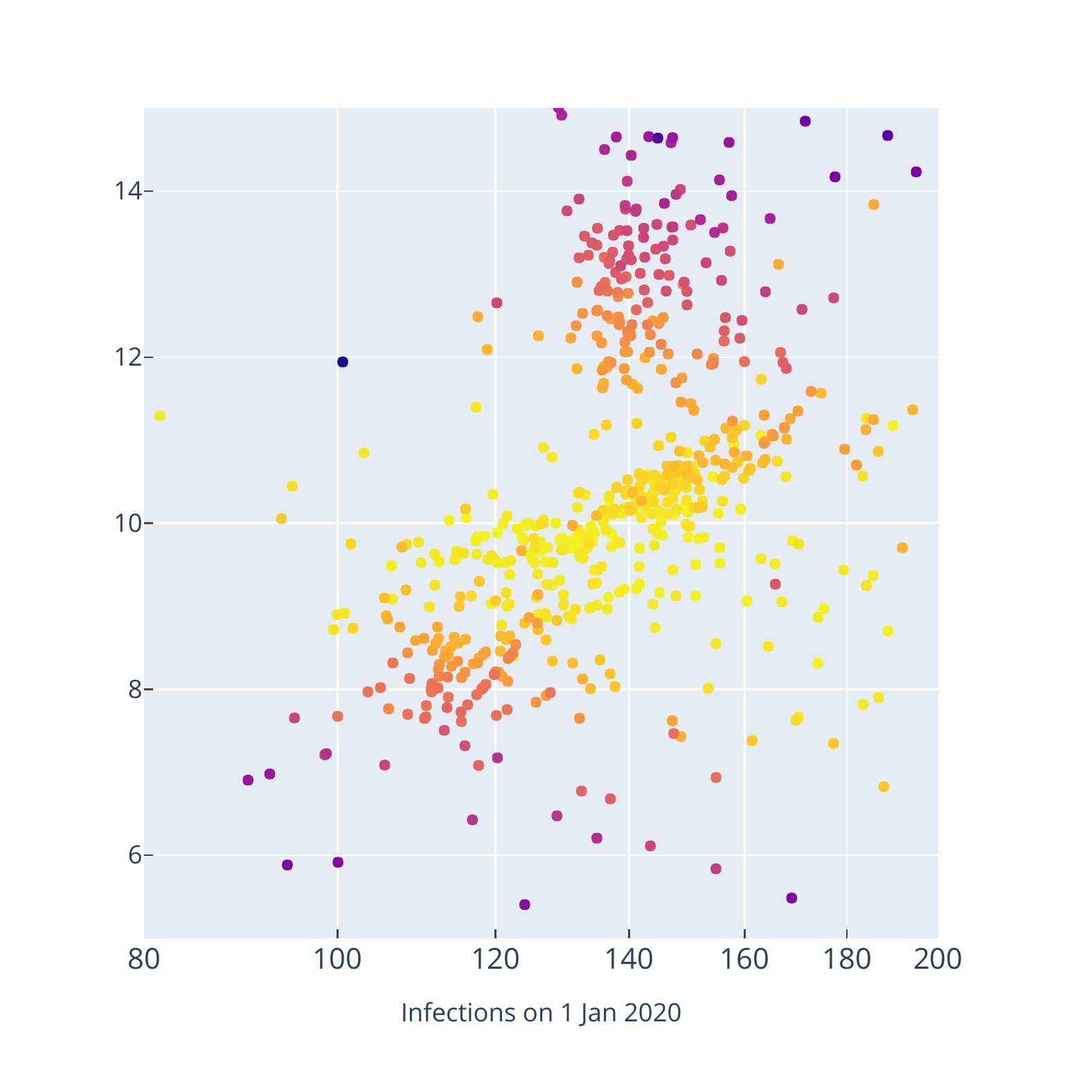}
\par\end{centering}
\begin{centering}
(b)
\par\end{centering}
\caption{SEIR model parameter estimation using RANSAC\label{fig:SEIR-model-parameter}}
\end{figure}

Due to the size of the populations and the short period of interest,
we can ignore the change of the population due to birth or death during
the process. Thus the basic reproductive number in this SEIR model
can be estimated as $R_{0}\approx\frac{\beta}{\gamma}$. Figure \ref{fig:SEIR-model-parameter}
shows the model parameters fitted to the minimum ($n_{s}=4$) random
samples in 1,000 RANSAC iterations. In the figure, the models are
specified by a pair of parameters: the basic reproductive $R_{0}$
and the estimated infection number in Wuhan on 1 January 2020, $I_{W}(0)$.
The numbers of inliers in the last 5 days in the recorded period (up
to 5 Feb 2020) is considered as the fitness of the corresponding models.
Fitness is indicated by the colour in the figure. The model producing
the greatest number of inliers is marked by a triangle in the figure. 

In Figure 1, as far as the available data is concerned, there is a
structure of two main clusters indicating candidates of valid models.
Intuitively, one cluster (\textquotedbl 1\textquotedbl ) corresponds
to the possibility of a highly infectious virus starting from a relatively
small group of subjects. The other cluster (\textquotedbl 2\textquotedbl )
indicates an R0 that is more consistent with existing estimations,
but the virus has started from a large number of individuals, which
is vastly exceeding the current expectation. The parameter set leading
to the greatest fitness in the RANSAC process is from cluster-2, 
\begin{align*}
\beta^{*} & =0.642\\
\gamma^{*} & =0.135\\
R_{0}^{*} & =4.76\\
I_{W}^{*}(0) & \approx641
\end{align*}
which has 256 out of 425 daily infection number (from 85 places in
the last 5 days) falling within the inlier-zone. 

It is too early to rule out either or both possibilities. It has become
evidential that the virus can show mild or no symptoms in a significant
portion of infections. Plus the fact that the virus was unknown to
human, it was not impossible that the virus had been circulating for
a period, even with sporadic severe cases being misdiagnosed for other
diseases, before a group of severe infection eventually broke and
called attention.

\subsection{Simulation of Infection in Metapopulation}

\begin{figure}
\begin{centering}
\hfill{}\includegraphics[width=0.48\textwidth]{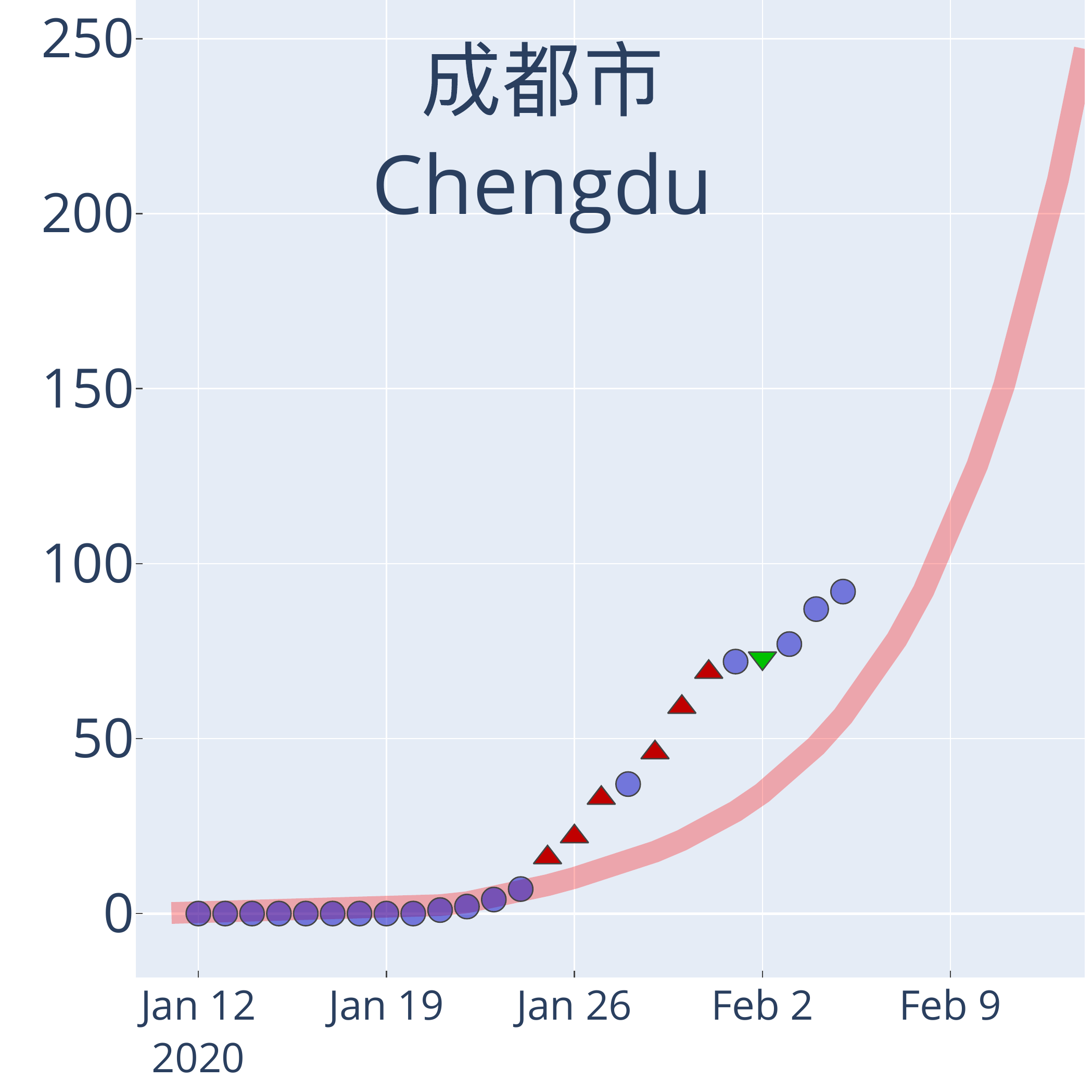}\hfill{}
\par\end{centering}
\caption{Simulation and forecasting of infections in a major China city, compared
with reported cases. \label{fig:sim-sample} The bold red curve represents
the predicted infection number by running simulation using the SEIR
model selected by the RANSAC algorithm. The markers correspond to
accumulated infection numbers up to the dates. Triangles represent
the newly reported infections of the corresponding days are classified
as outliers given the predicted Poisson distributions. Red up-triangles
${\color{red}\blacktriangle}$ represent the recorded value exceeds
the upper bound of the CI (infection number is too high according
to the model). Green down-triangles ${\color{green}\blacktriangledown}$
represent the opposite cases. Blue circles ${\color{blue}\bullet}$
represent inliers.}
\end{figure}

We have built metapopulation SEIR model using the parameters $\beta^{*},\gamma^{*}$
and $I_{W}^{*}(0)$ selected by the RANSAC algorithm above. We then
run simulations using the fitted SEIR model and compare the model
prediction with empirical data of infection recorded in different
cities over China. Figure \ref{fig:sim-sample} shows the simulation
result and the accumulated infection data for one major China city
Chengdu. The model simulation has explained the newly identified infections
in a significant number of days during the period of interest. See
figure caption for detailed interpretation of the curves and marks
in the plots.

\begin{figure*}[t]
\begin{centering}
\hfill{}\includegraphics[width=0.24\textwidth]{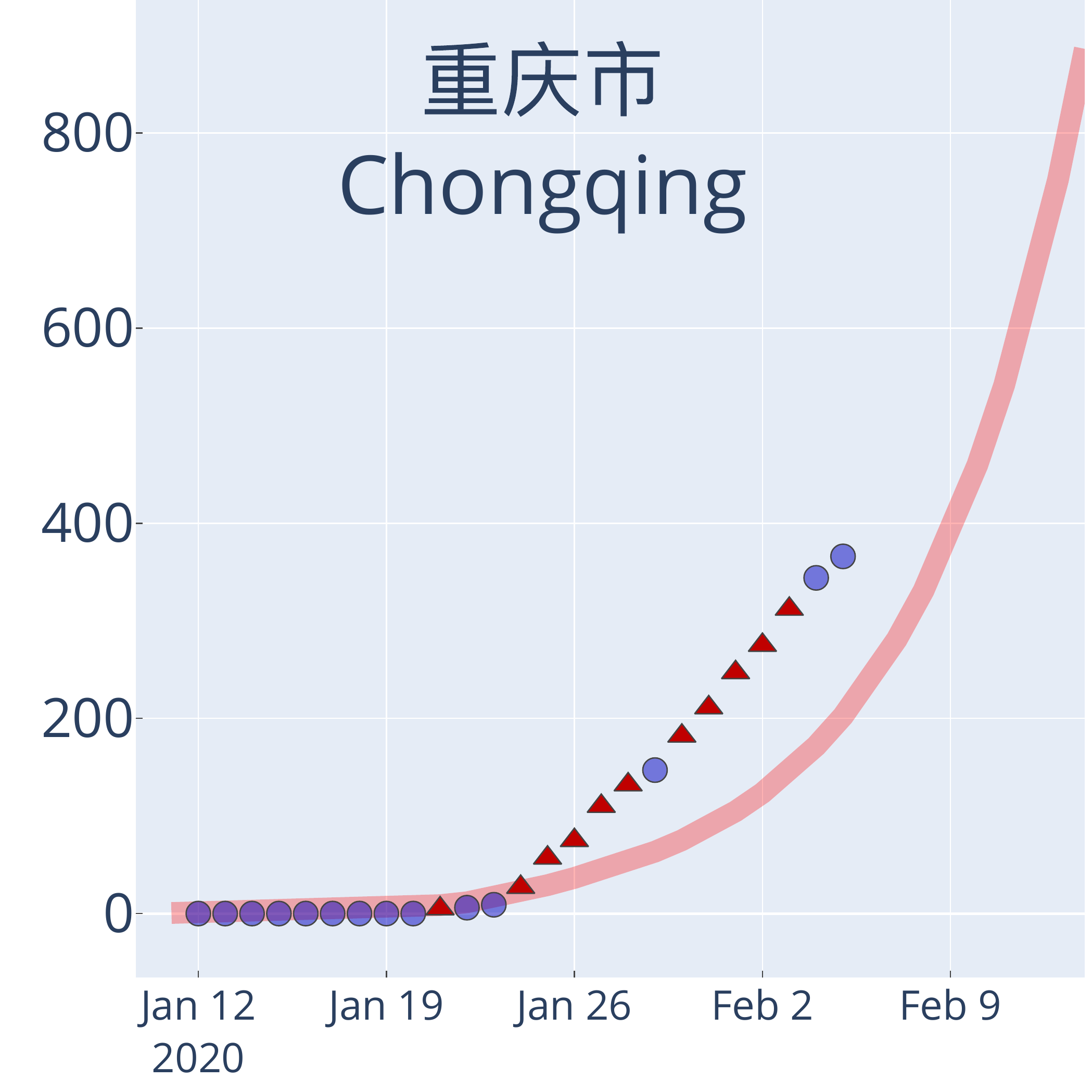}\hfill{}\includegraphics[width=0.24\textwidth]{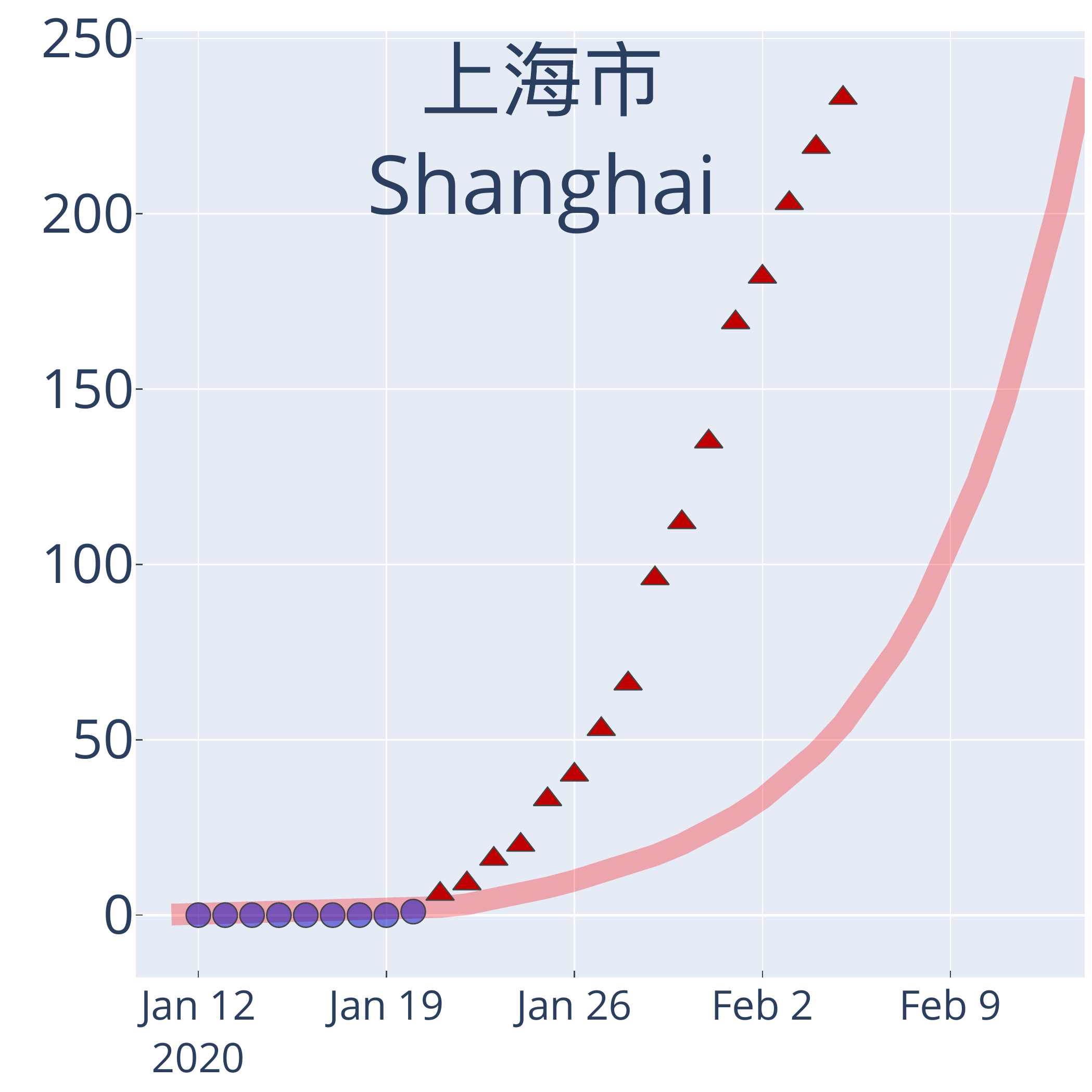}\hfill{}\includegraphics[width=0.24\textwidth]{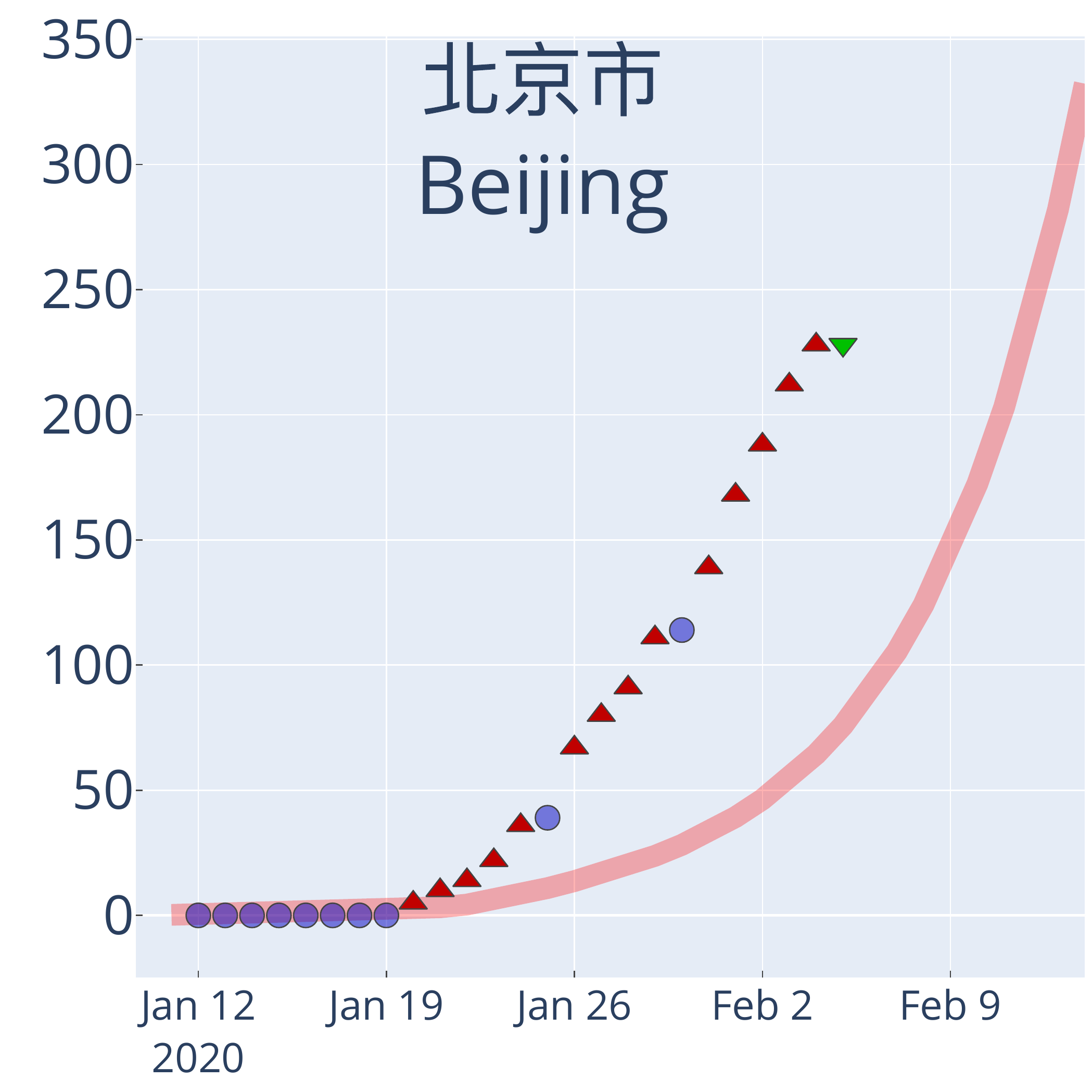}\hfill{}\includegraphics[width=0.24\textwidth]{checkpoints/pred_num_figures_r1/wd003}\hfill{}
\par\end{centering}
\begin{centering}
\hfill{}\includegraphics[width=0.24\textwidth]{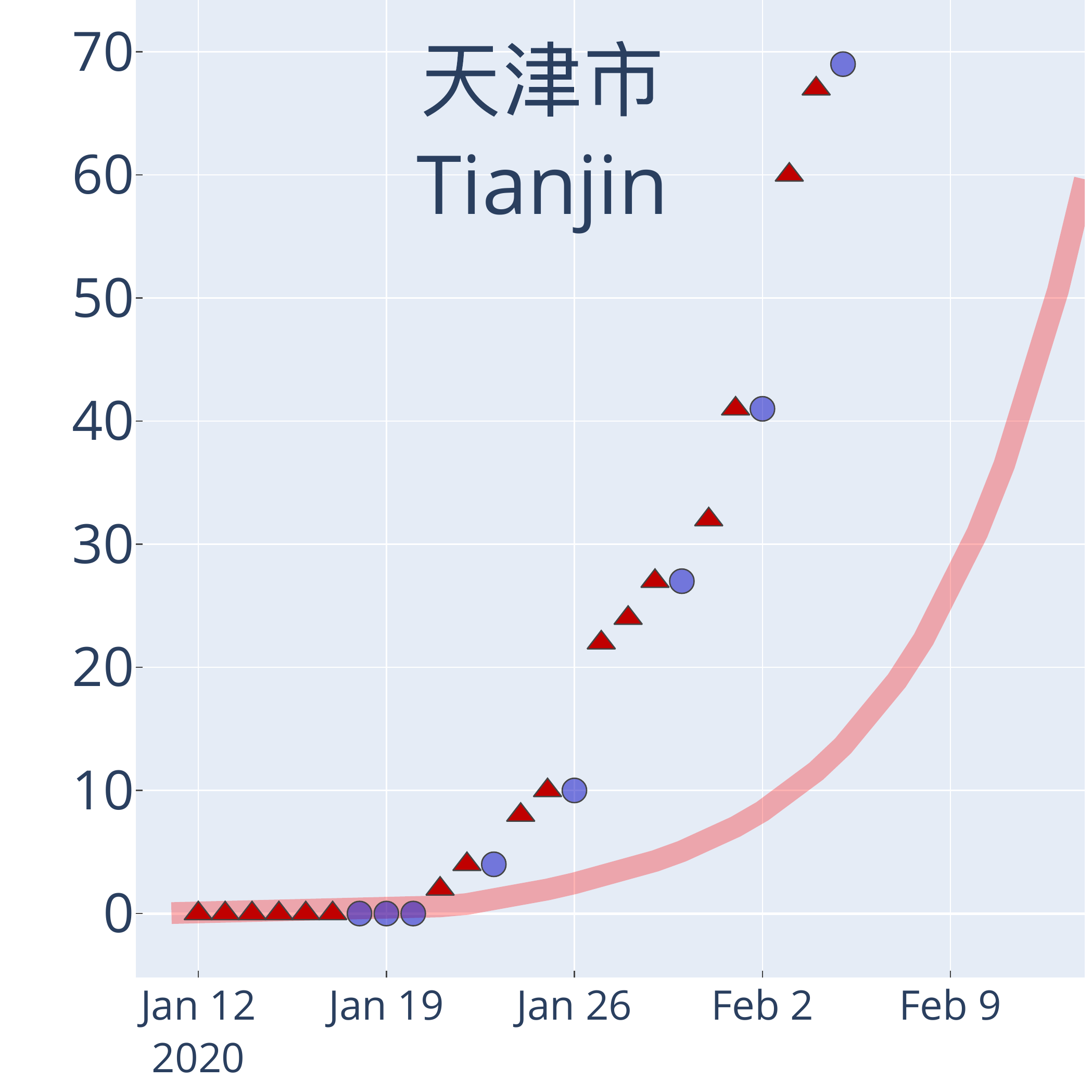}\hfill{}\includegraphics[width=0.24\textwidth]{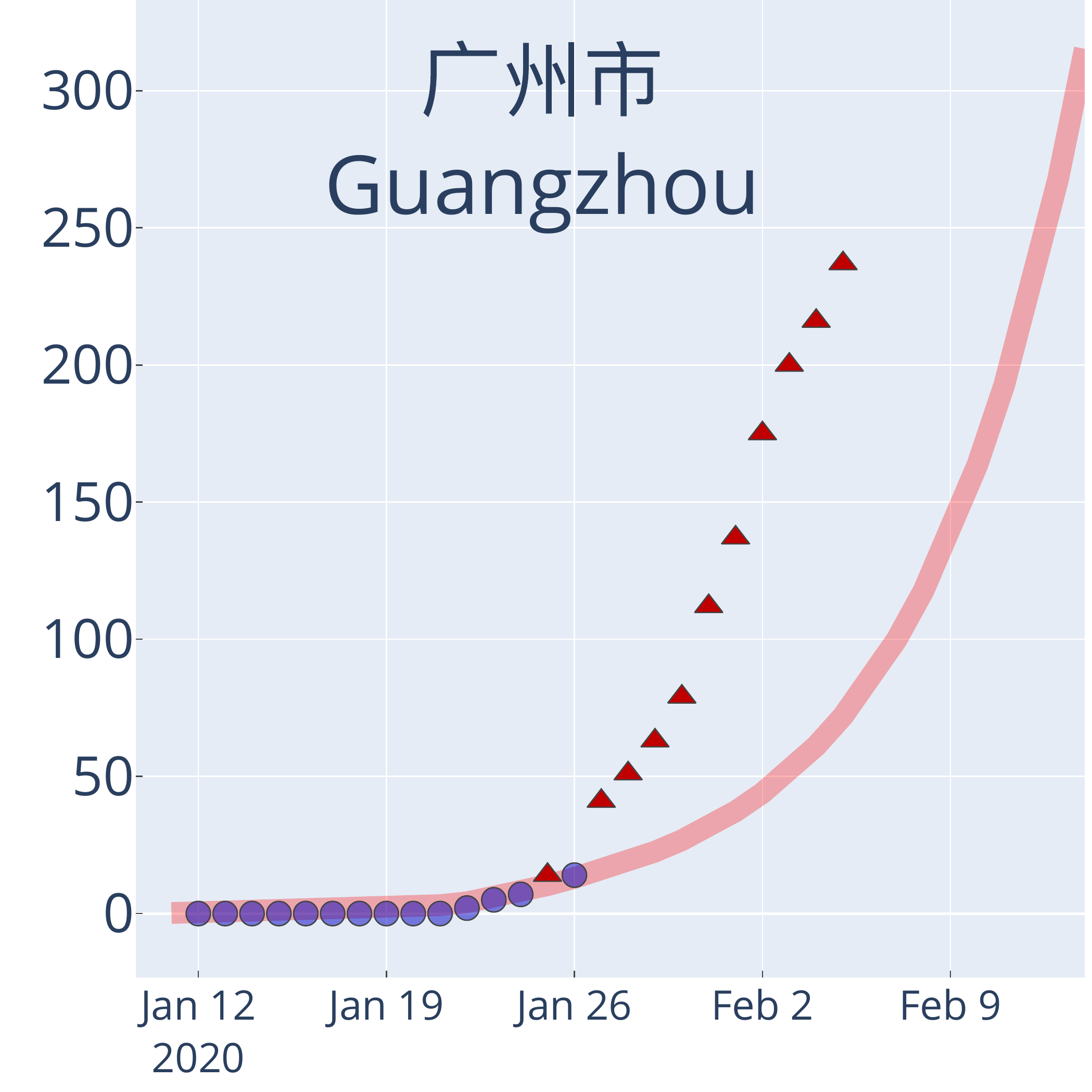}\hfill{}\includegraphics[width=0.24\textwidth]{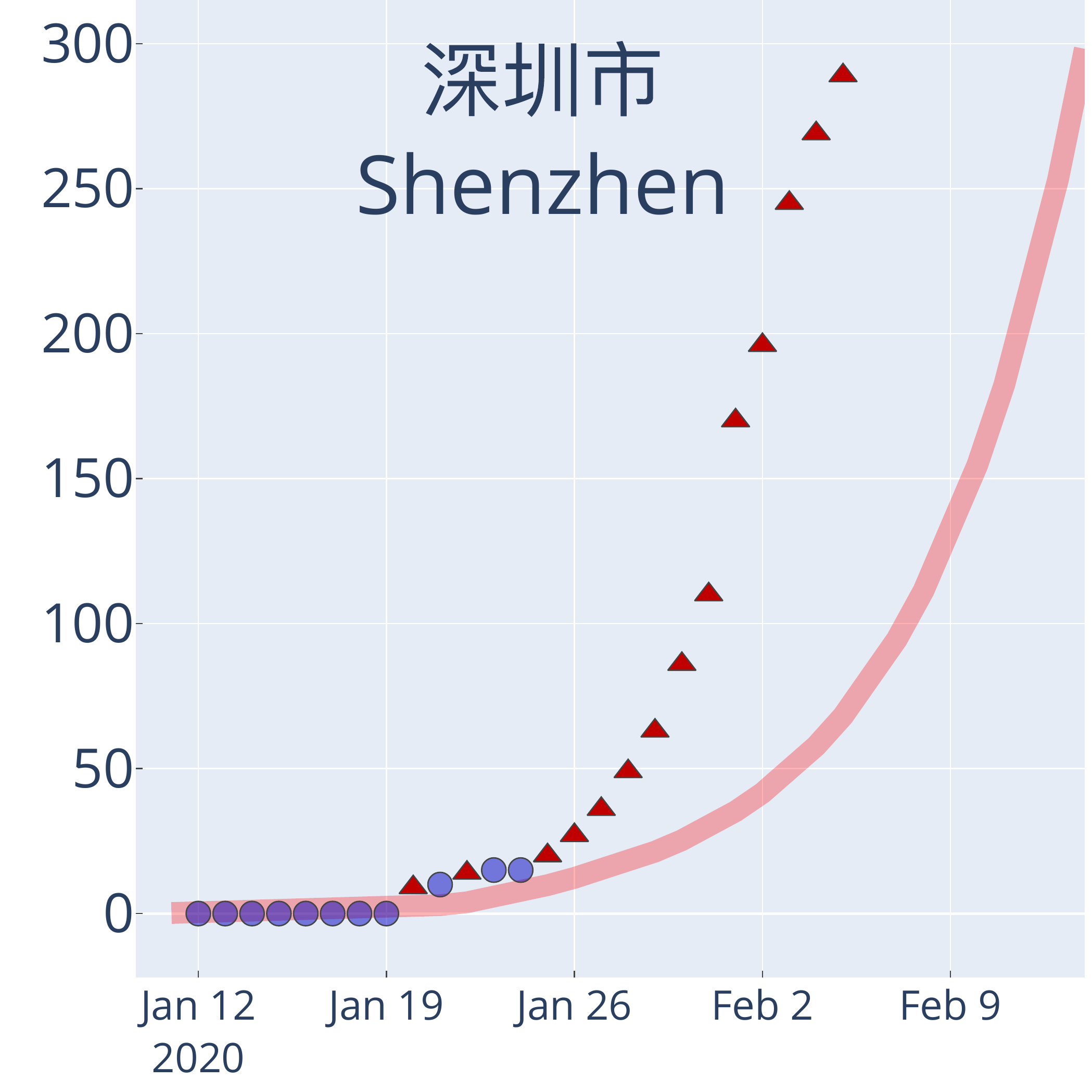}\hfill{}\includegraphics[width=0.24\textwidth]{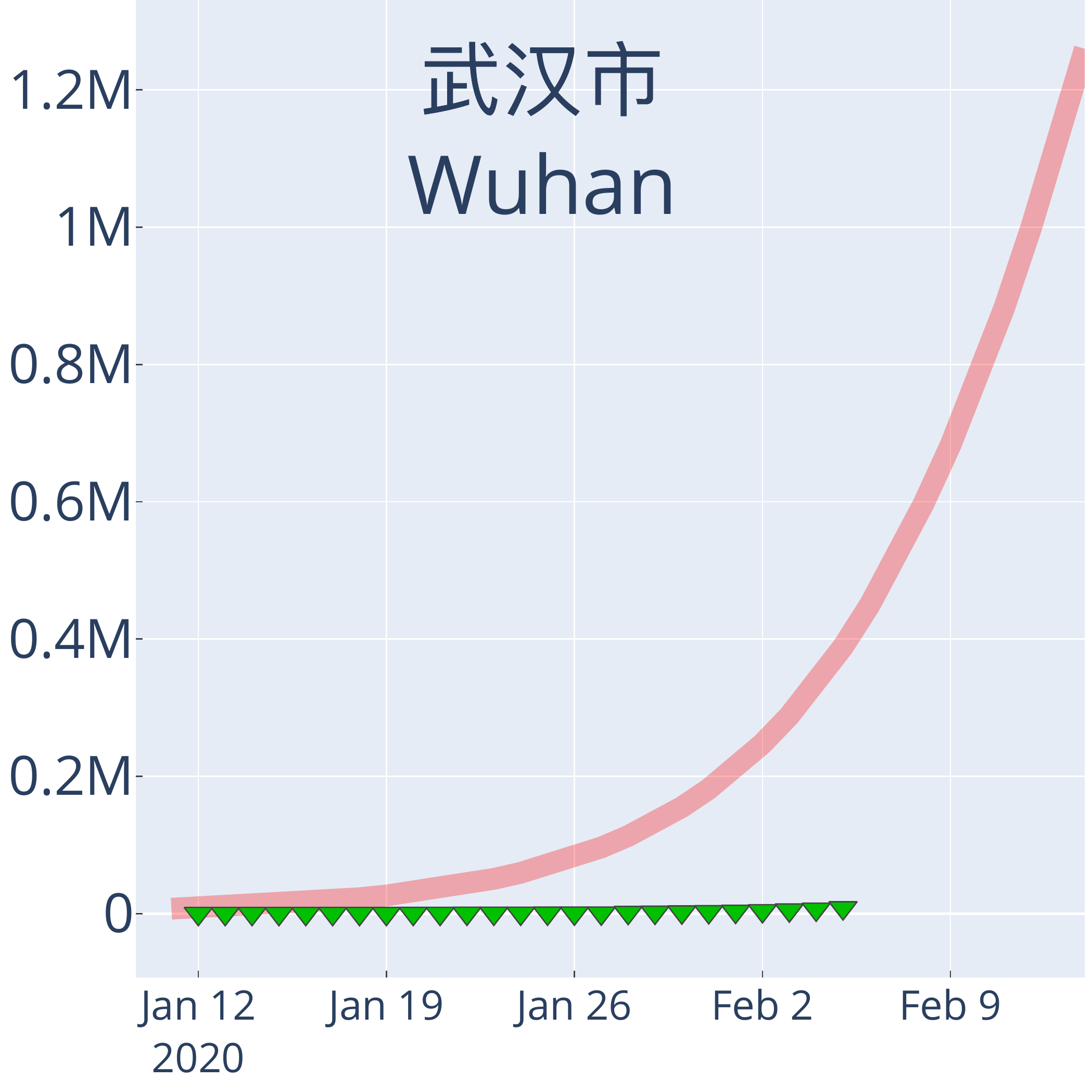}\hfill{}
\par\end{centering}
\begin{centering}
\hfill{}\includegraphics[width=0.24\textwidth]{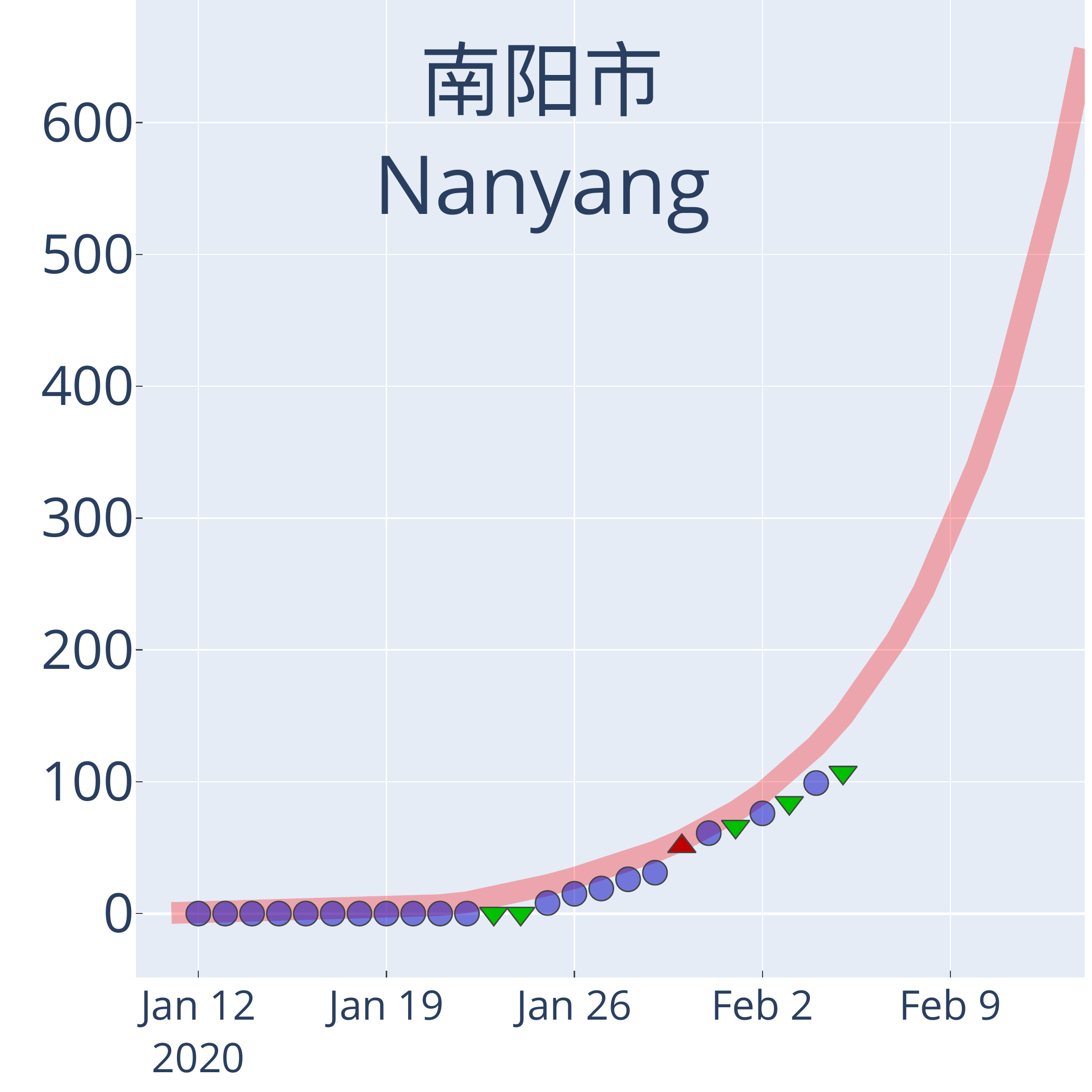}\hfill{}\includegraphics[width=0.24\textwidth]{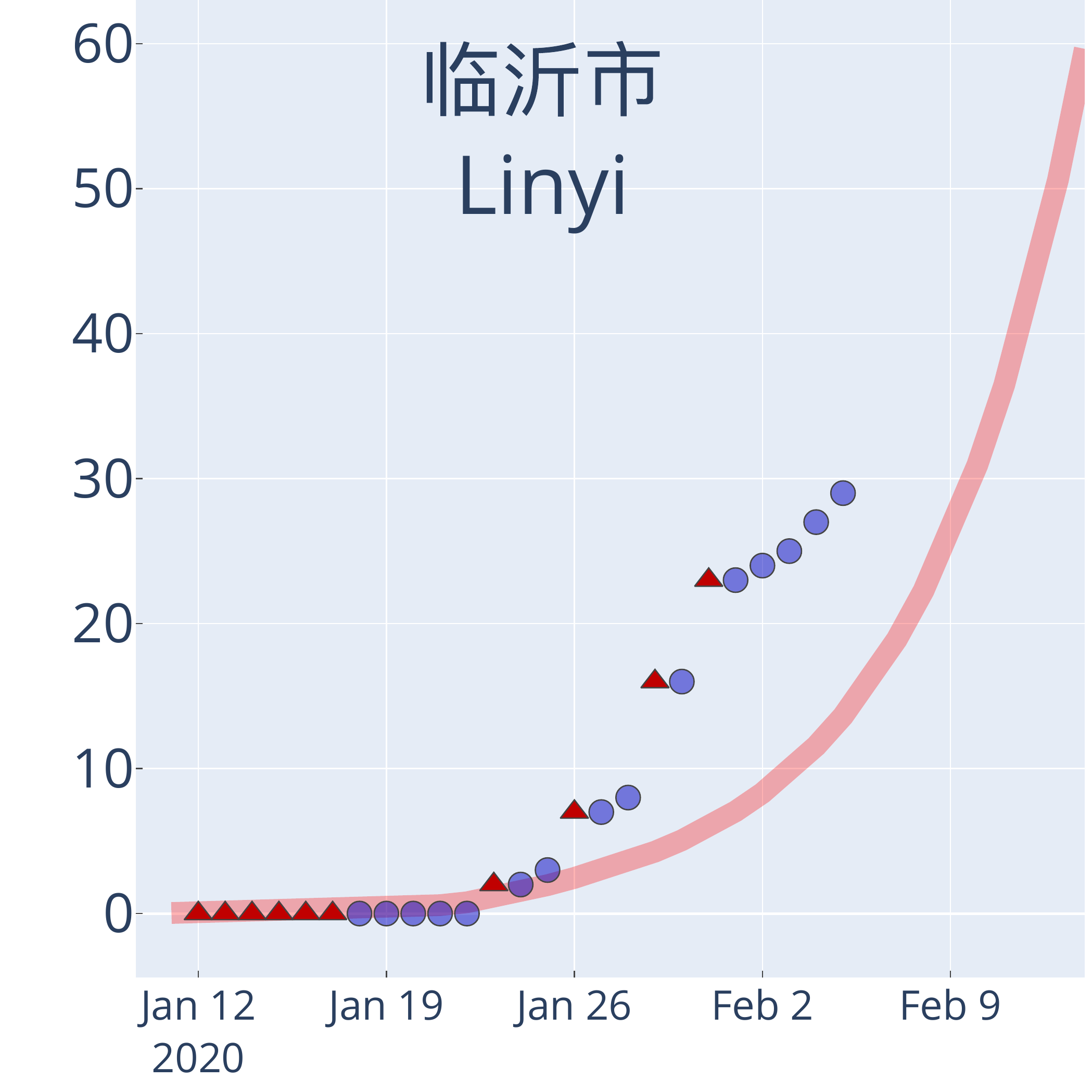}\hfill{}\includegraphics[width=0.24\textwidth]{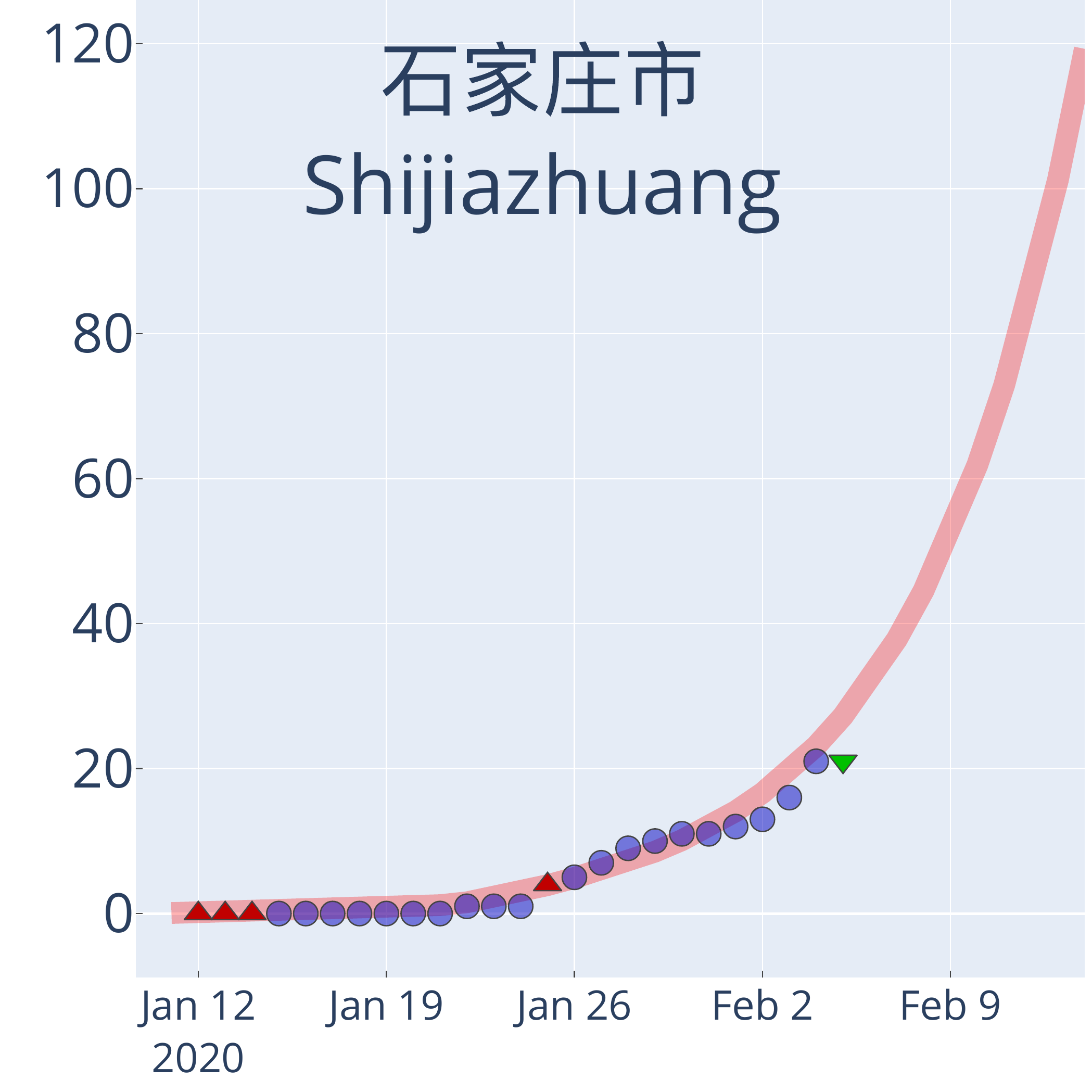}\hfill{}\includegraphics[width=0.24\textwidth]{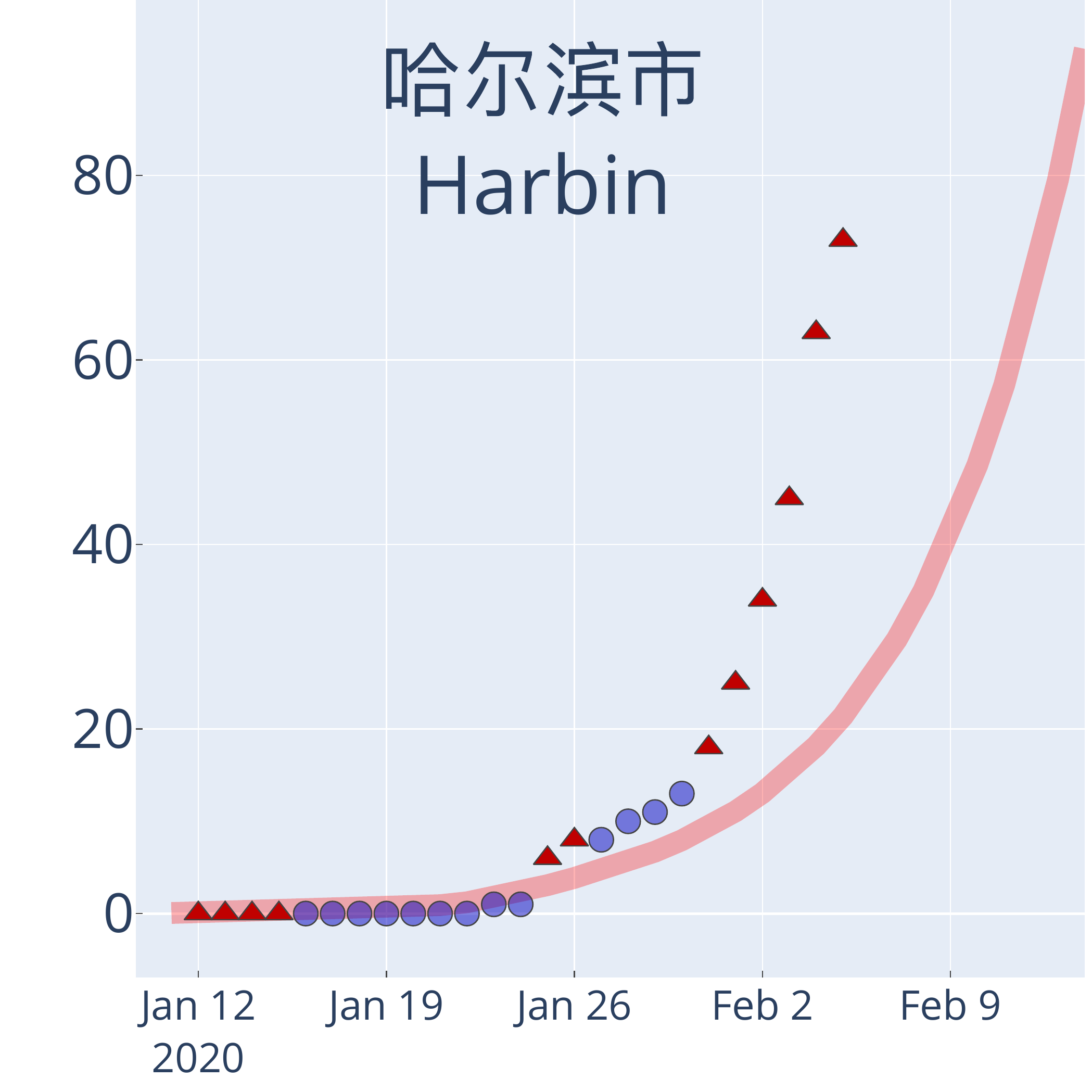}\hfill{}
\par\end{centering}
\begin{centering}
\hfill{}\includegraphics[width=0.24\textwidth]{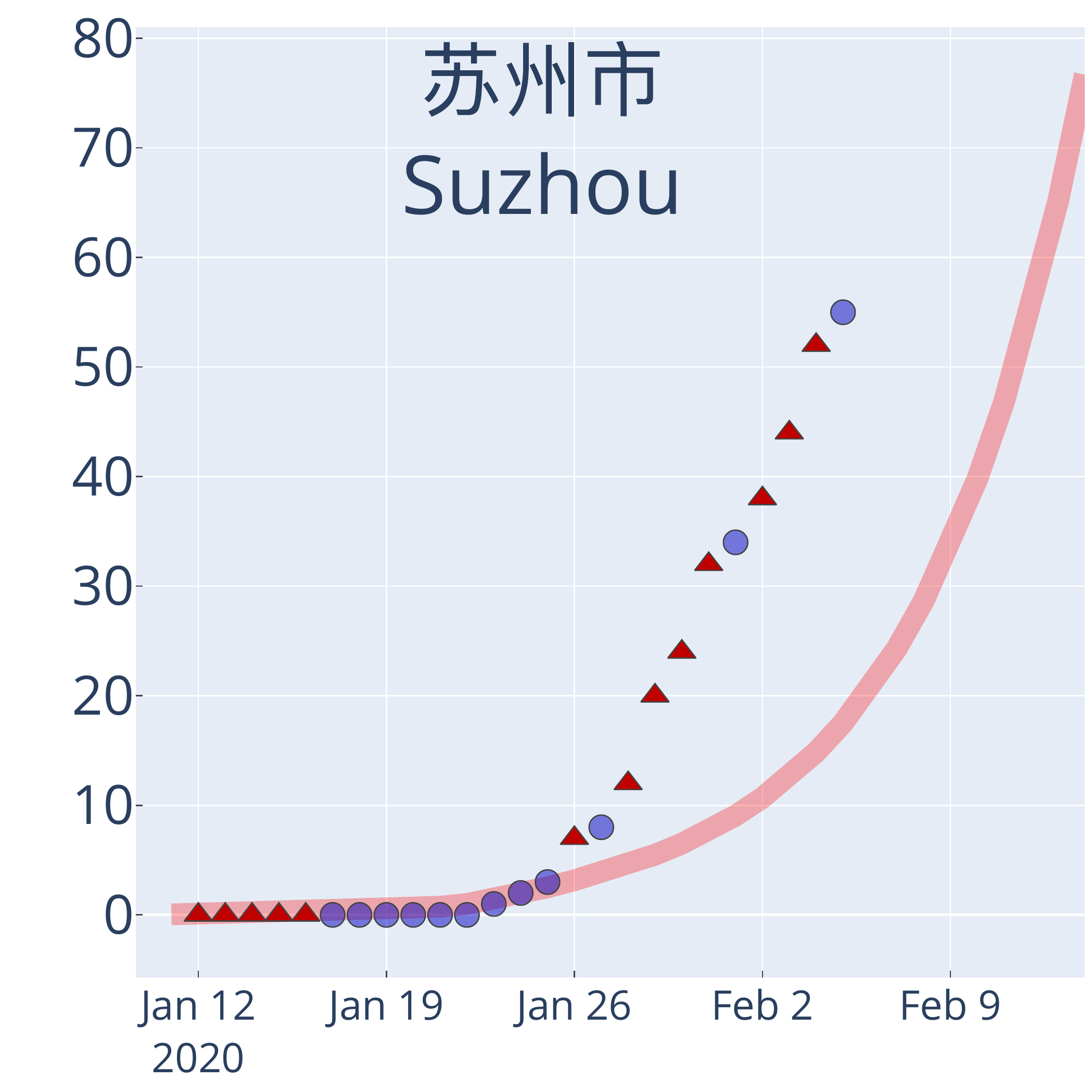}\hfill{}\includegraphics[width=0.24\textwidth]{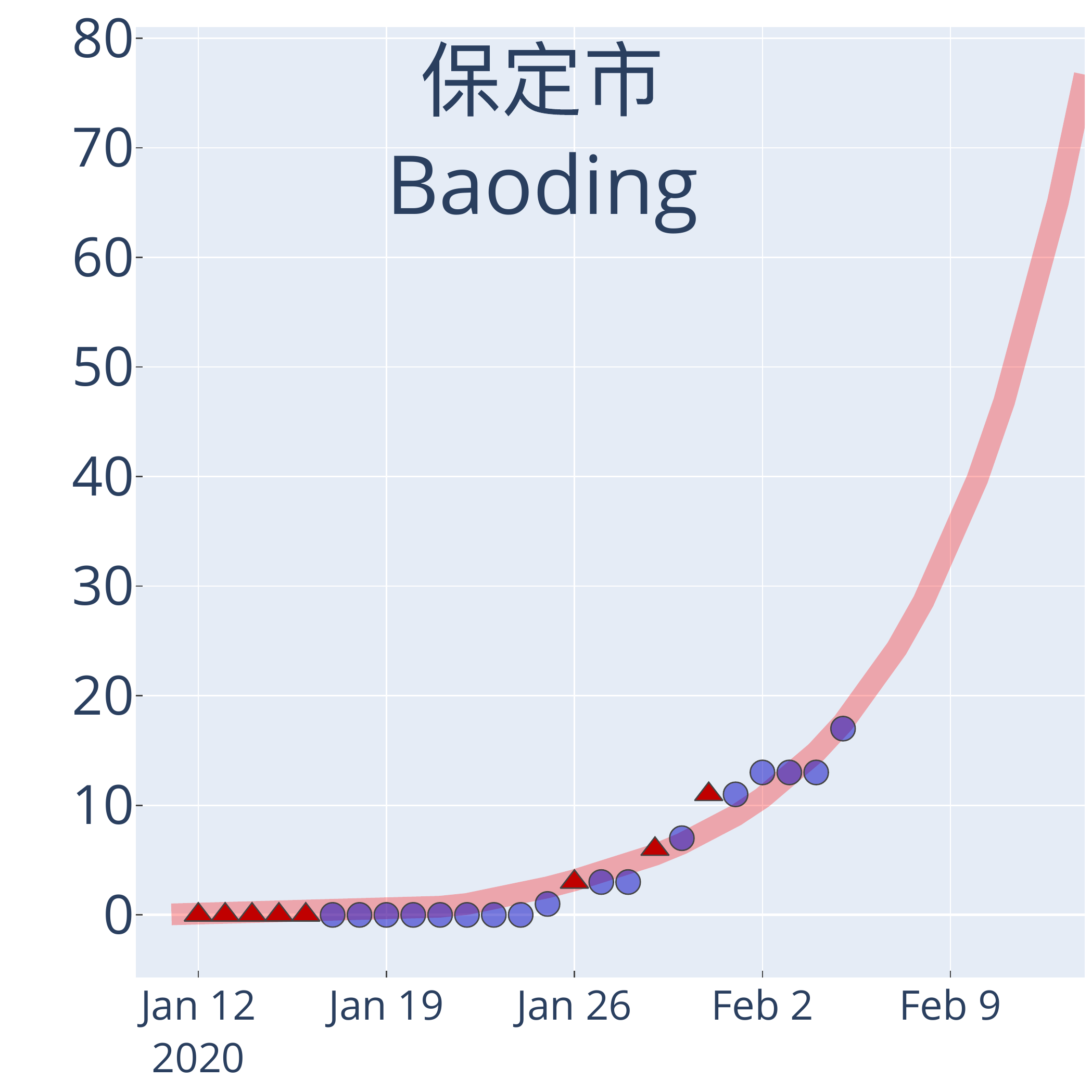}\hfill{}\includegraphics[width=0.24\textwidth]{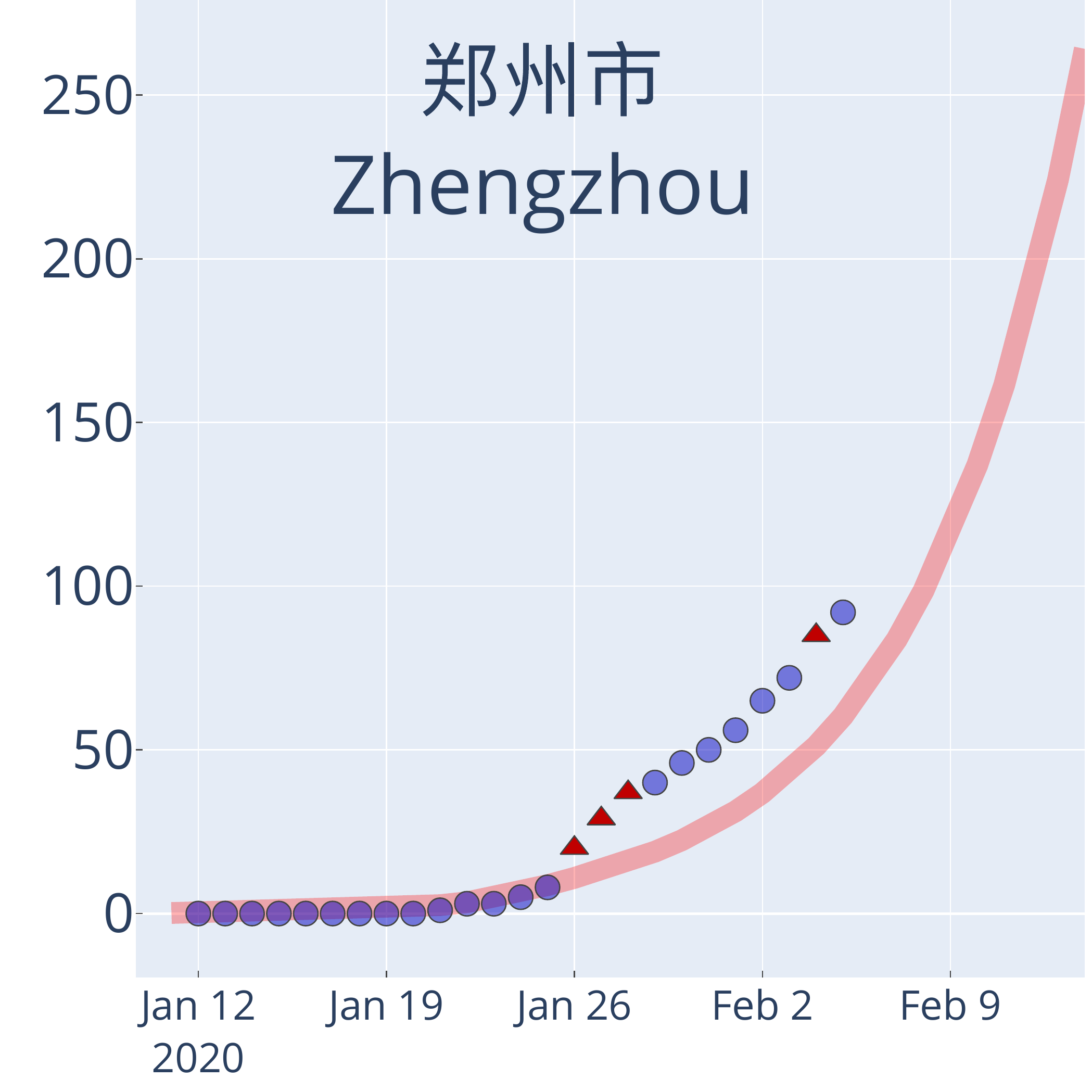}\hfill{}\includegraphics[width=0.24\textwidth]{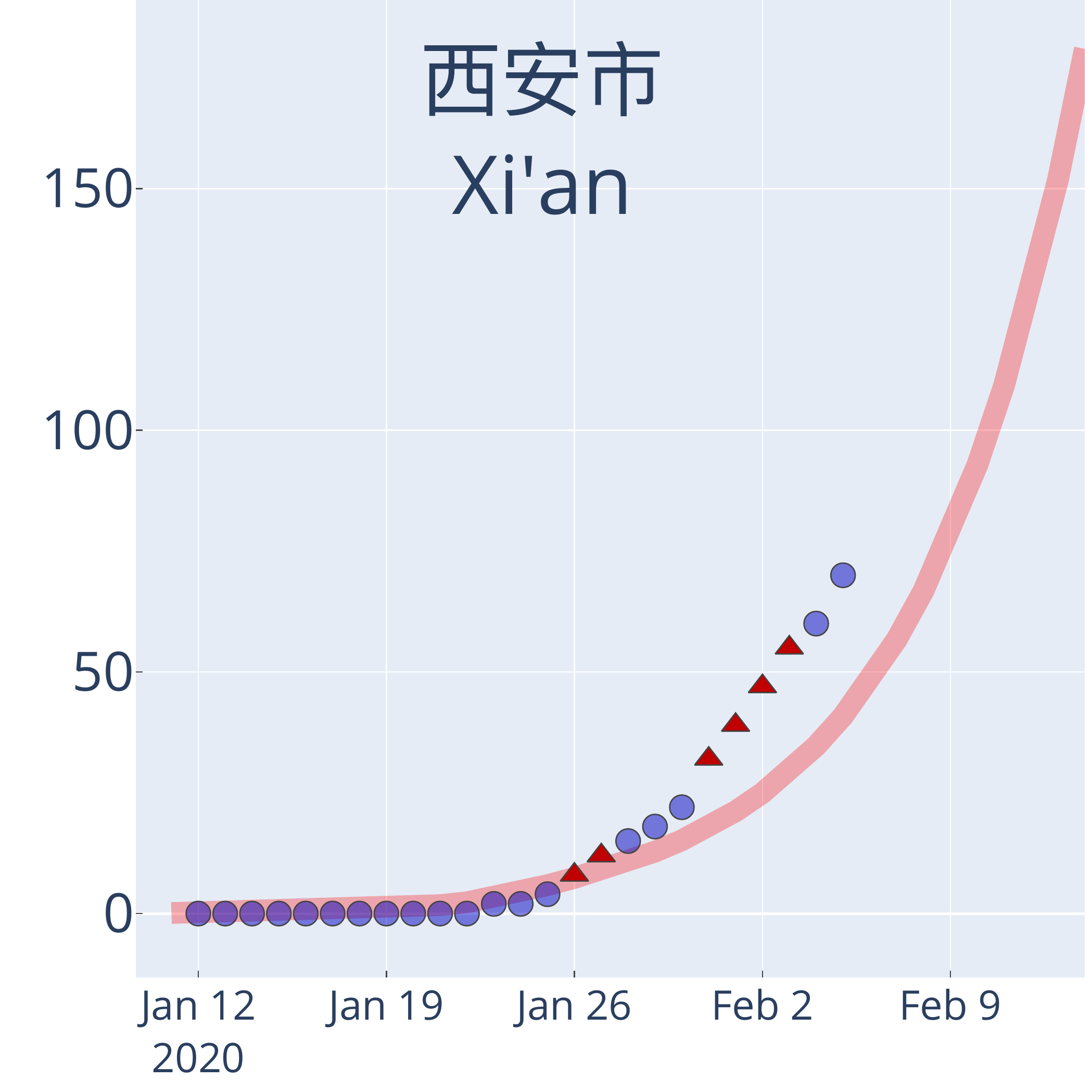}\hfill{}
\par\end{centering}
\begin{centering}
\hfill{}\includegraphics[width=0.24\textwidth]{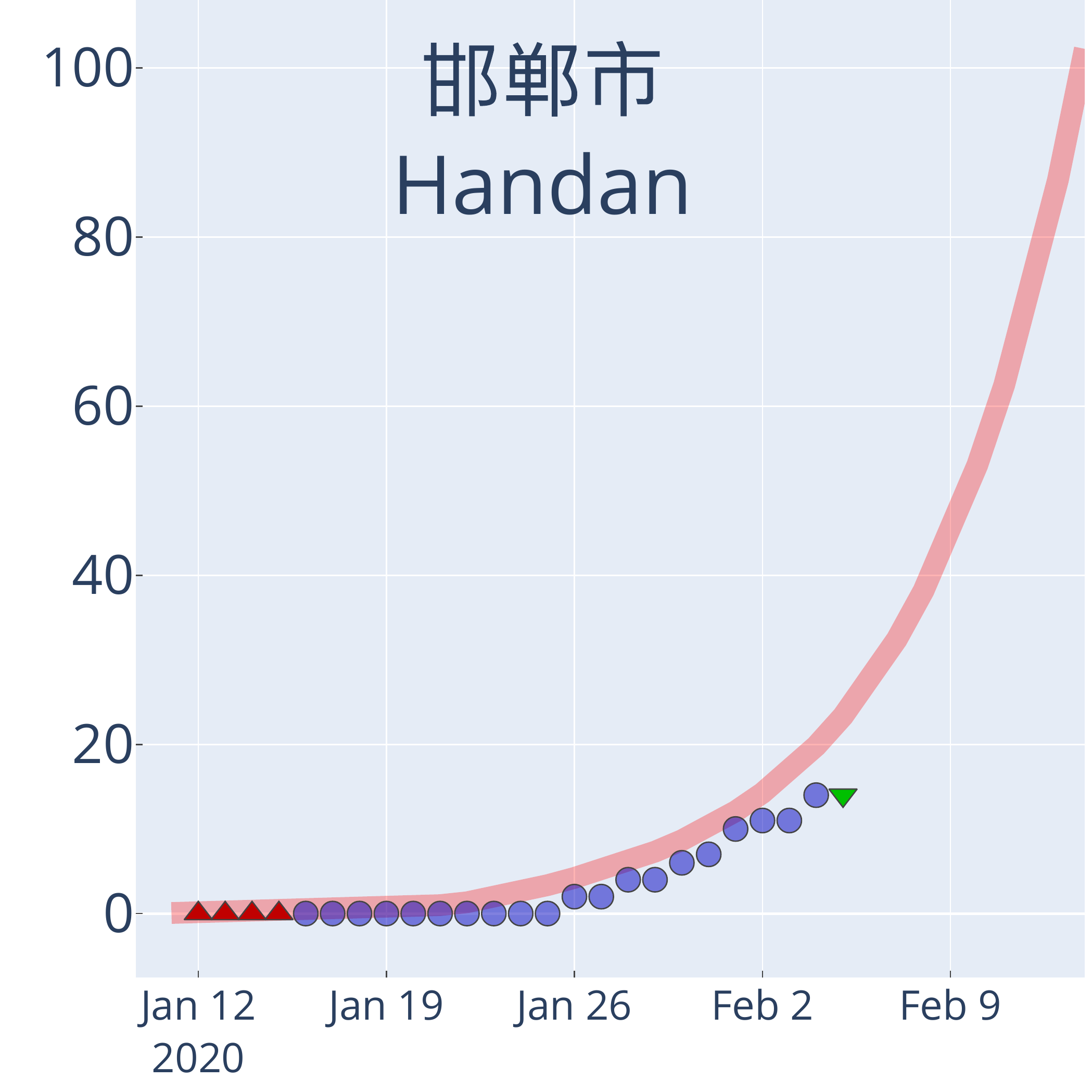}\hfill{}\includegraphics[width=0.24\textwidth]{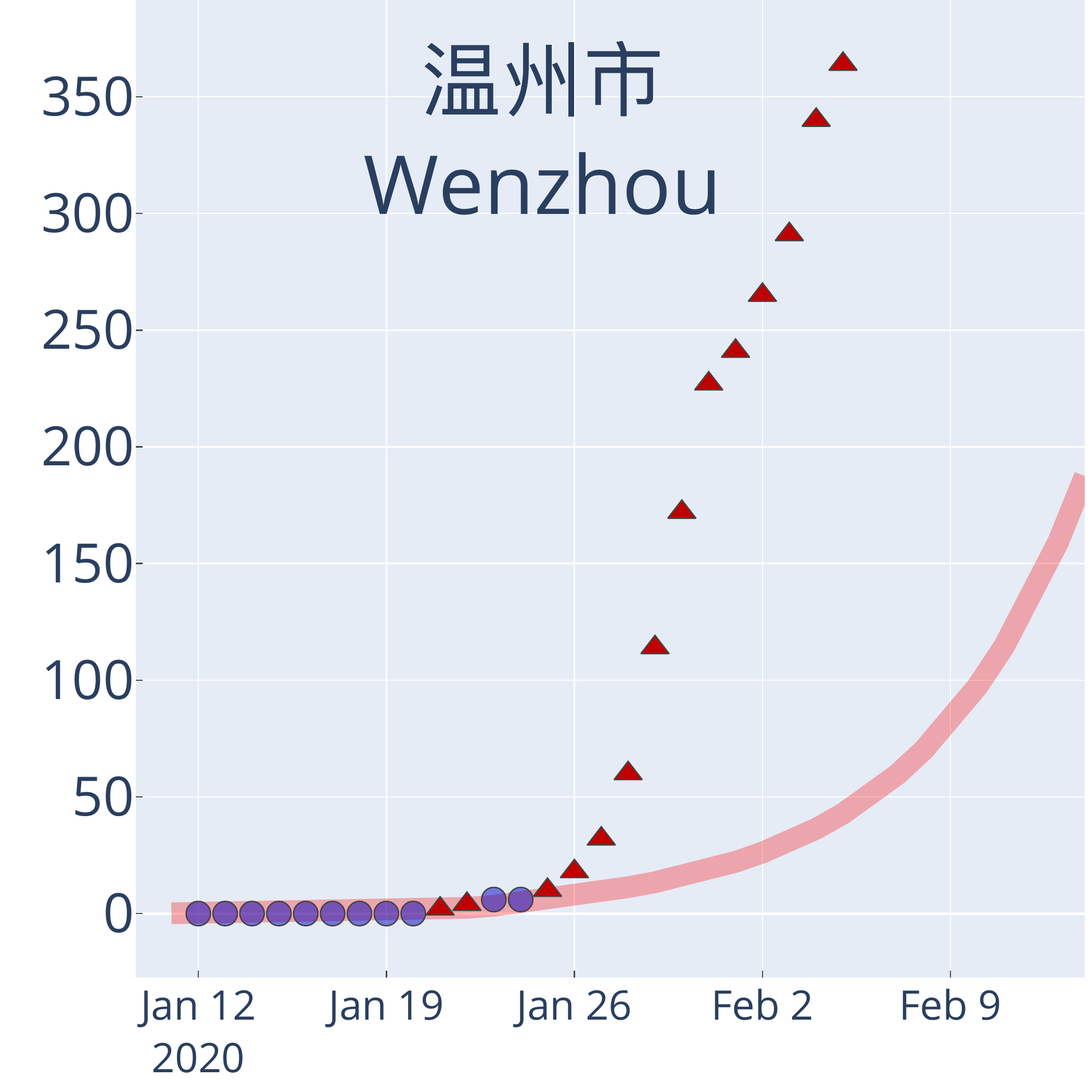}\hfill{}\includegraphics[width=0.24\textwidth]{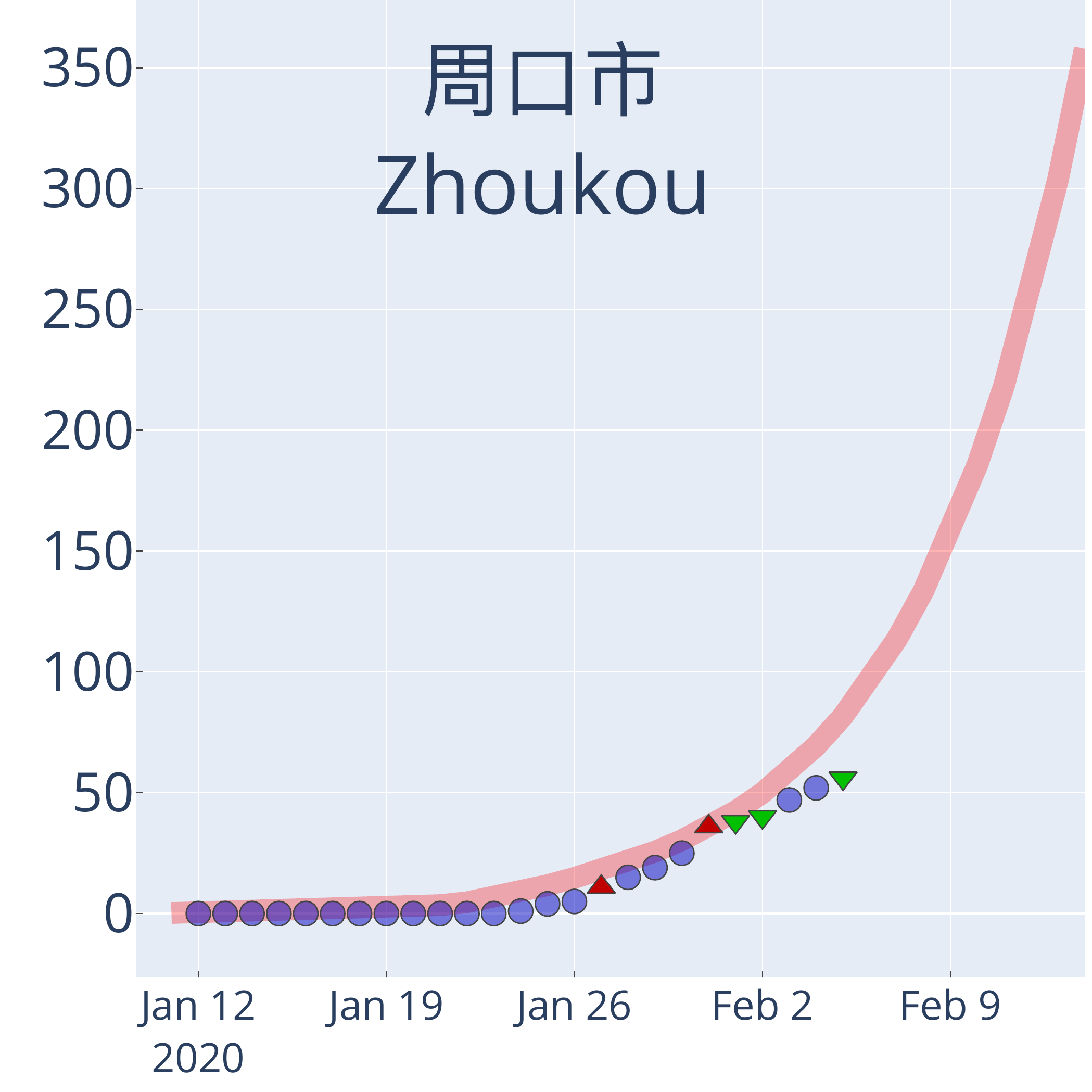}\hfill{}\includegraphics[width=0.24\textwidth]{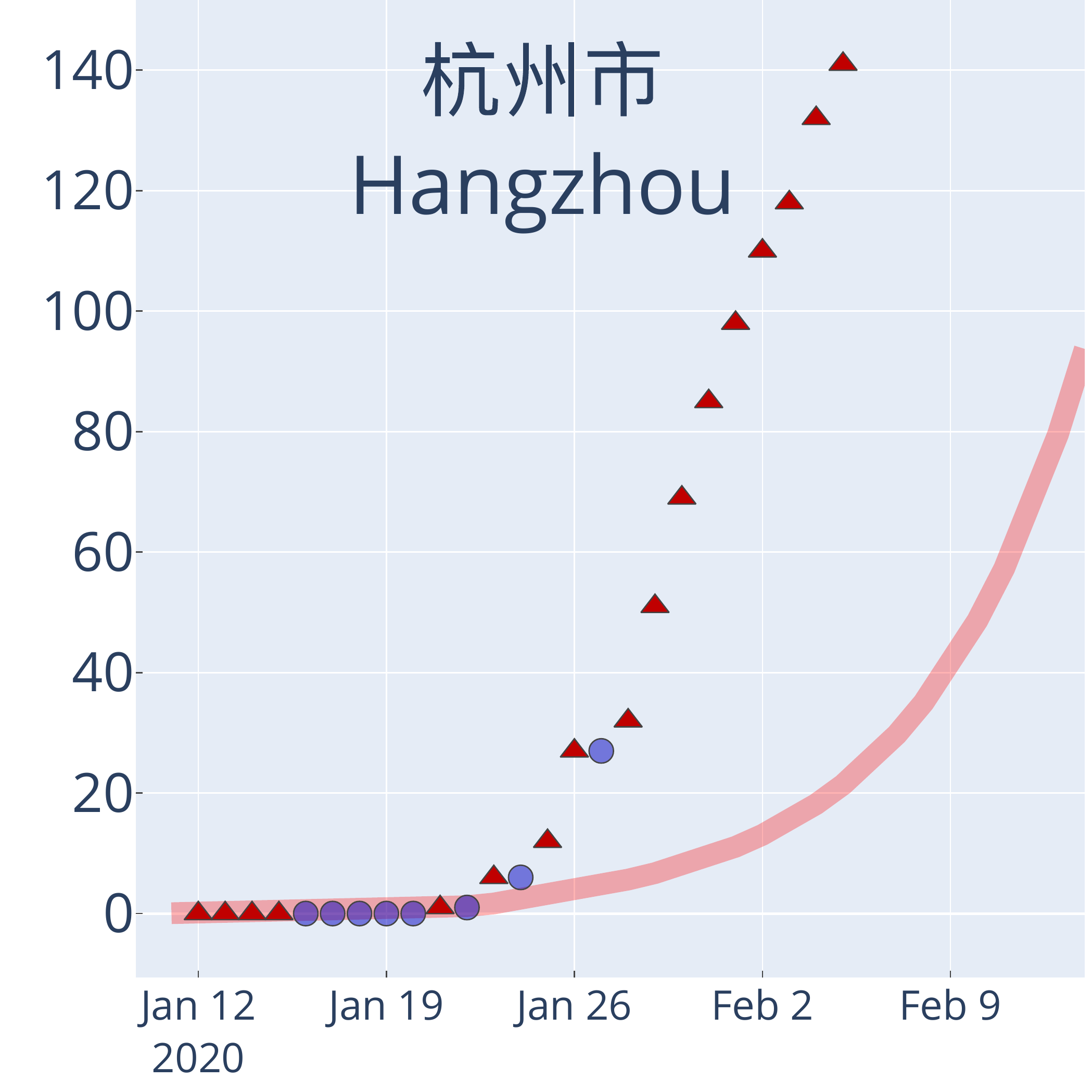}\hfill{}
\par\end{centering}
\caption{Simulation and forecasting of infections in major China cities and
comparison to accumulated cases. \label{fig:sim} See Figure \ref{fig:sim-sample}
for detailed interpretation of the marks and legends used in the plots.}
\end{figure*}

Simulation results for 80+ major China cities of strong connections
with Wuhan are available in the figures (Figure \ref{fig:sim}\nobreakdash-\ref{fig:sim-1})
at the end of this document. The simulation results suggest the spreading
of 2019-nCoV in China megapolitans (e.g. Beijing, Shanghai, Guangzhou
and Shenzhen) is exceeding the expectation of the overall SEIR model.
The model simulation matches the observation in a range of large China
cities, such as the capital cities, Shijiazhuang, Zhengzhou and Xi'an
of the middle west provinces. However, the spreading rate is greater
than the expectation in cities connected to Wuhan closely.

On the other hand, for satellite cities with closest connections with
Wuhan the recorded infection cases are significantly lower than expected.
For Wuhan herself, the record is lower than what has been expected,
in terms of several orders of magnitude. We will discuss possible
explanations in the next section.

\section{Conclusion, Limits and Future Research}

In this study, we adopt a robust model fitting method, random sample
consensus, which has enabled us to establish stable SEIR model families
and identify outliers in the infection data of 2019-nCoV. The random
sample consensus is made possible by employing traffic network dynamics
in the SEIR model to handle the infection in cities connected to Wuhan.

\subsection{Improve Data Quality}

Domestic and international airline traffic: We did \emph{not} include
international cities and air-traffic in the current analysis. One
reason is that our focus is on the China populous cities, while the
volume of travellers by train vastly exceeds that by air. The airline
data can be added in future research. 

Traffic networks: The current transportation matrices K 's have only
one row of values corresponding to the traveller's departing Wuhan.
This would not be a major issue in the period when the first generation
of human-to-human transmission is our main concern. The inter-city
traffic would play a more significant role in the spreading of the
virus after cities other than Wuhan had accumulated an infected population.

Early infection data: a phenomenon demanding explanation is that:
the SEIR has failed to capture the variations of the infection data
within Wuhan and nearby cities. What is fairly surprising is that
the SEIR model overestimated the infection numbers. This is counter-intuitive
because it is those cities that are mostly affected by the virus and
have a large number of infections. This could be the result of poor
data quality, or the spreading mode has changed in different stages
of the spreading.

\subsection{Modelling Tools}

We used SEIR model to represent the characteristics of the infection
data. The model is effective and simple to fit, thanks to the simplicity
of the parameter structure in the model (3 only). On the other hand,
ODE based modelling is simultaneously stiff and sensitive. Modern
end-to-end learning based models can be considered in future research.

\begin{figure*}[t]
\begin{centering}
\hfill{}\includegraphics[width=0.24\textwidth]{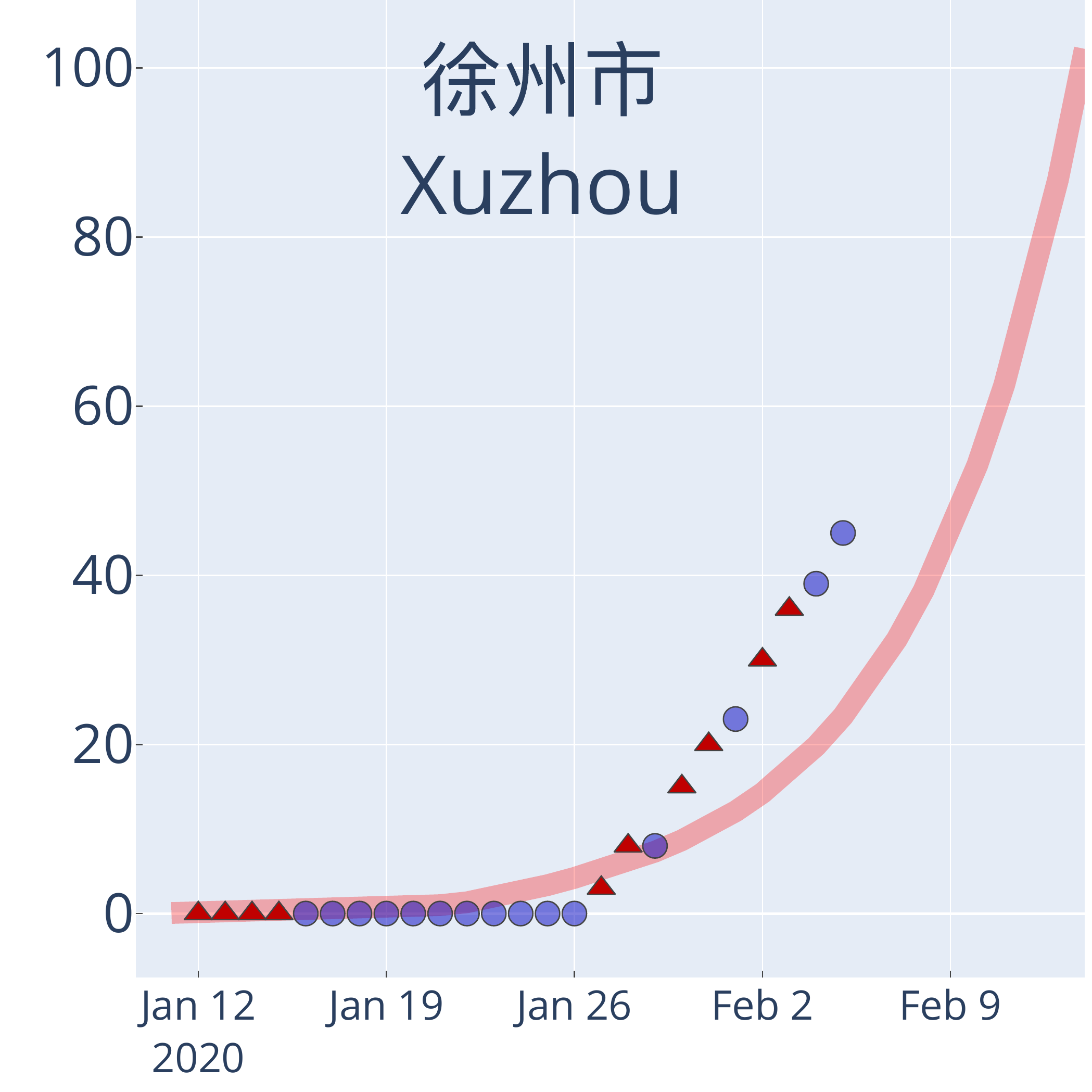}\hfill{}\includegraphics[width=0.24\textwidth]{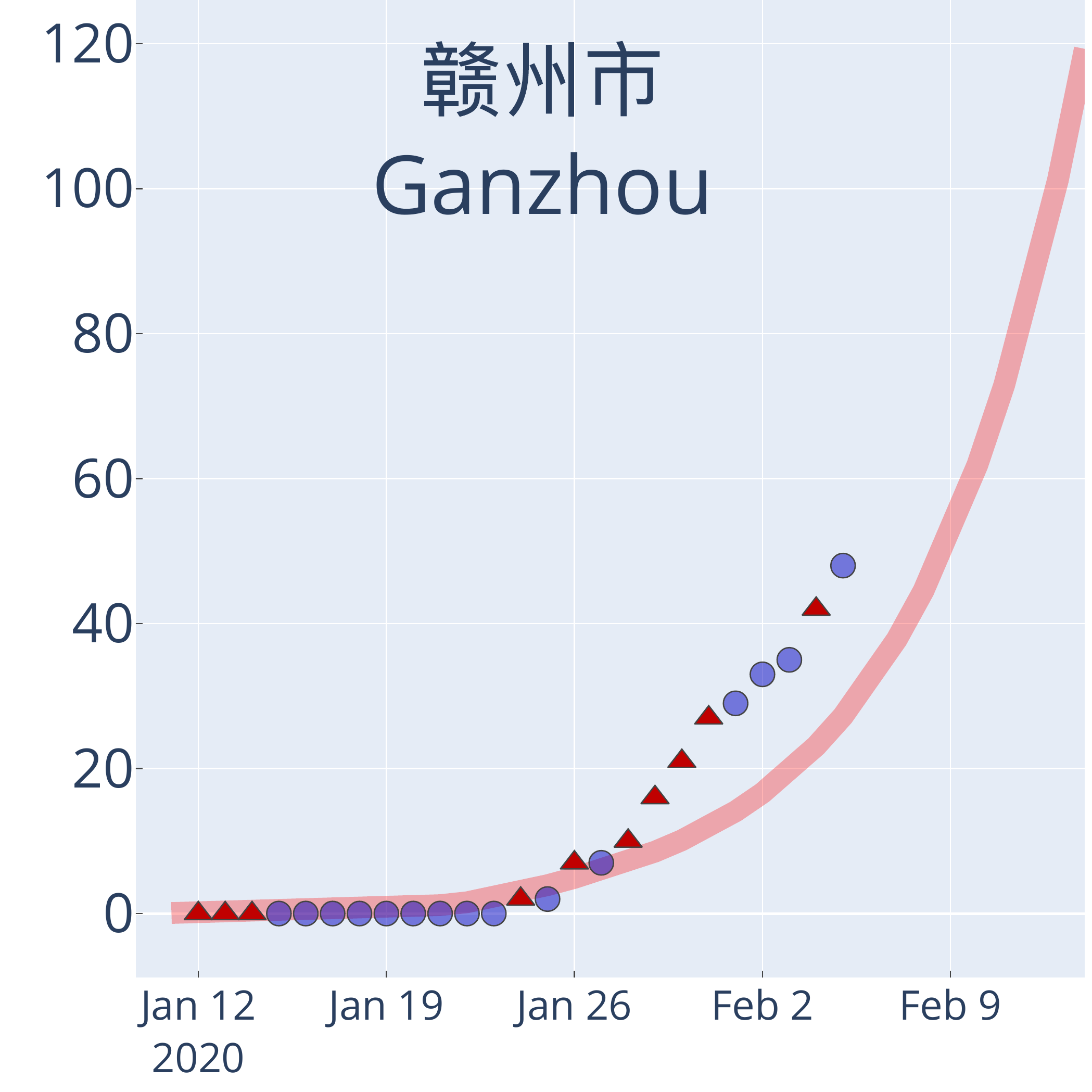}\hfill{}\includegraphics[width=0.24\textwidth]{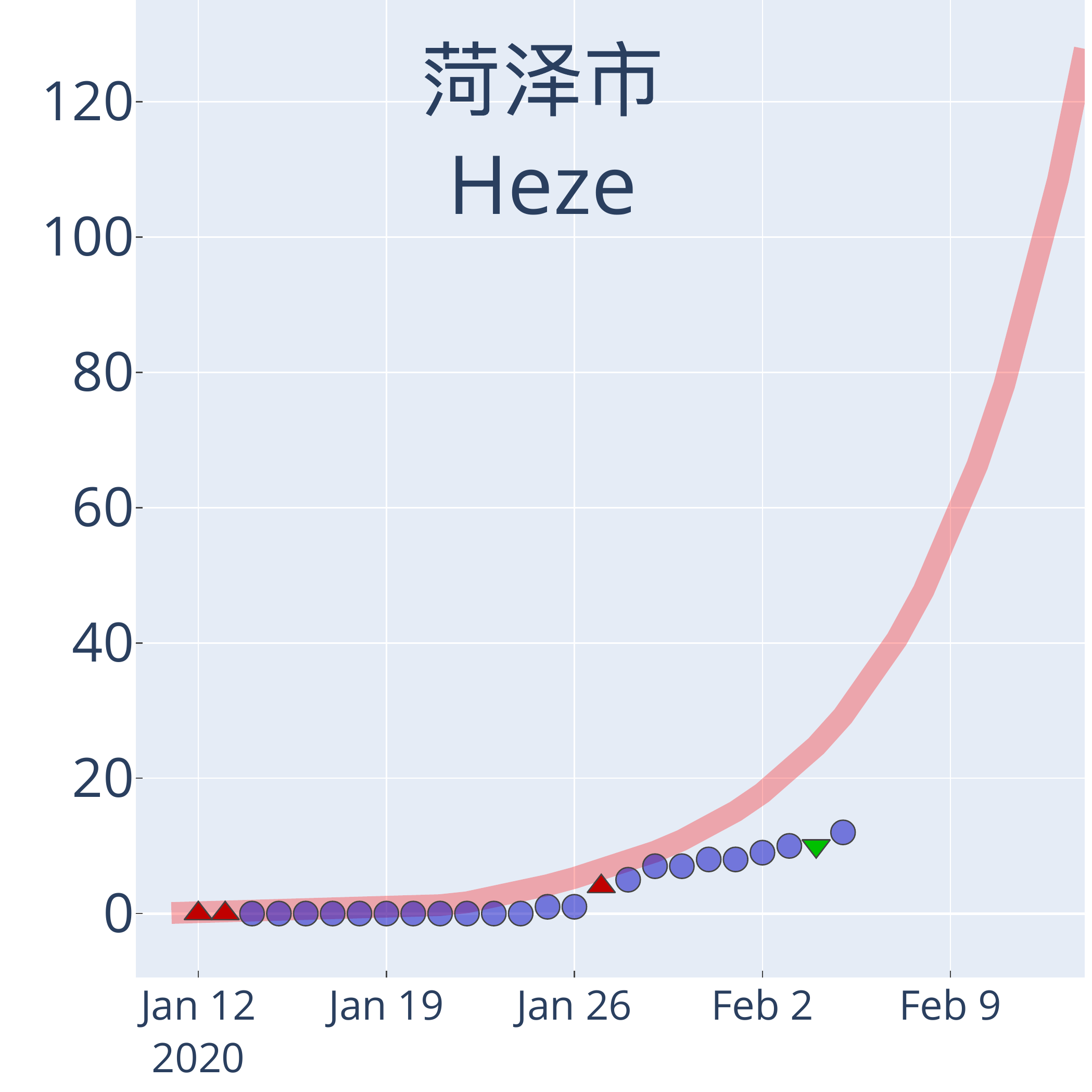}\hfill{}\includegraphics[width=0.24\textwidth]{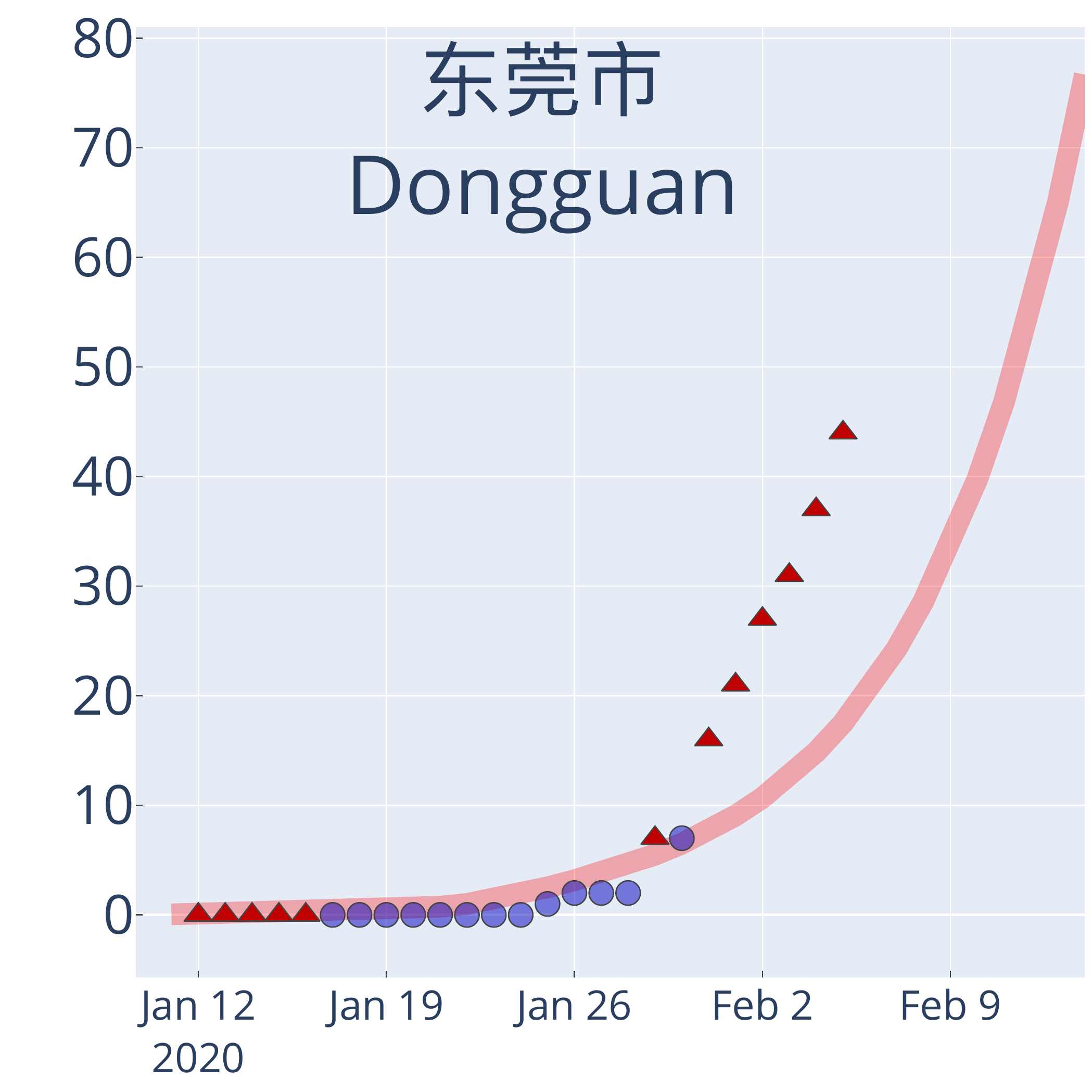}\hfill{}
\par\end{centering}
\begin{centering}
\hfill{}\includegraphics[width=0.24\textwidth]{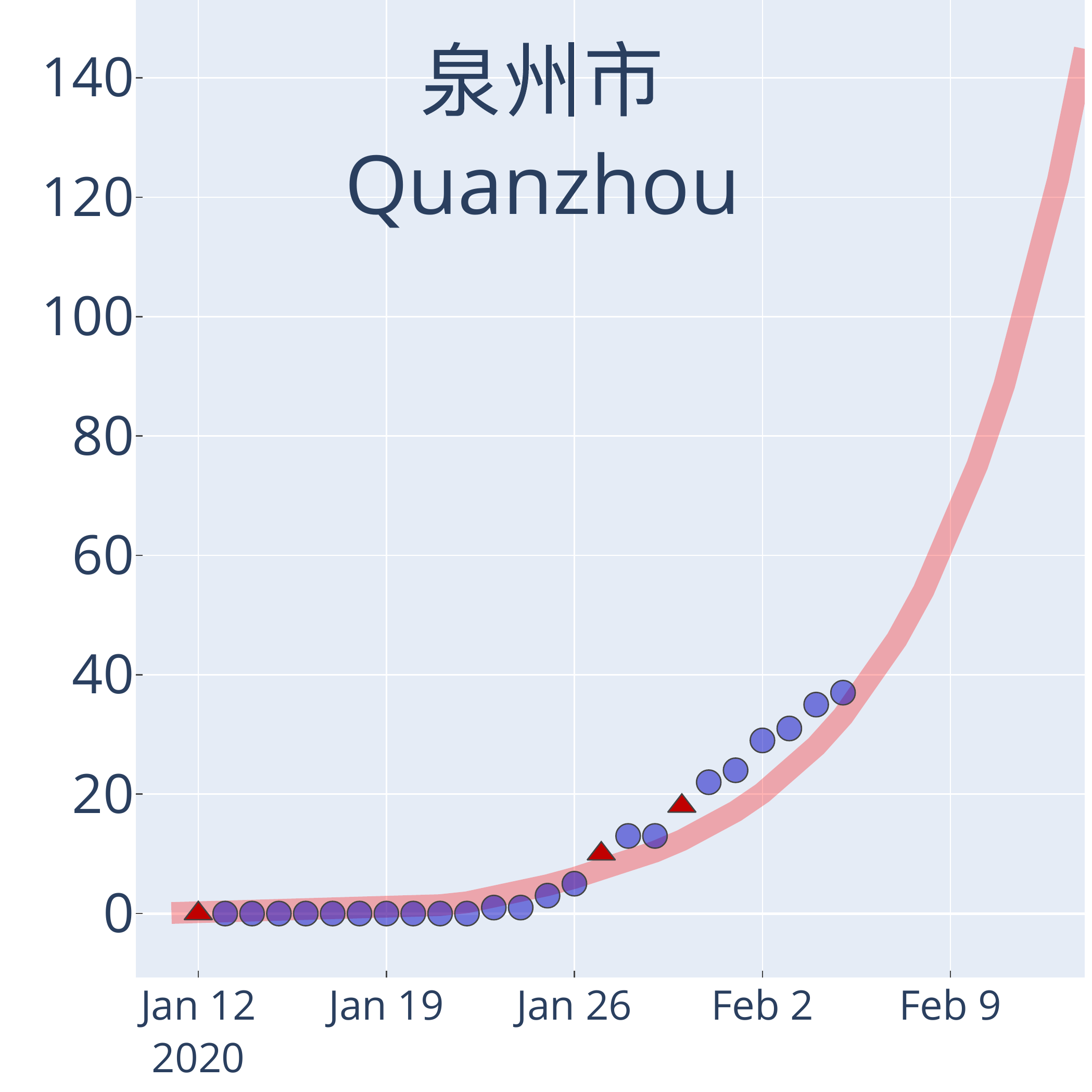}\hfill{}\includegraphics[width=0.24\textwidth]{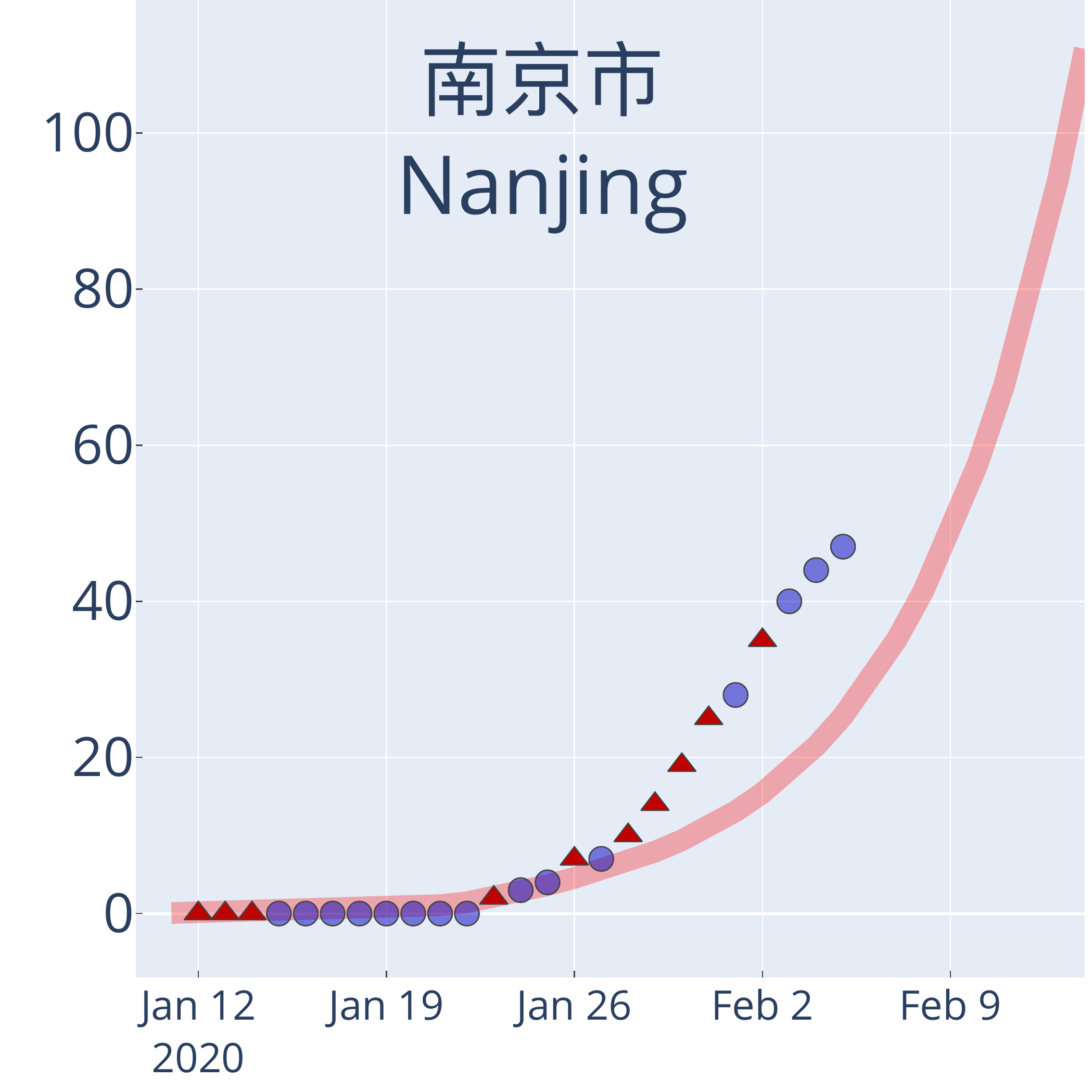}\hfill{}\includegraphics[width=0.24\textwidth]{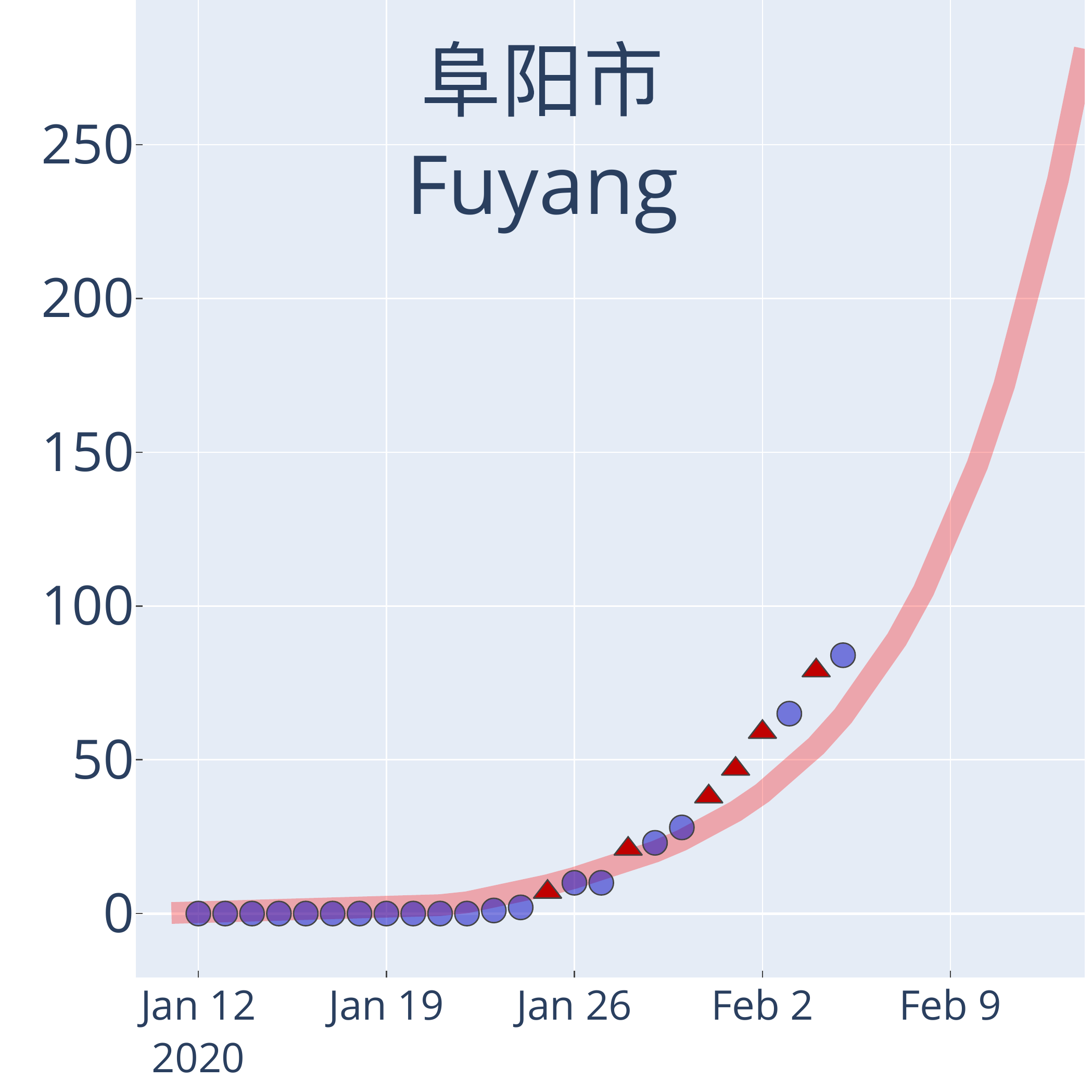}\hfill{}\includegraphics[width=0.24\textwidth]{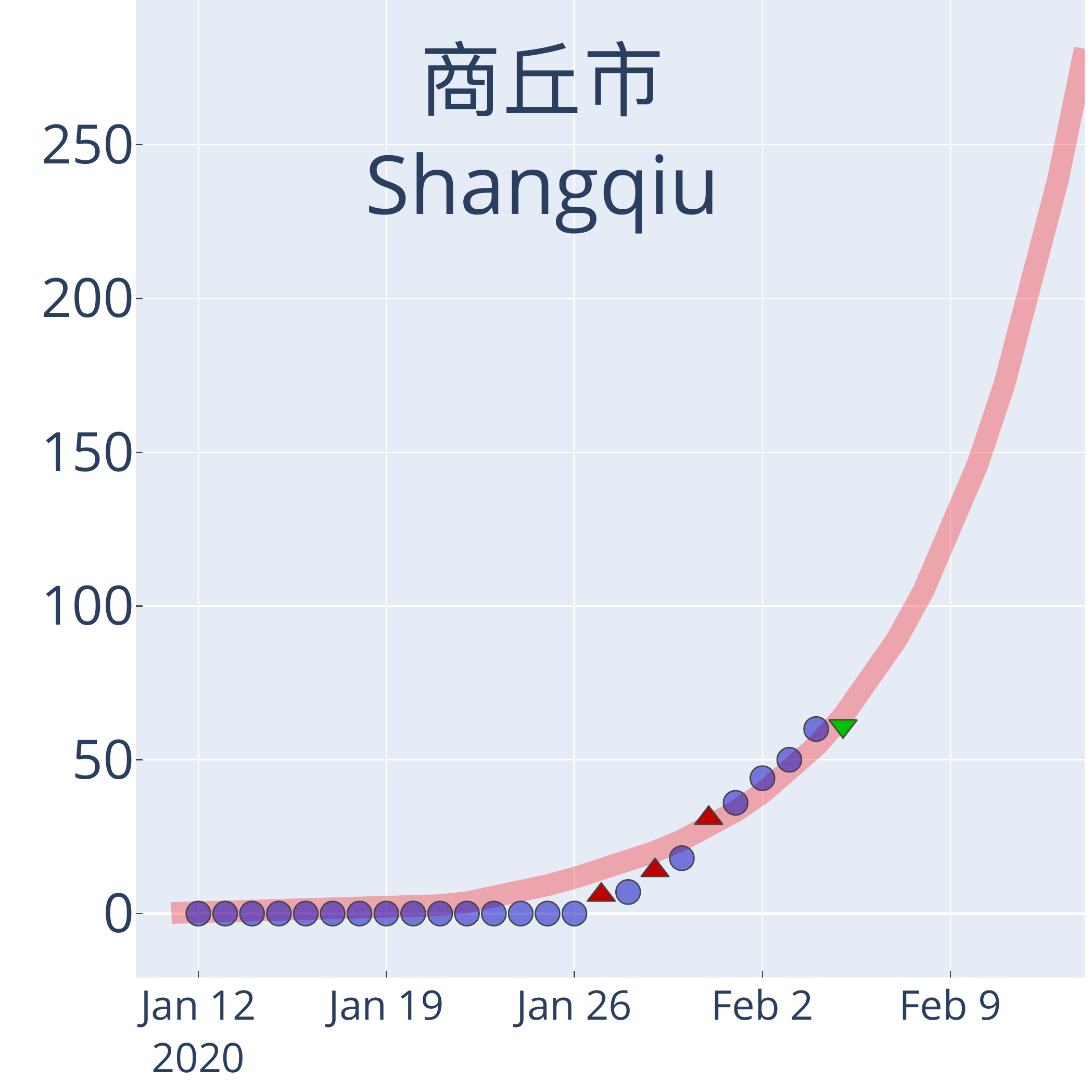}\hfill{}
\par\end{centering}
\begin{centering}
\hfill{}\includegraphics[width=0.24\textwidth]{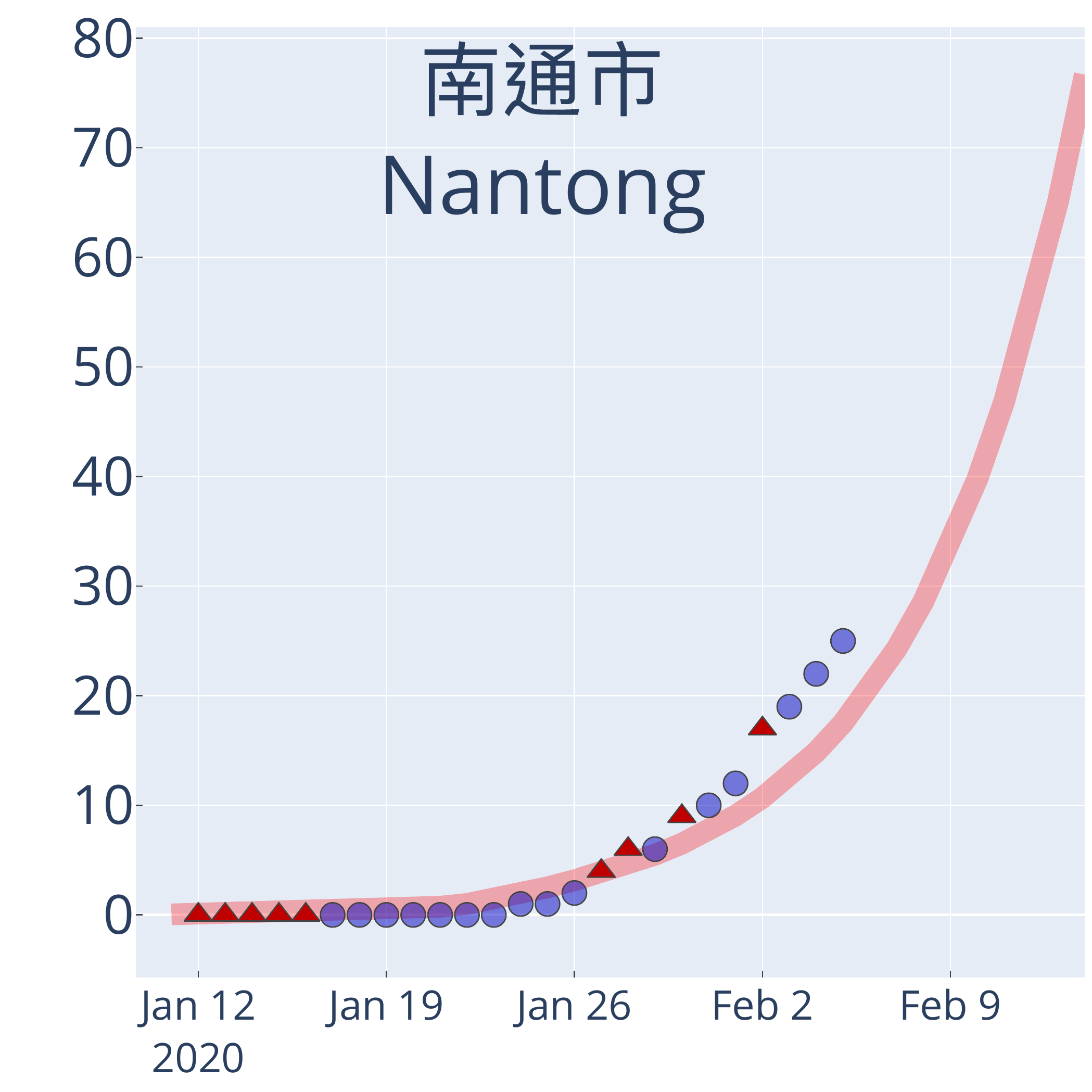}\hfill{}\includegraphics[width=0.24\textwidth]{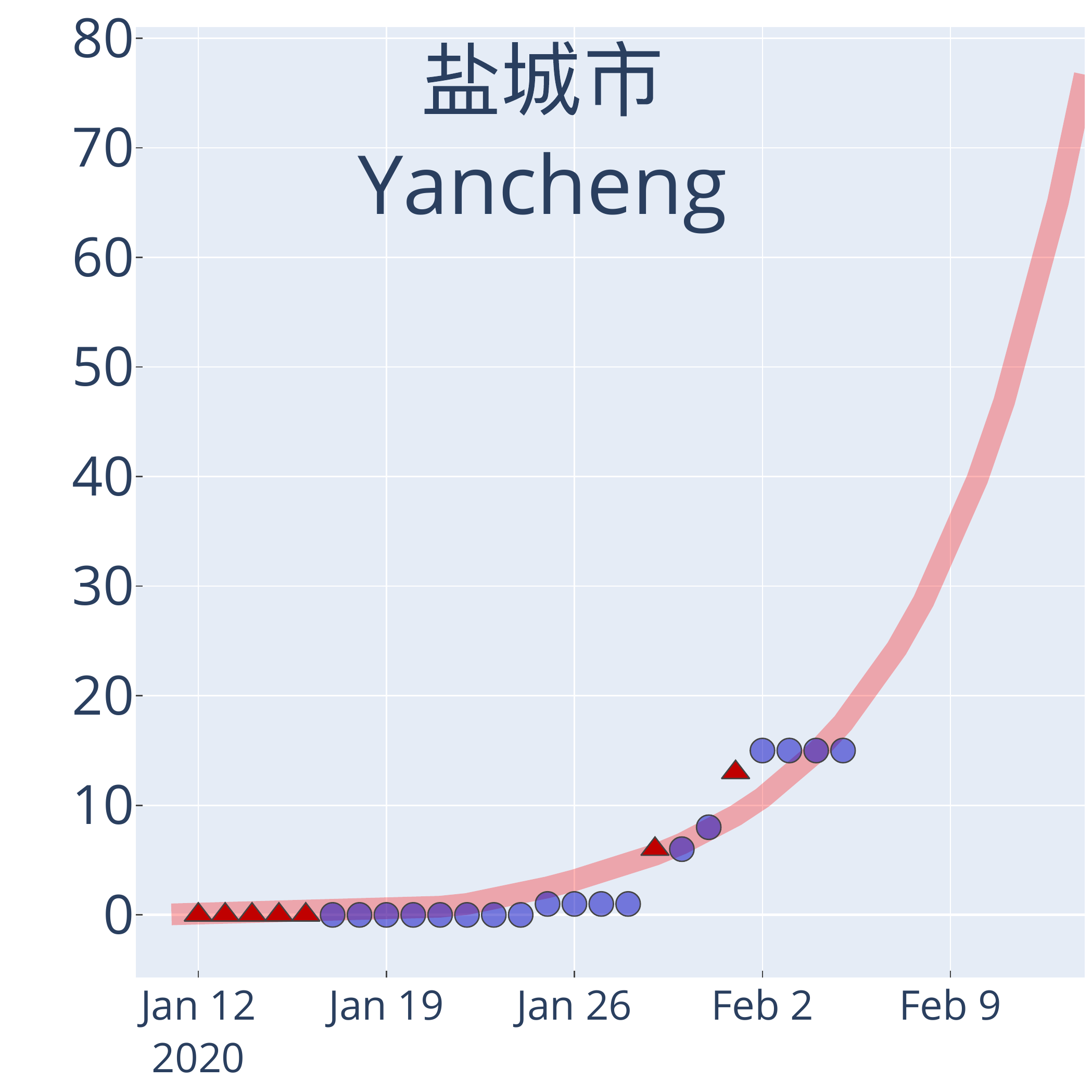}\hfill{}\includegraphics[width=0.24\textwidth]{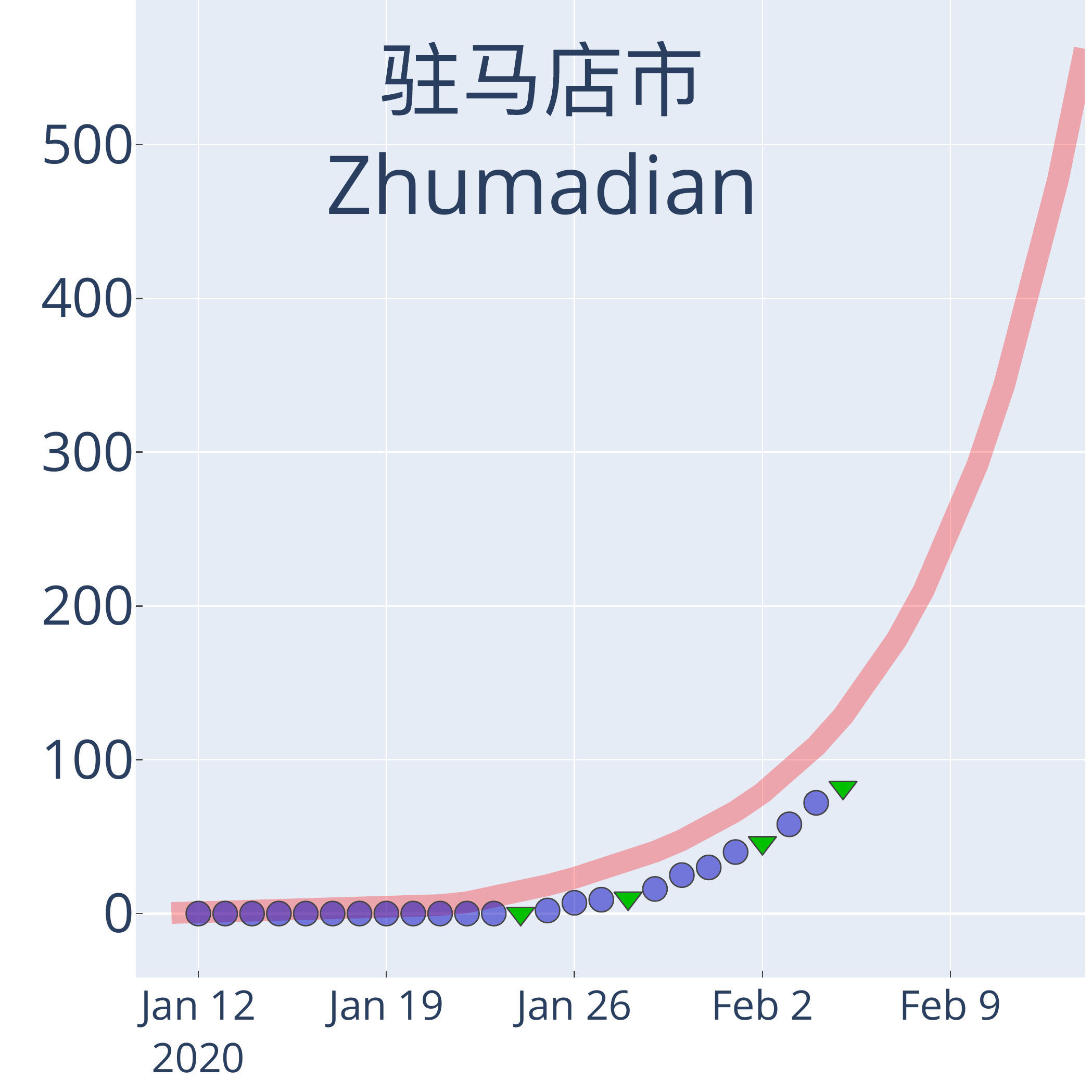}\hfill{}\includegraphics[width=0.24\textwidth]{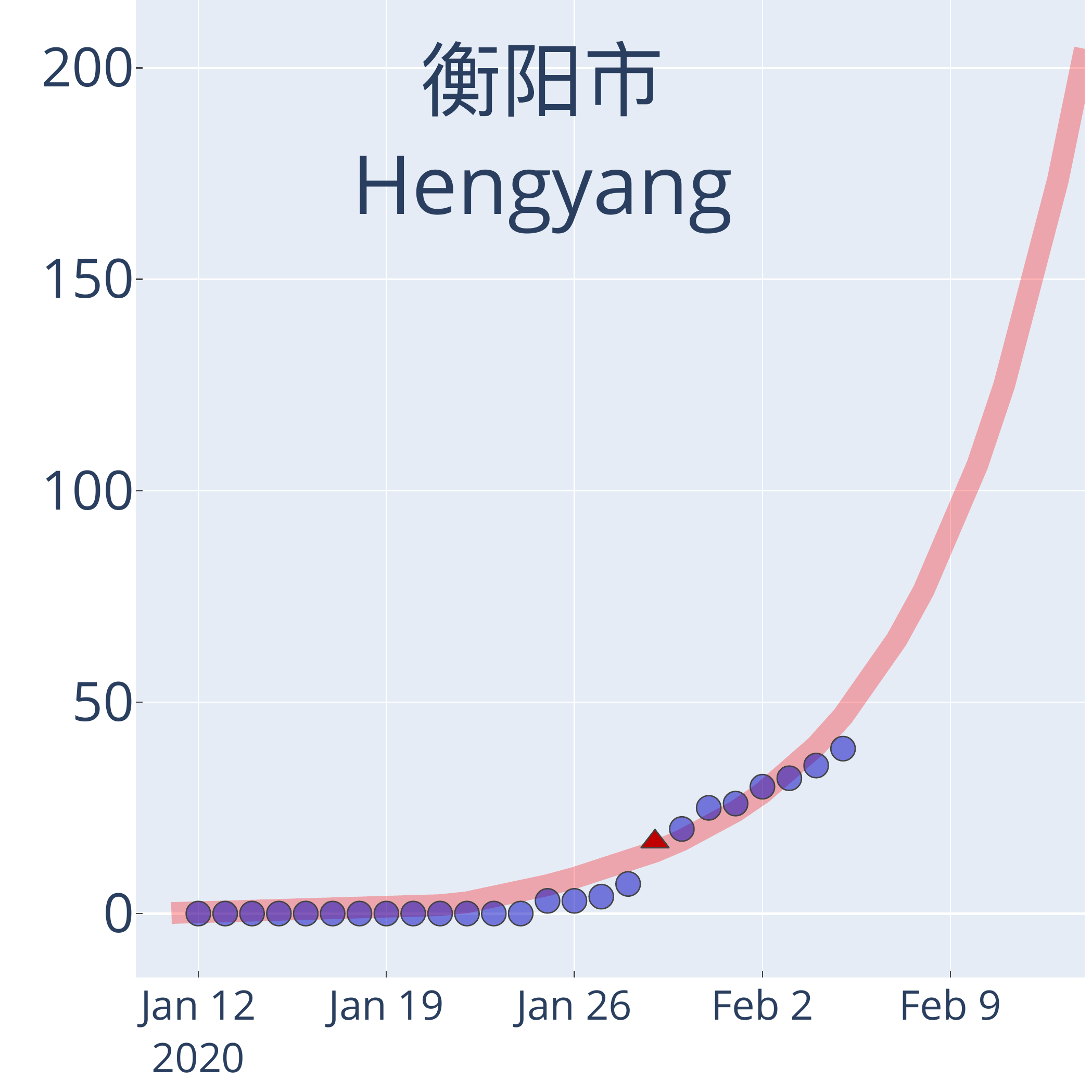}\hfill{}
\par\end{centering}
\begin{centering}
\hfill{}\includegraphics[width=0.24\textwidth]{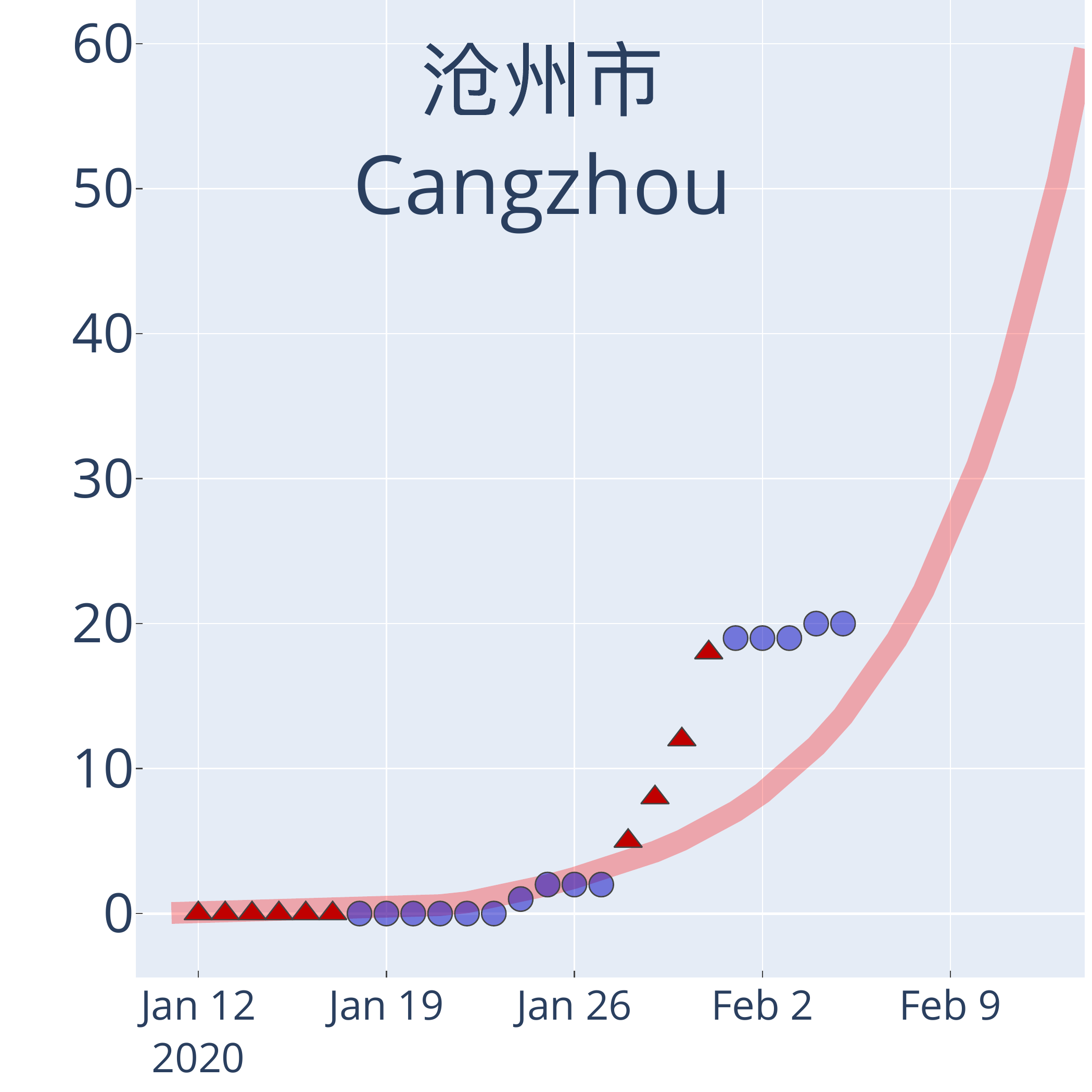}\hfill{}\includegraphics[width=0.24\textwidth]{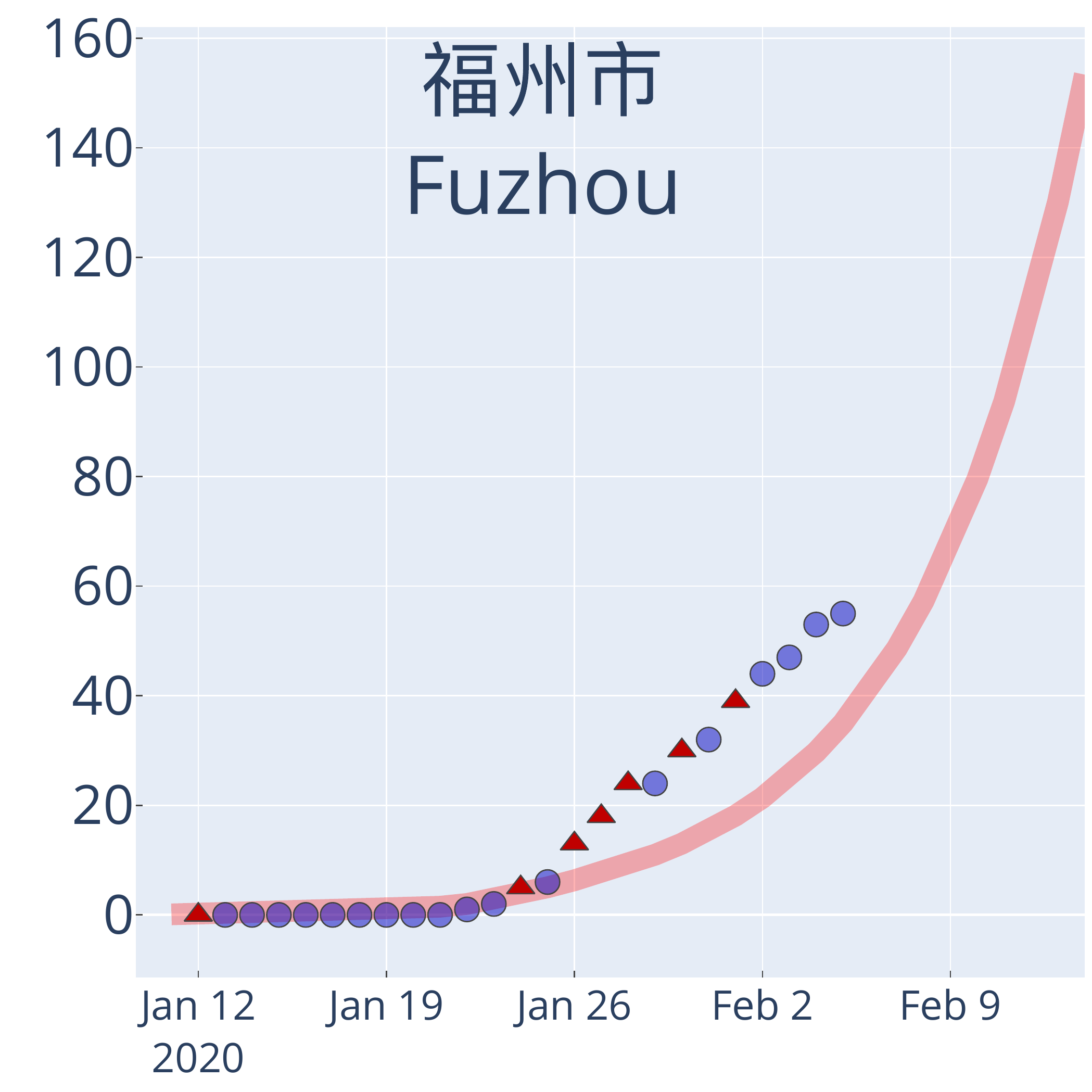}\hfill{}\includegraphics[width=0.24\textwidth]{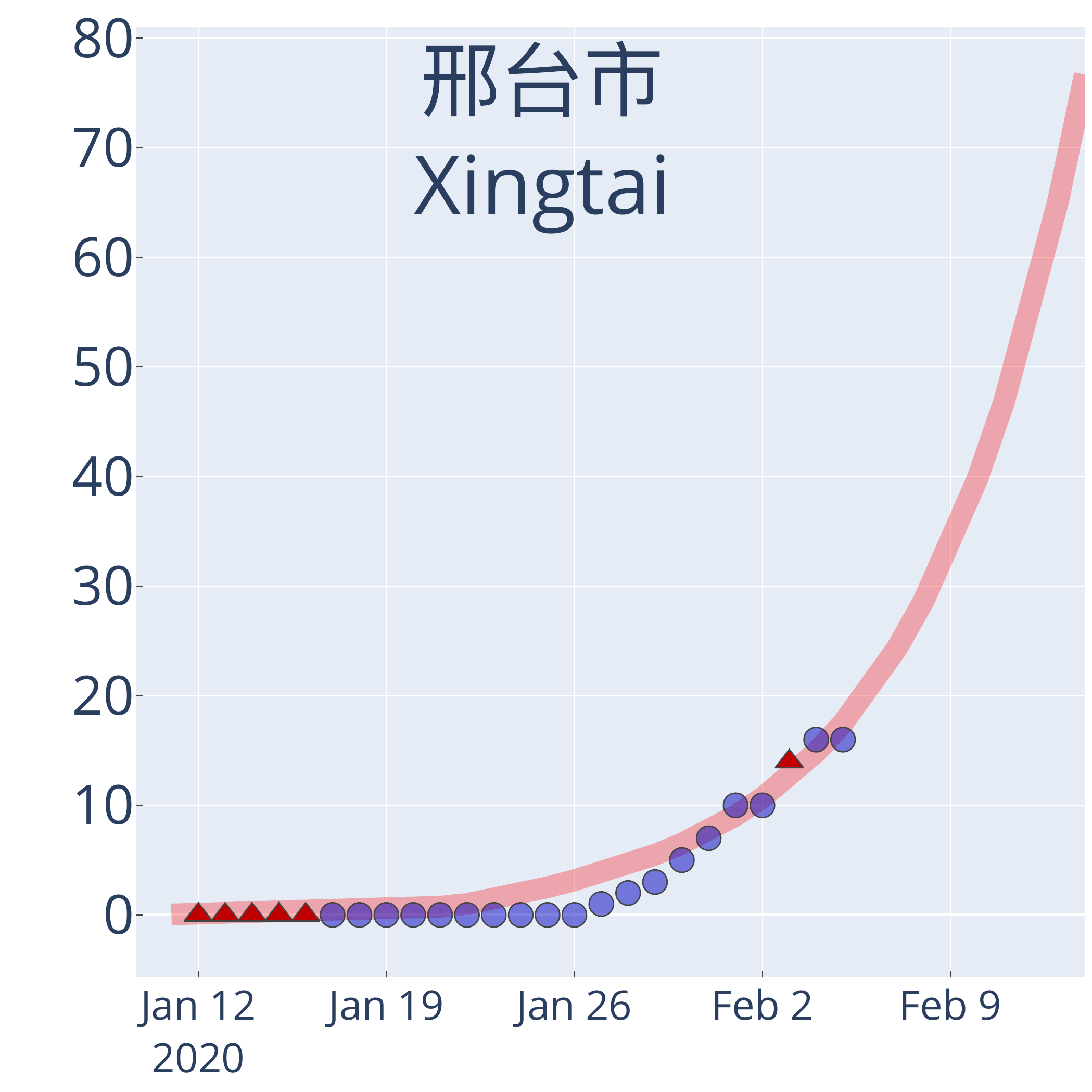}\hfill{}\includegraphics[width=0.24\textwidth]{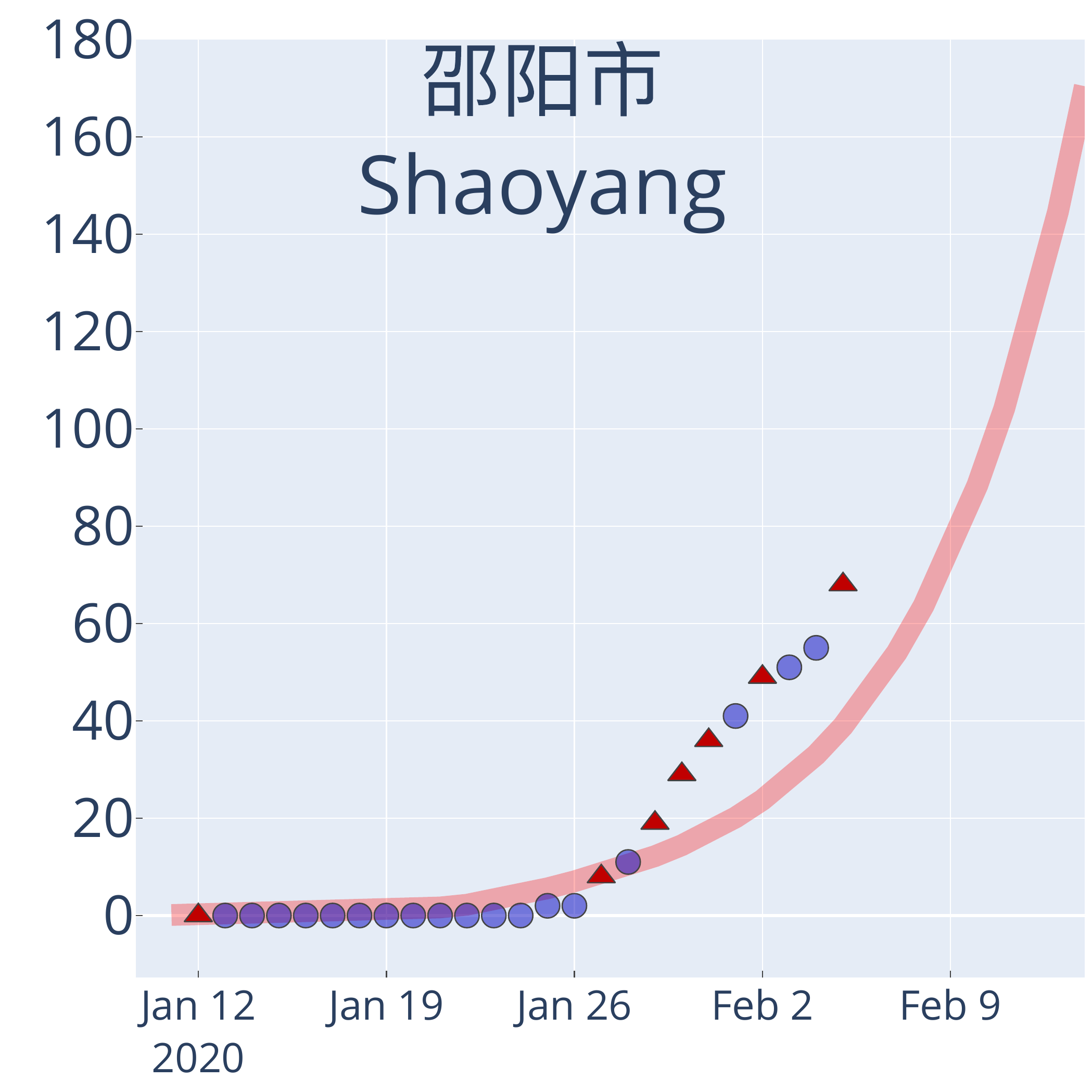}\hfill{}
\par\end{centering}
\begin{centering}
\hfill{}\includegraphics[width=0.24\textwidth]{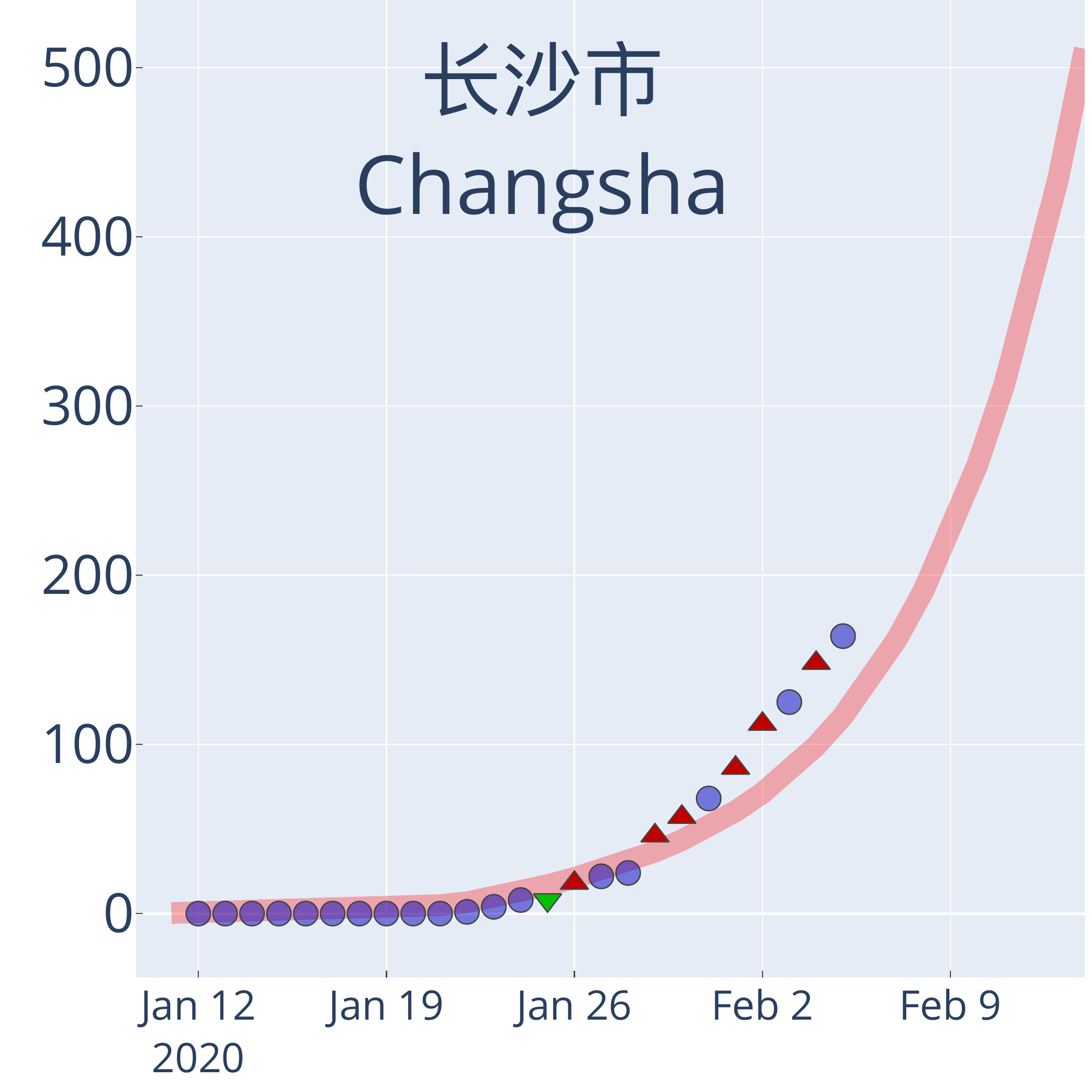}\hfill{}\includegraphics[width=0.24\textwidth]{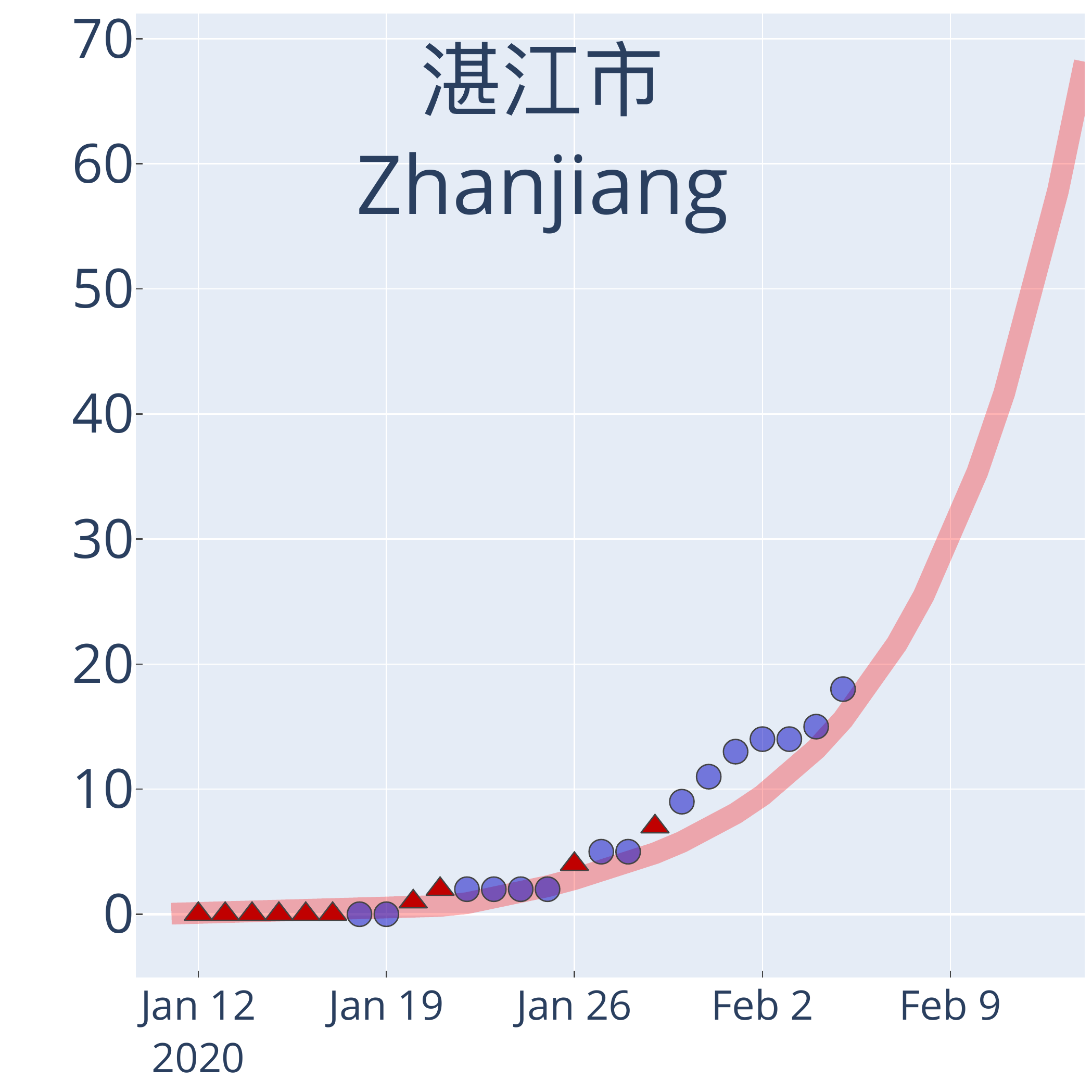}\hfill{}\includegraphics[width=0.24\textwidth]{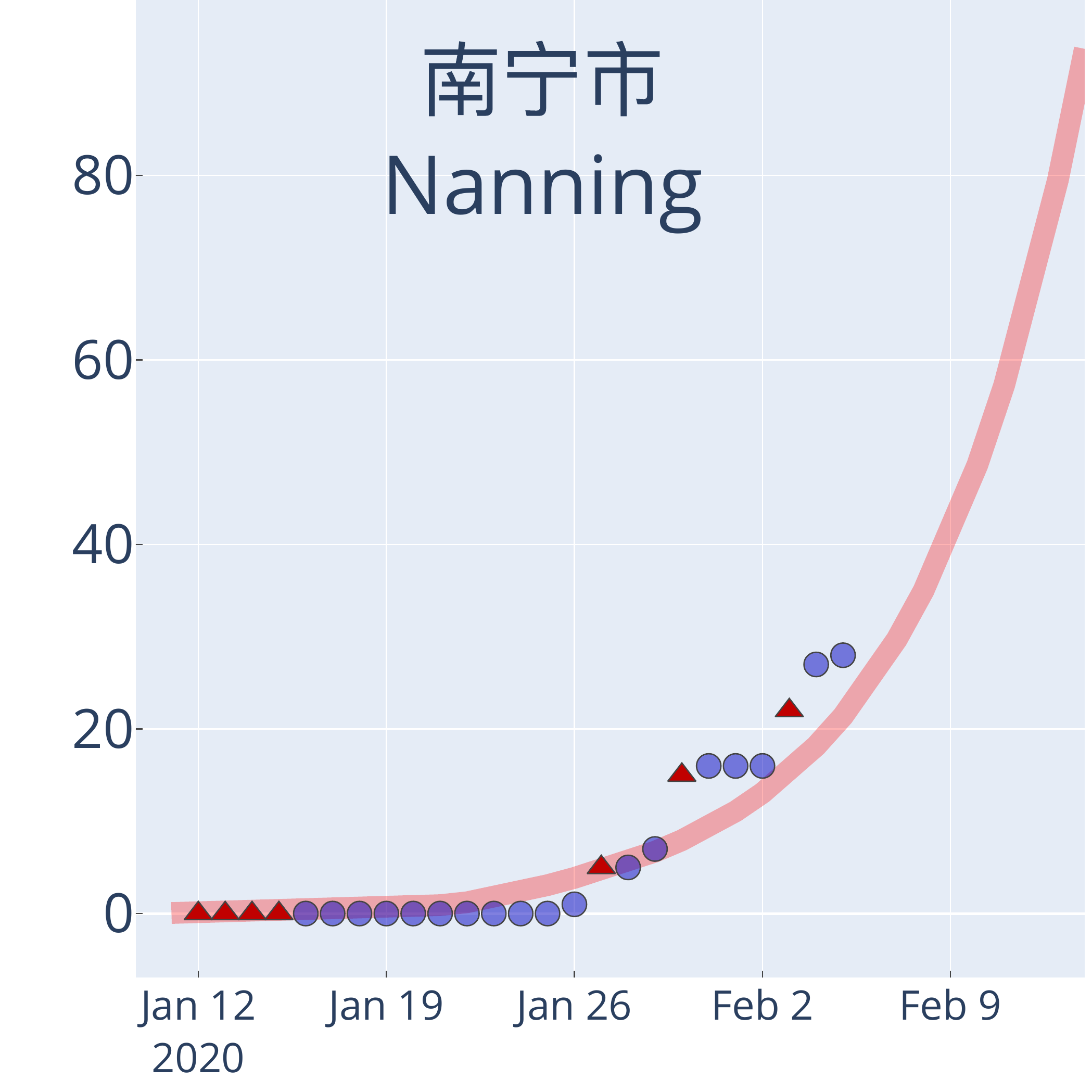}\hfill{}\includegraphics[width=0.24\textwidth]{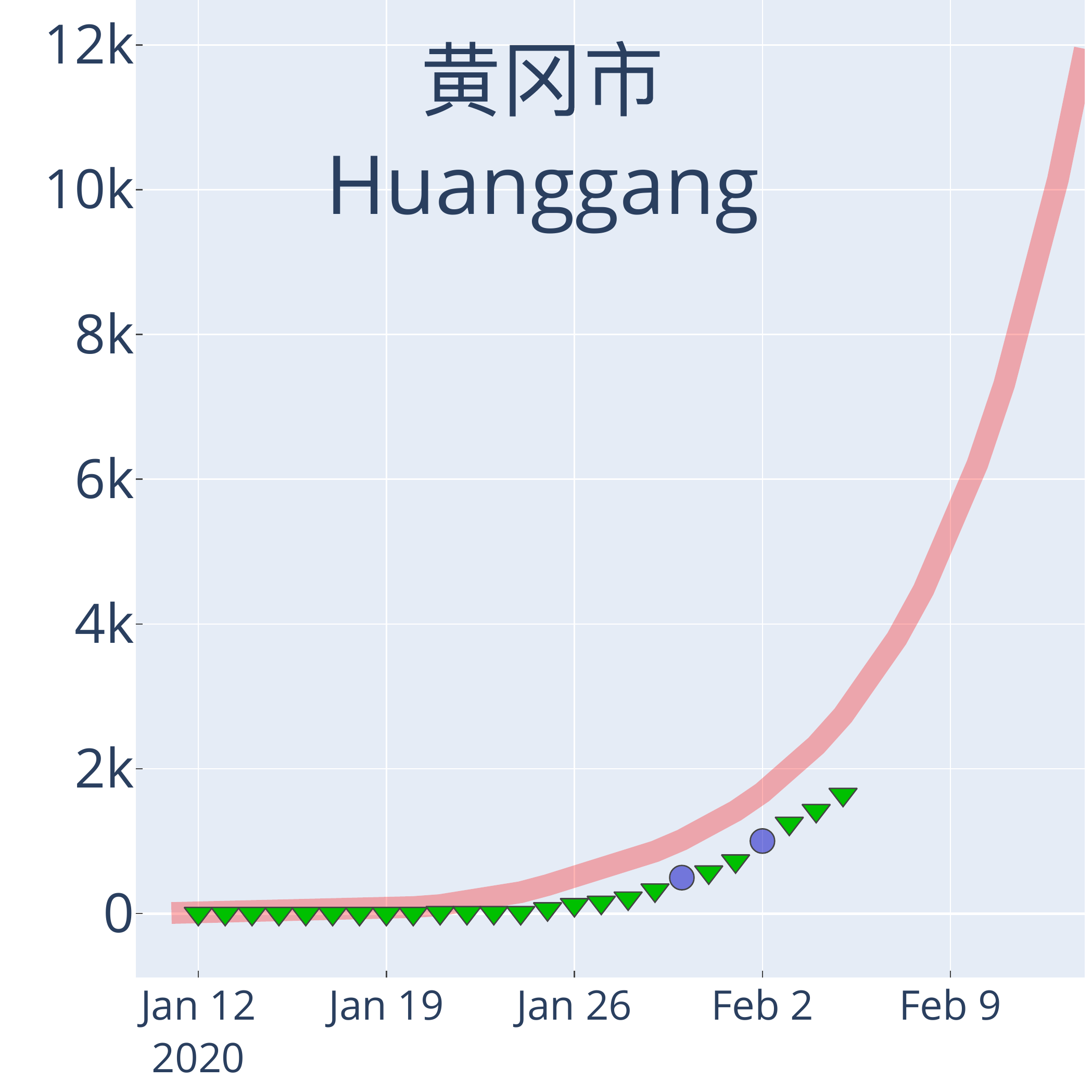}\hfill{}
\par\end{centering}
\caption{Simulation and forecasting of infections in major China cities and
comparison to accumulated cases. See Figure \ref{fig:sim-sample}
for detailed interpretation of the marks and legends used in the plots.}
\end{figure*}

\begin{figure*}[t]
\begin{centering}
\hfill{}\includegraphics[width=0.24\textwidth]{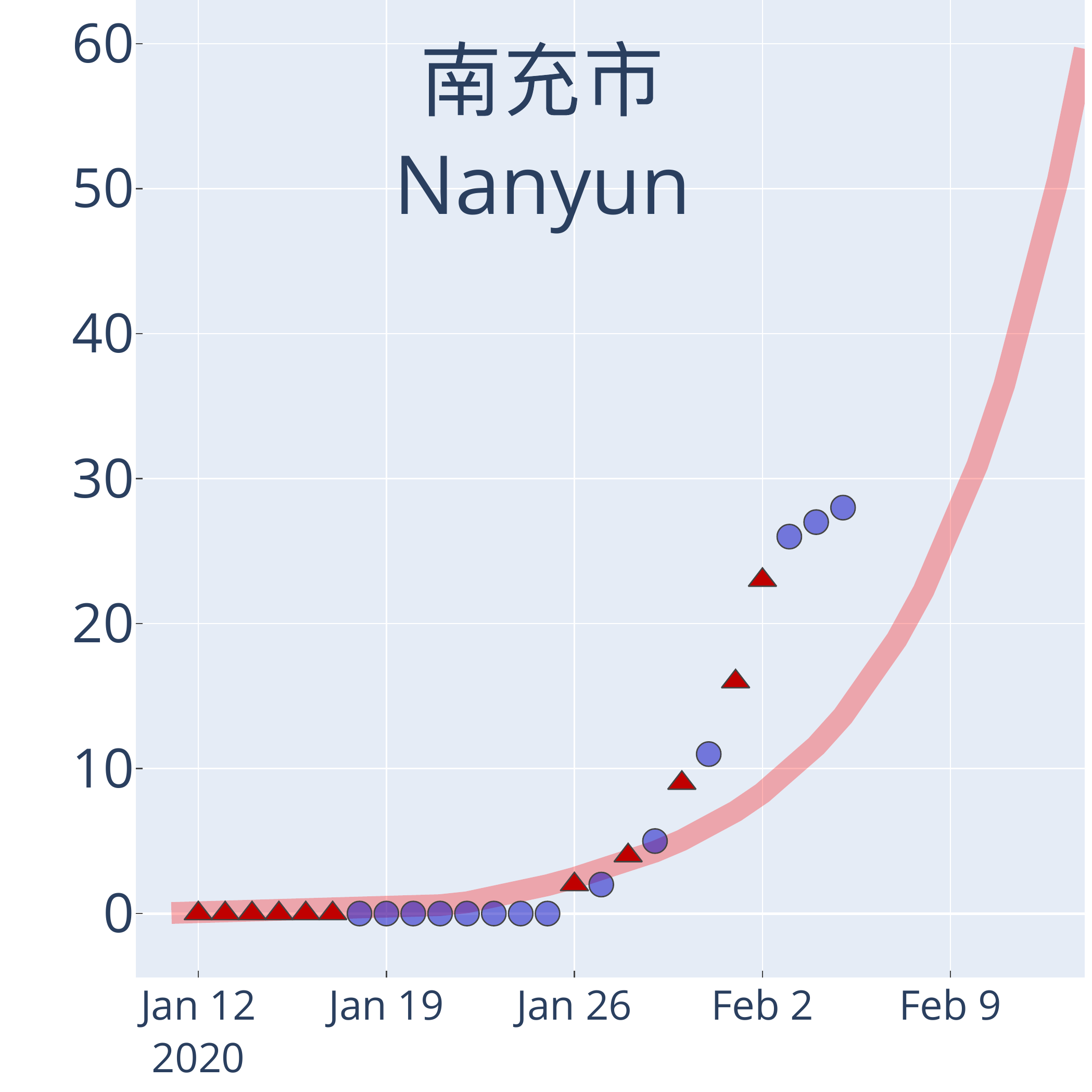}\hfill{}\includegraphics[width=0.24\textwidth]{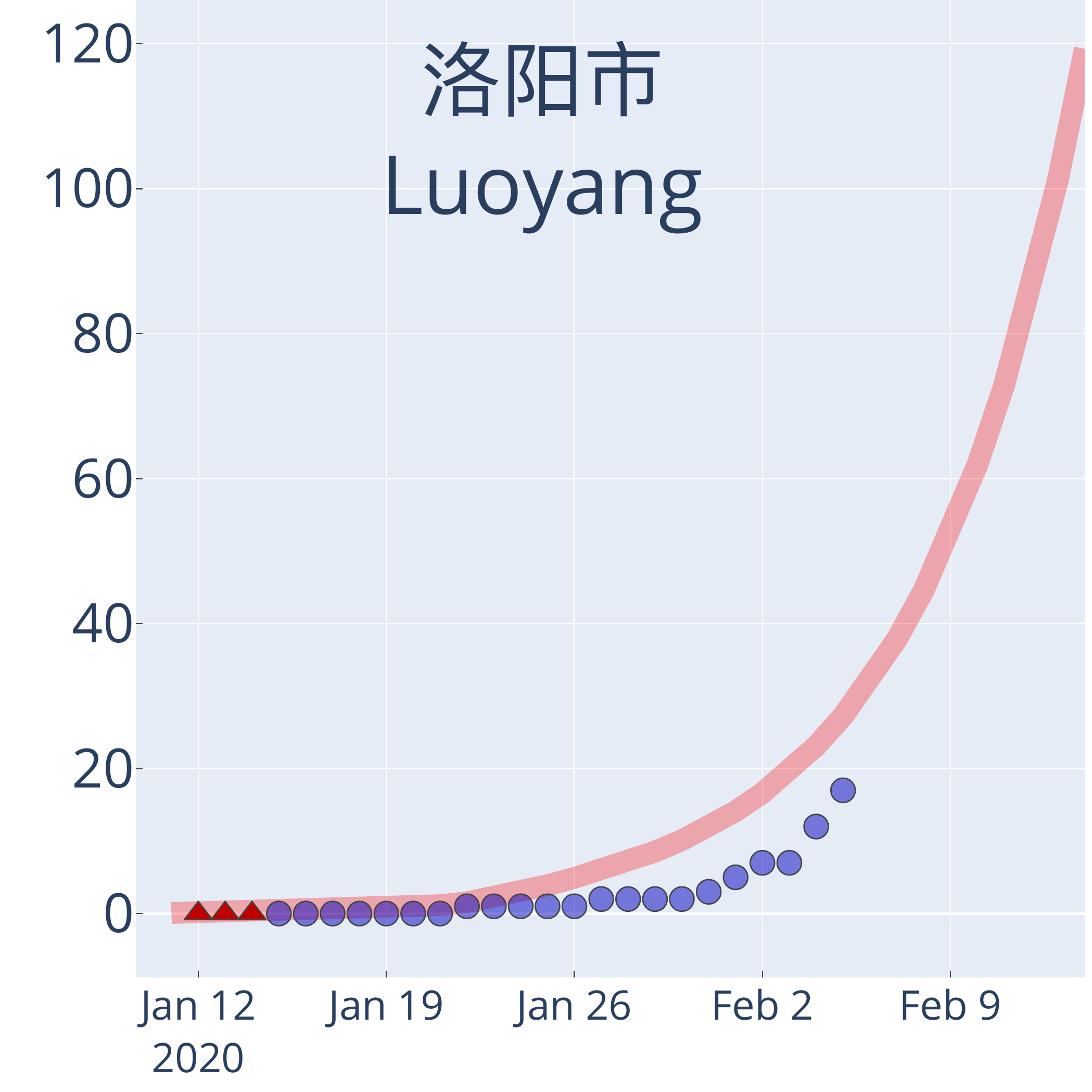}\hfill{}\includegraphics[width=0.24\textwidth]{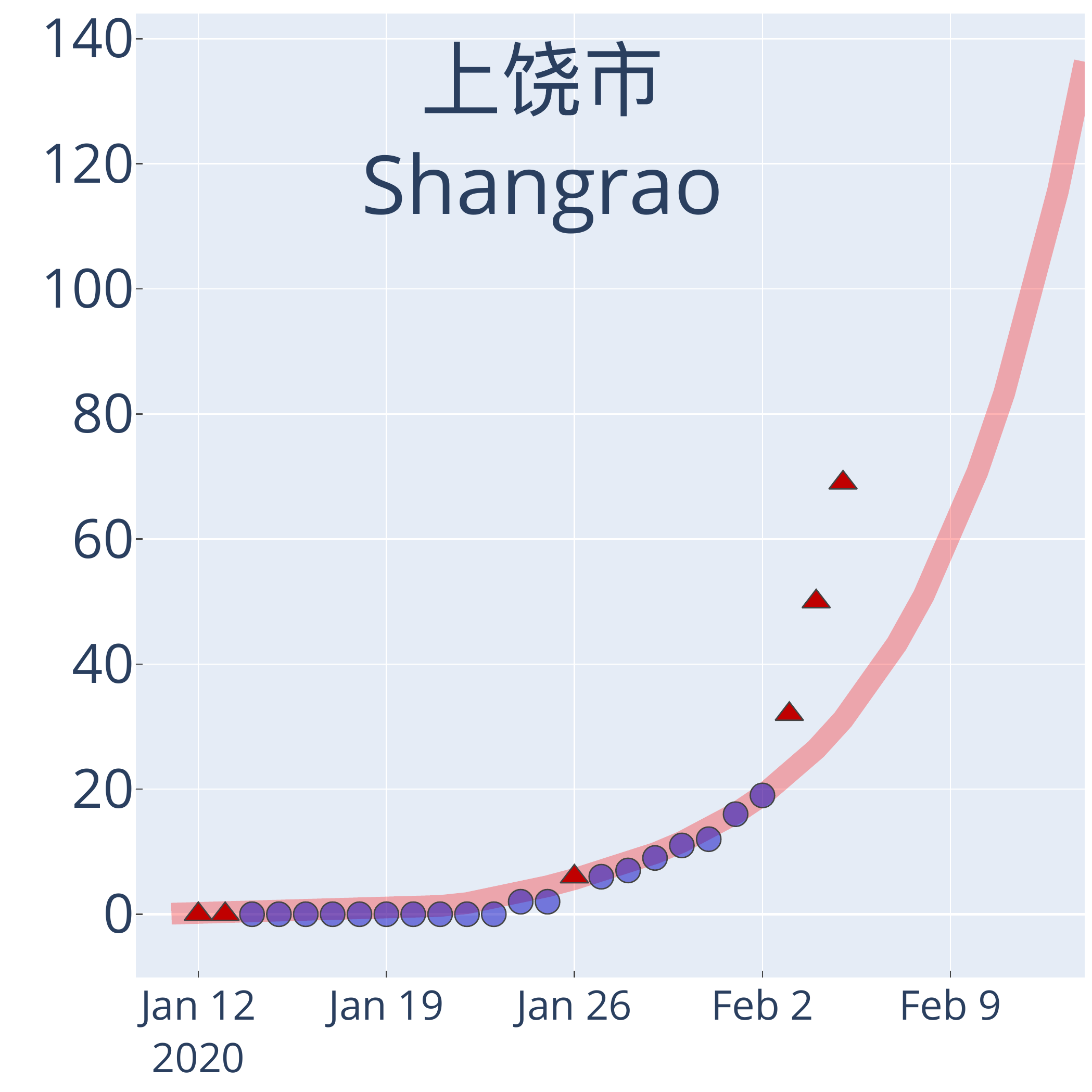}\hfill{}\includegraphics[width=0.24\textwidth]{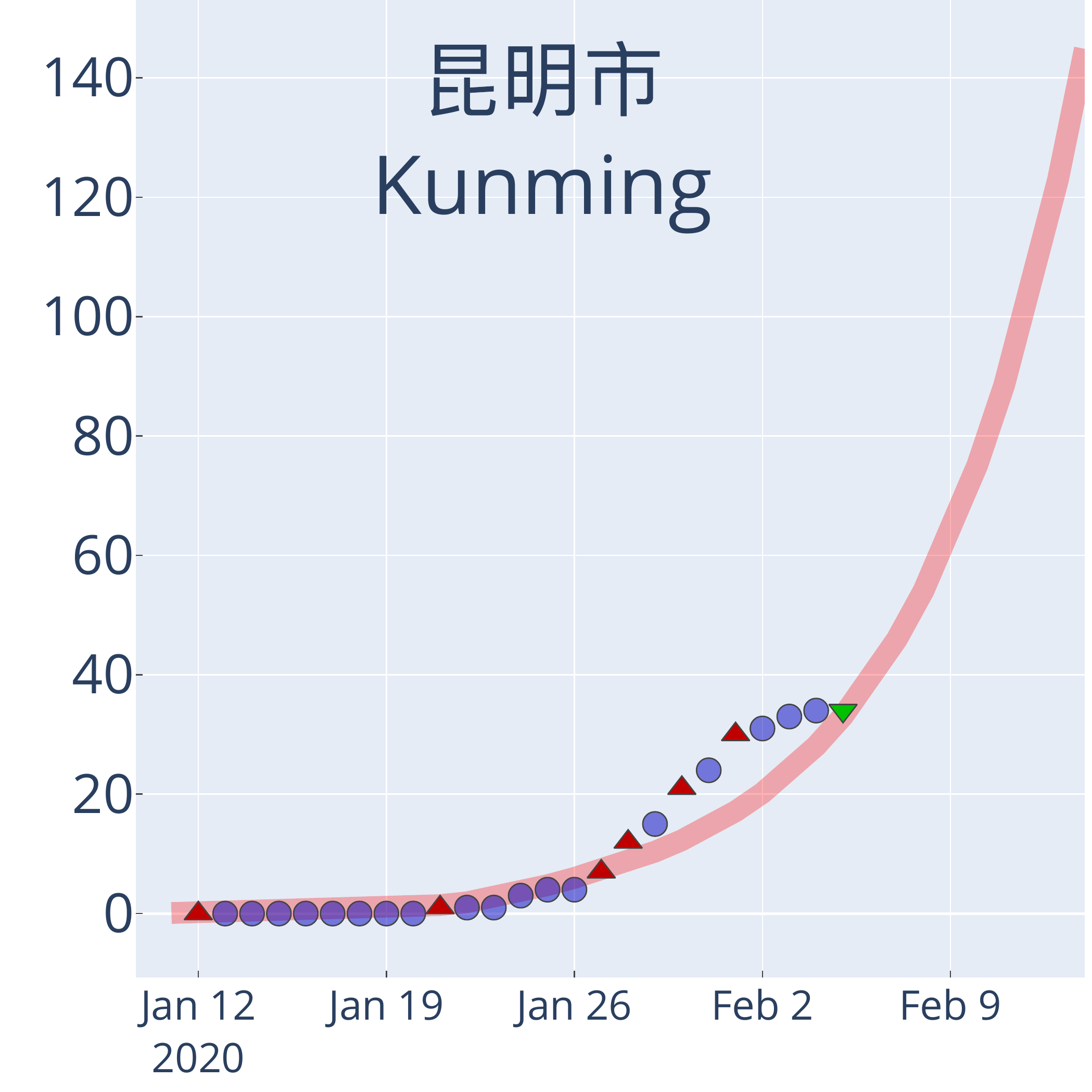}\hfill{}
\par\end{centering}
\begin{centering}
\hfill{}\includegraphics[width=0.24\textwidth]{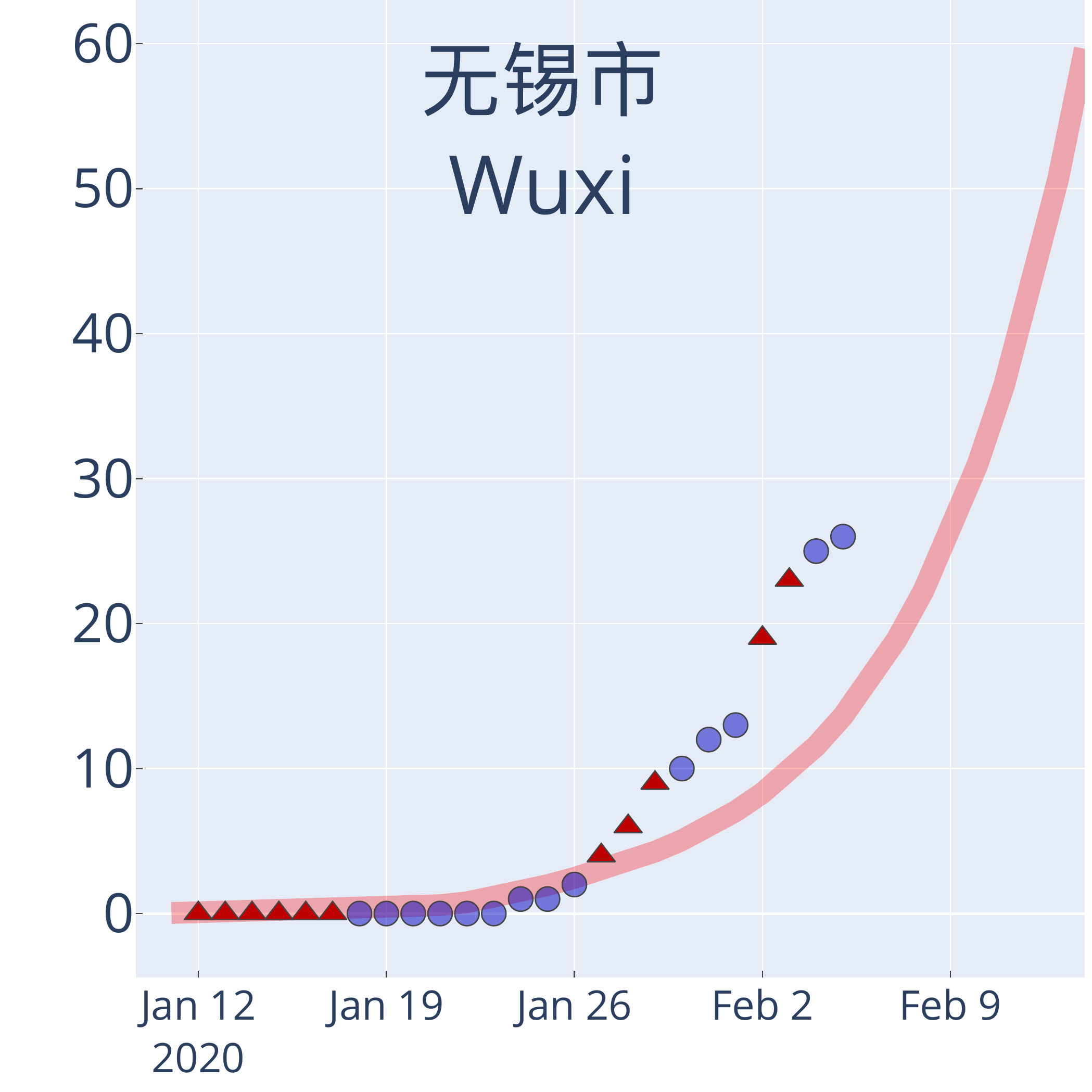}\hfill{}\includegraphics[width=0.24\textwidth]{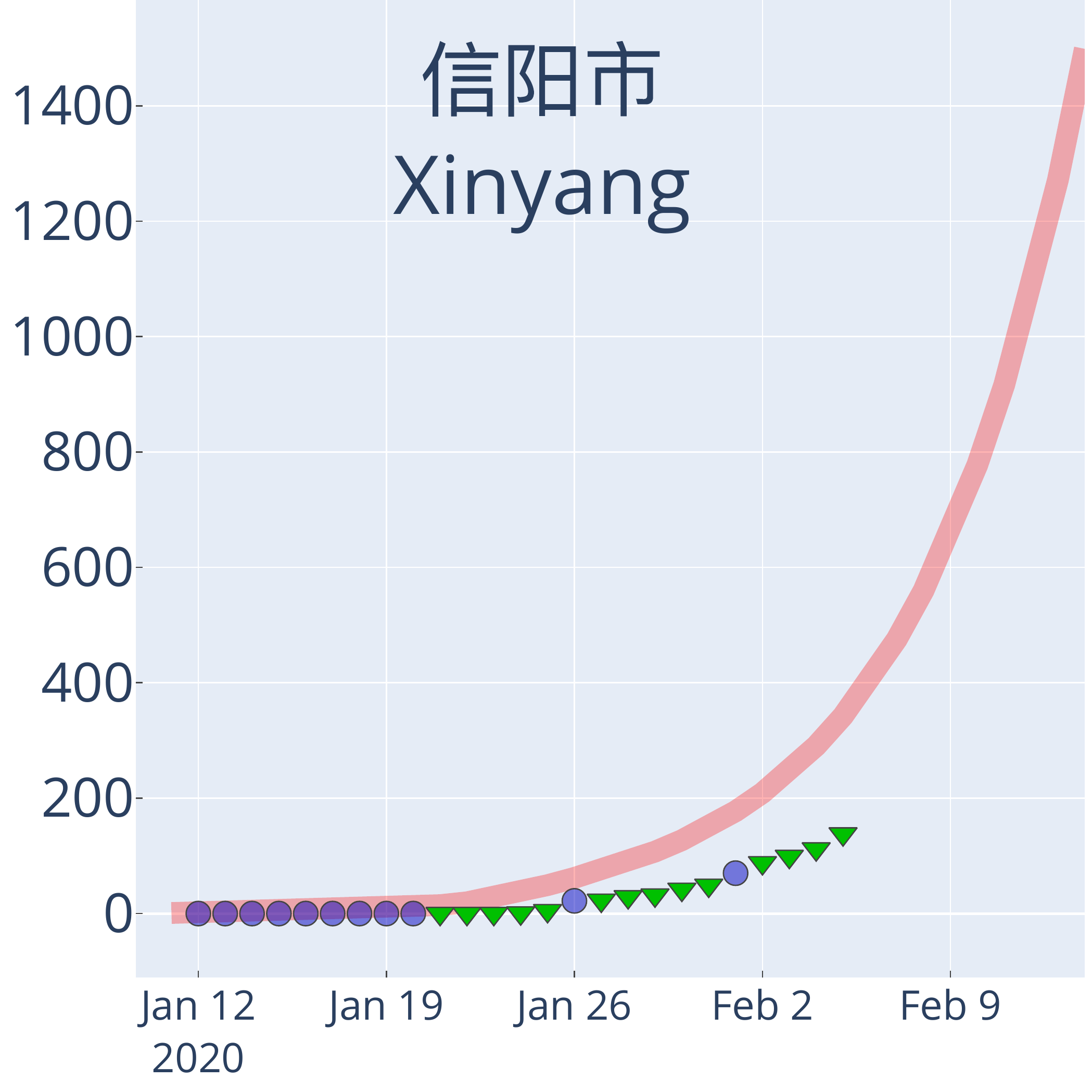}\hfill{}\includegraphics[width=0.24\textwidth]{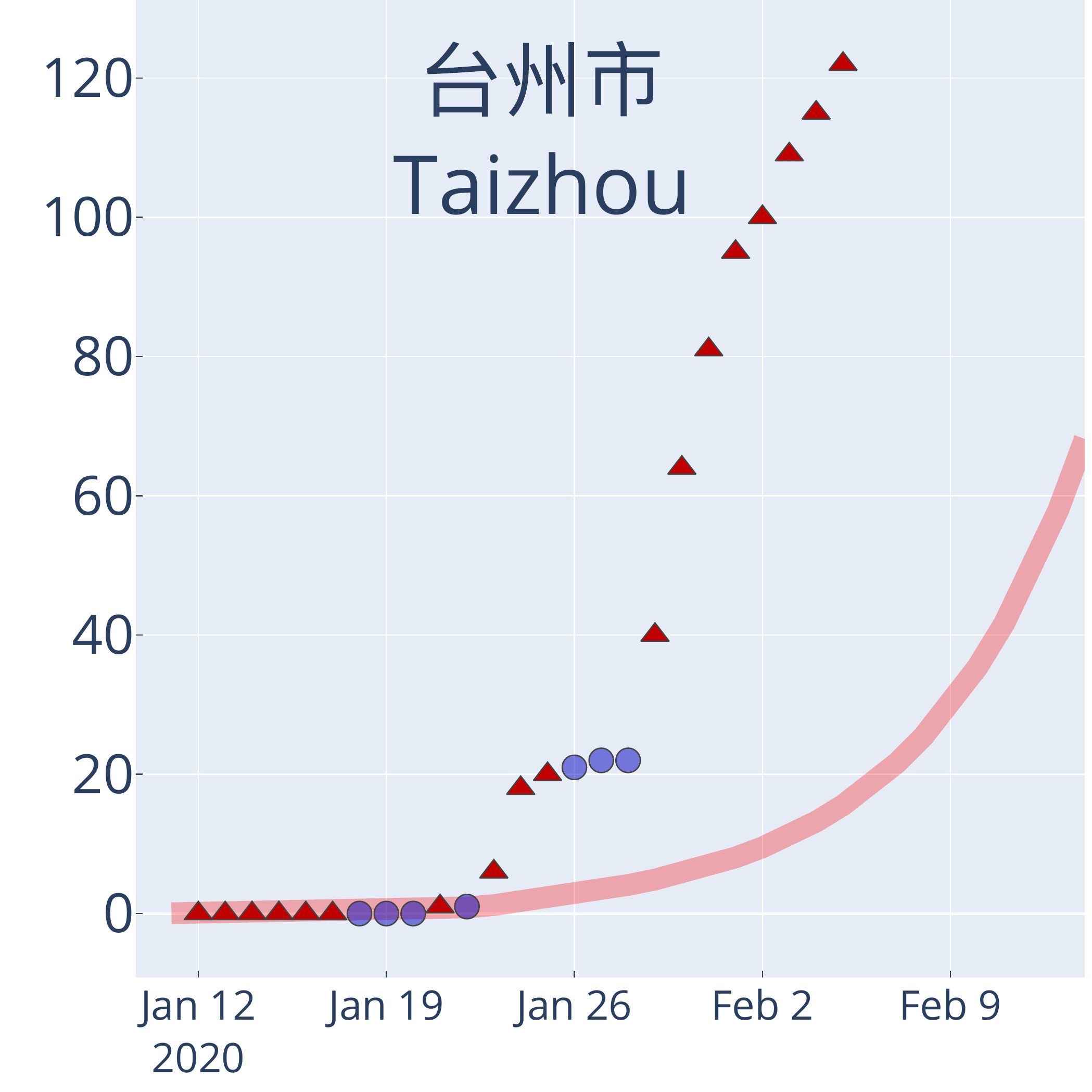}\hfill{}\includegraphics[width=0.24\textwidth]{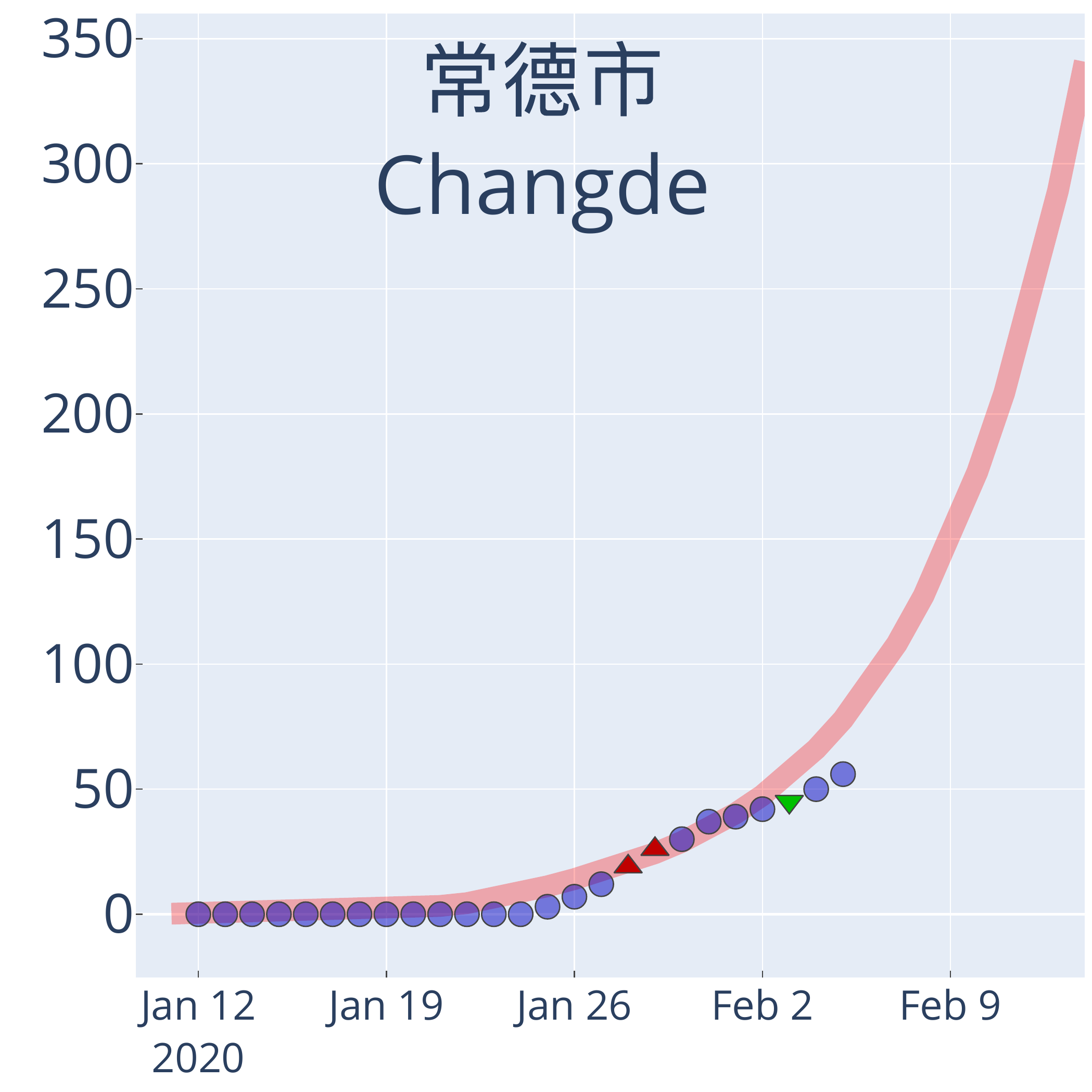}\hfill{}
\par\end{centering}
\begin{centering}
\hfill{}\includegraphics[width=0.24\textwidth]{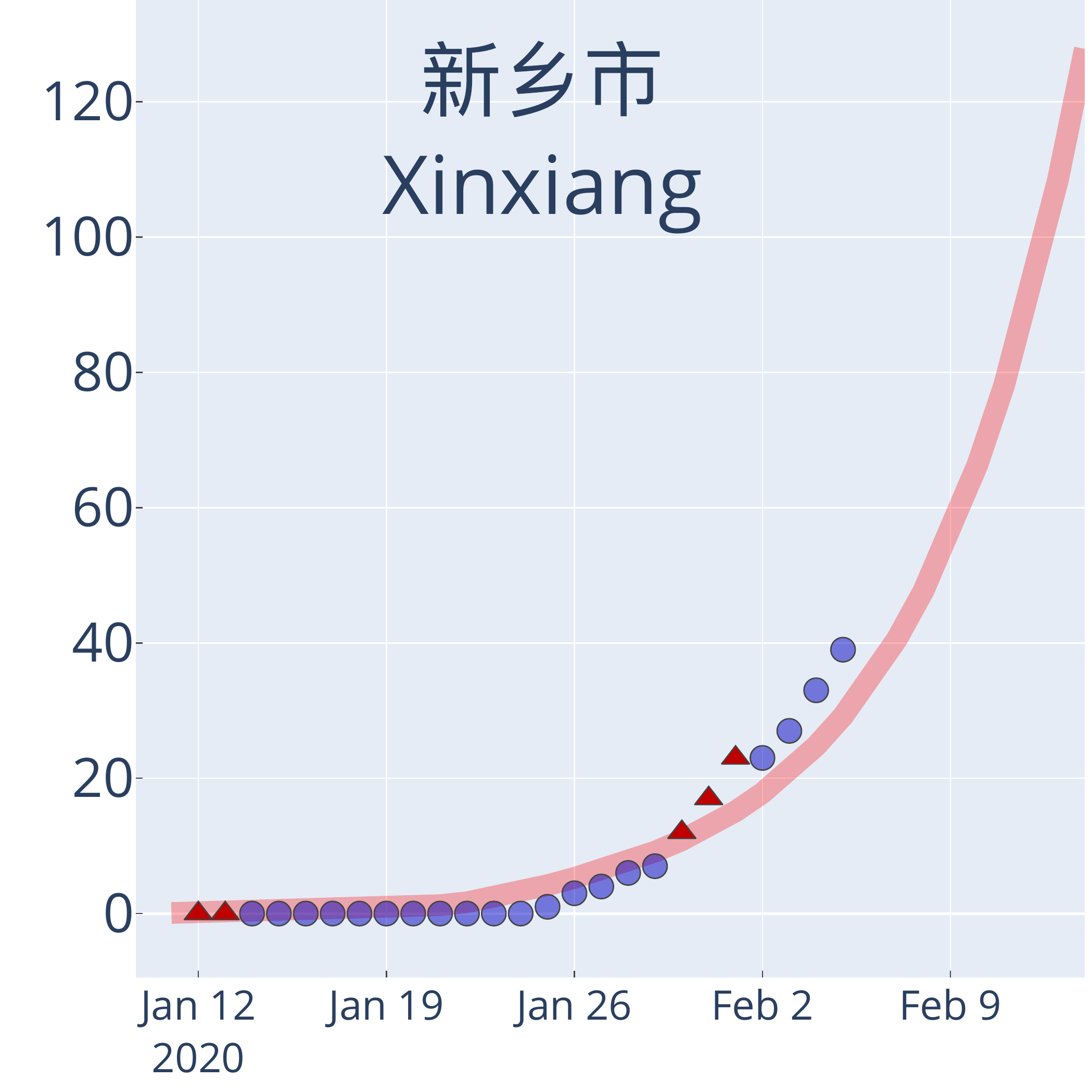}\hfill{}\includegraphics[width=0.24\textwidth]{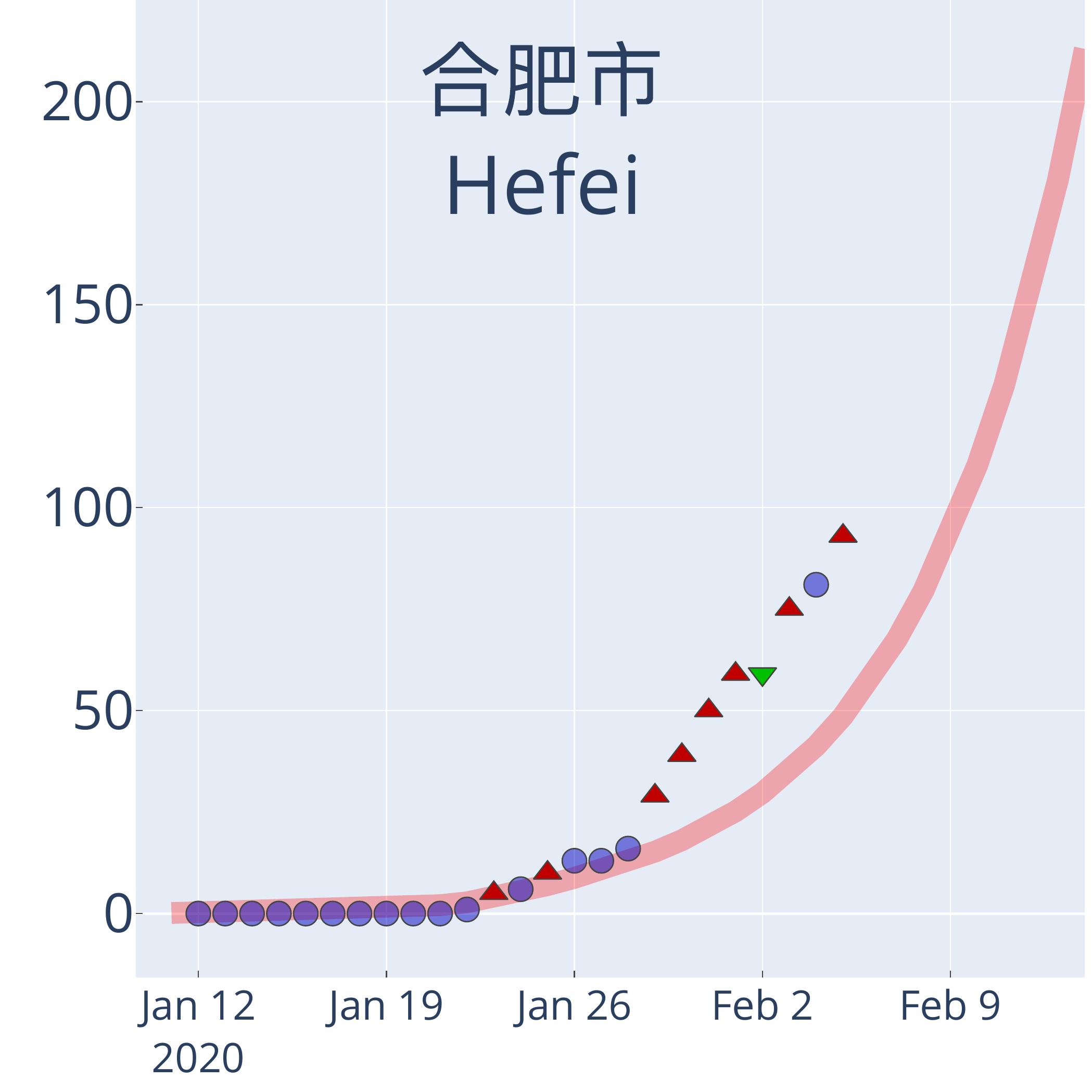}\hfill{}\includegraphics[width=0.24\textwidth]{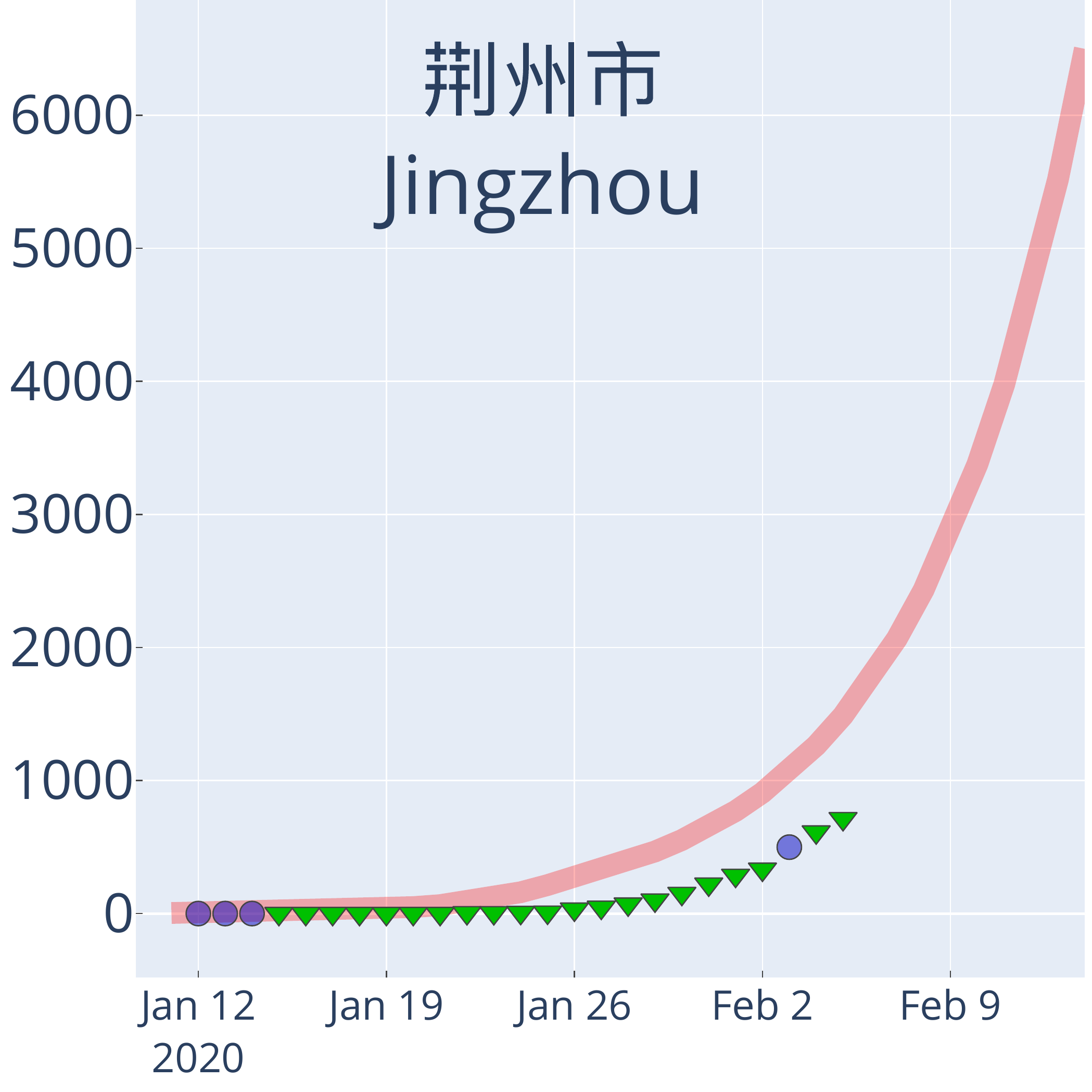}\hfill{}\includegraphics[width=0.24\textwidth]{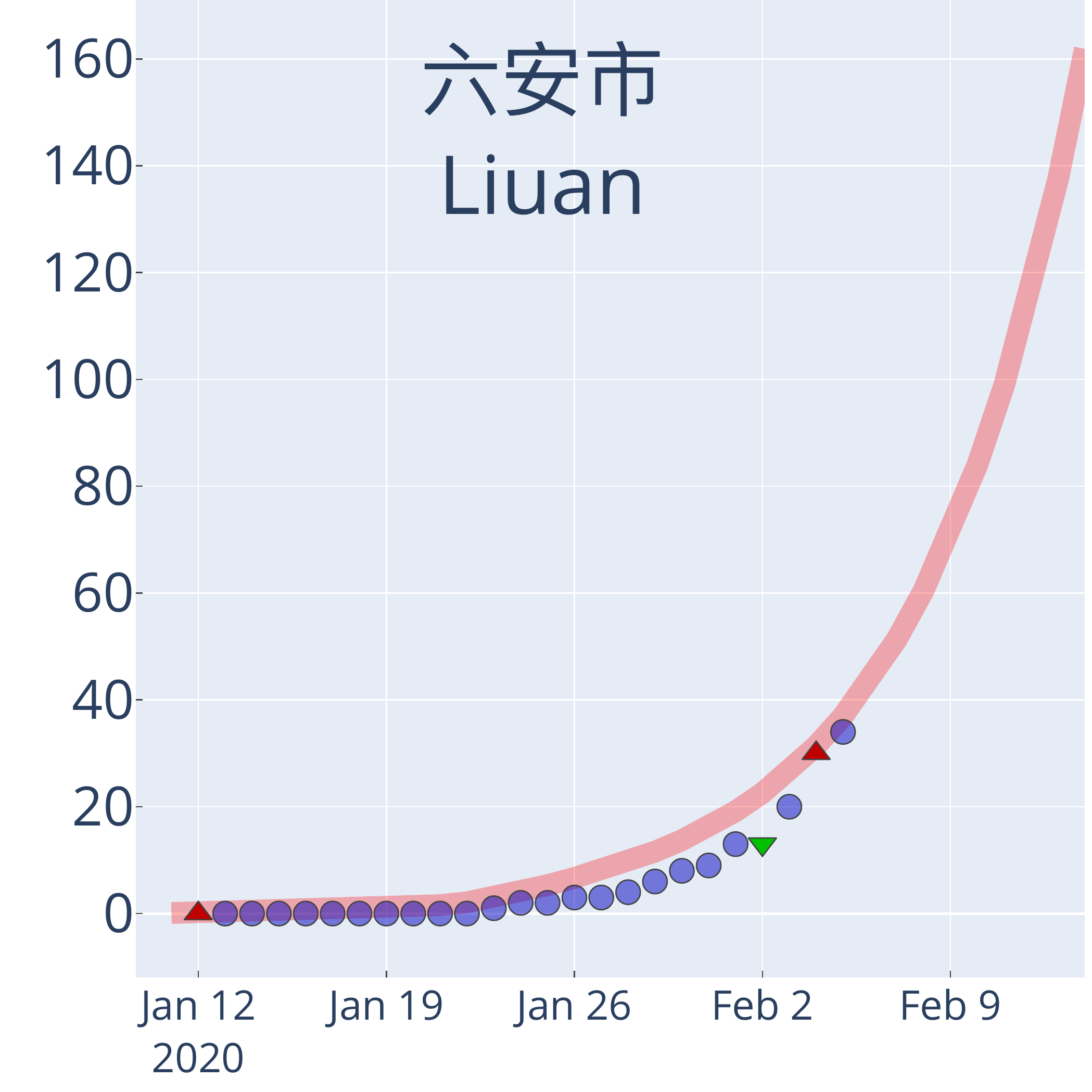}\hfill{}
\par\end{centering}
\begin{centering}
\hfill{}\includegraphics[width=0.24\textwidth]{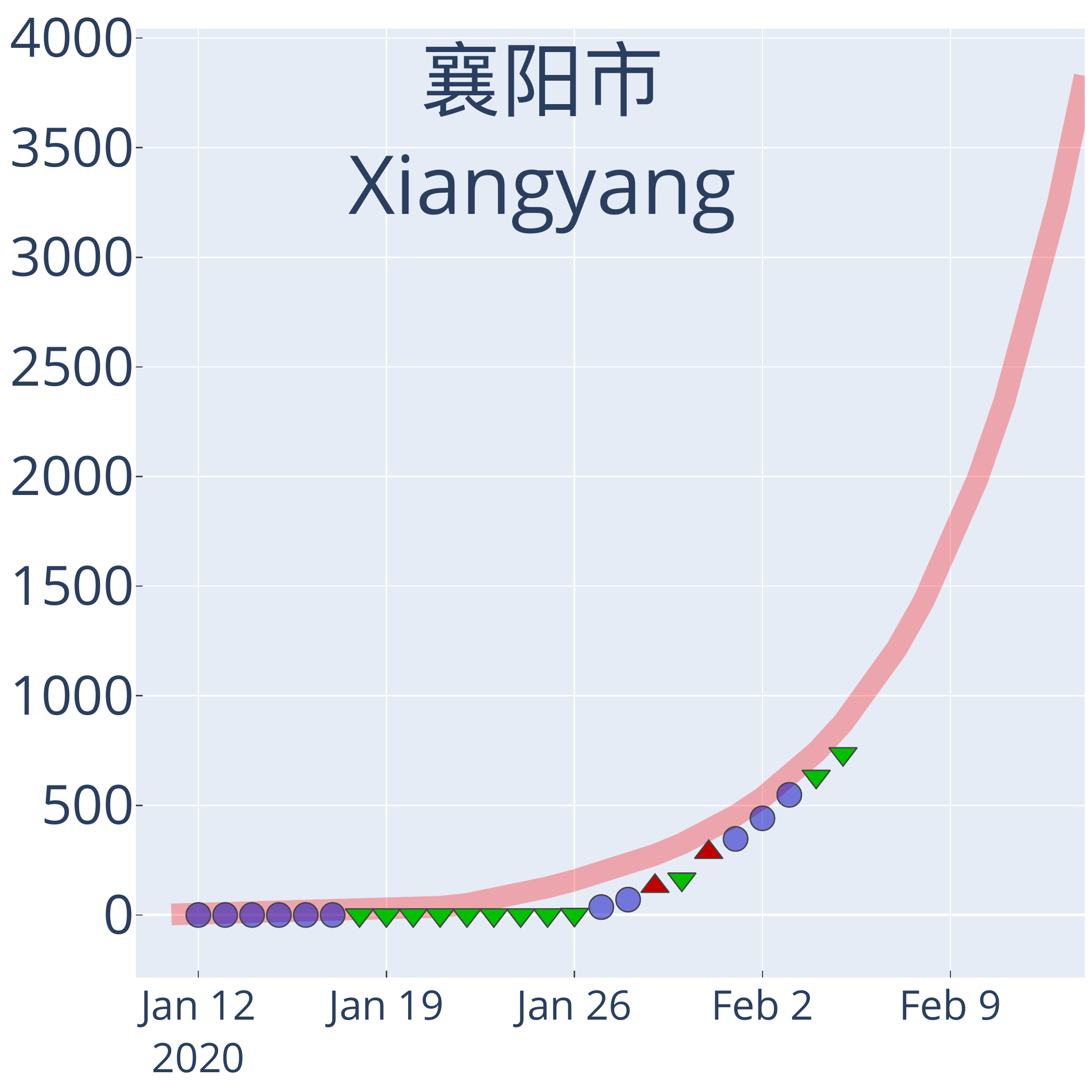}\hfill{}\includegraphics[width=0.24\textwidth]{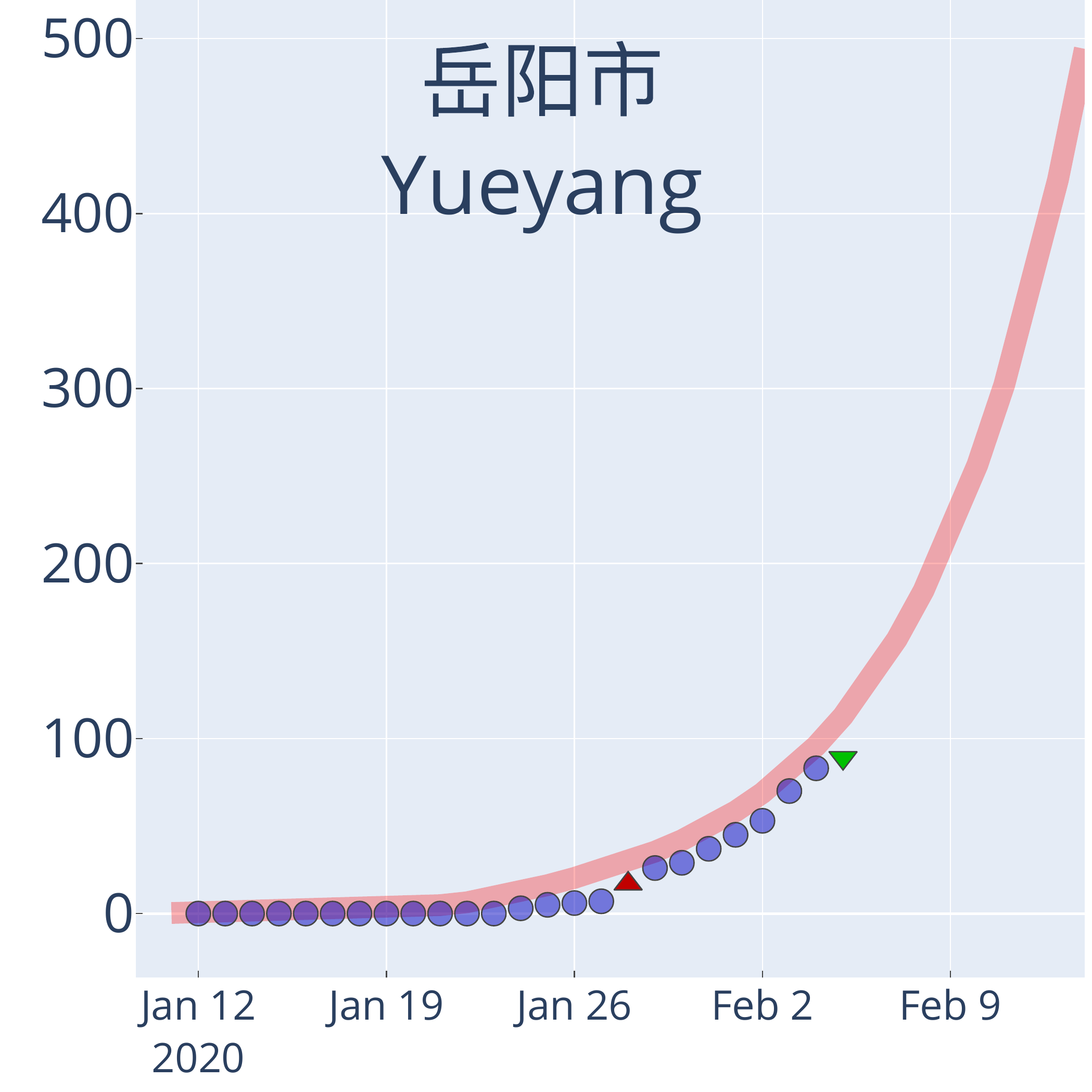}\hfill{}\includegraphics[width=0.24\textwidth]{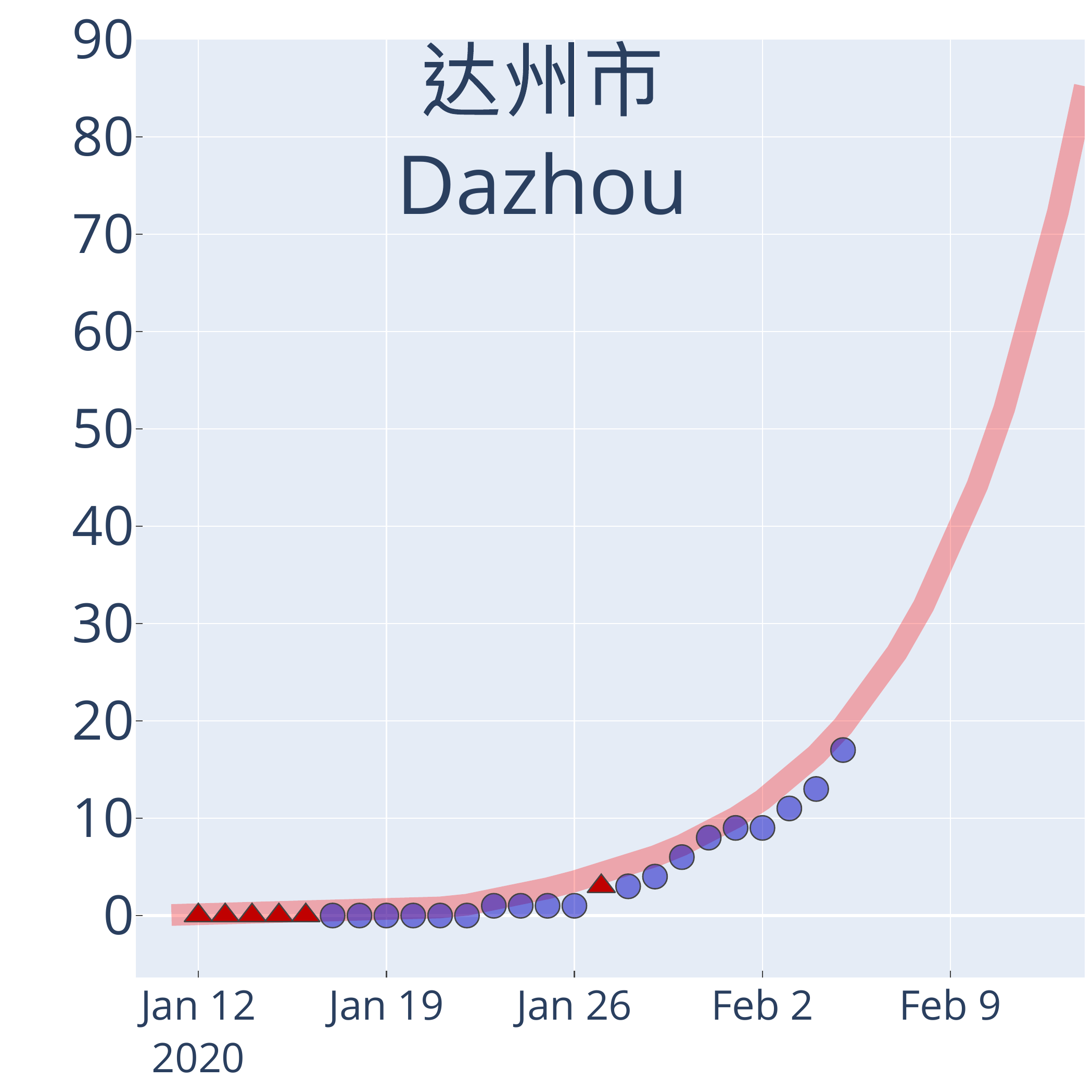}\hfill{}\includegraphics[width=0.24\textwidth]{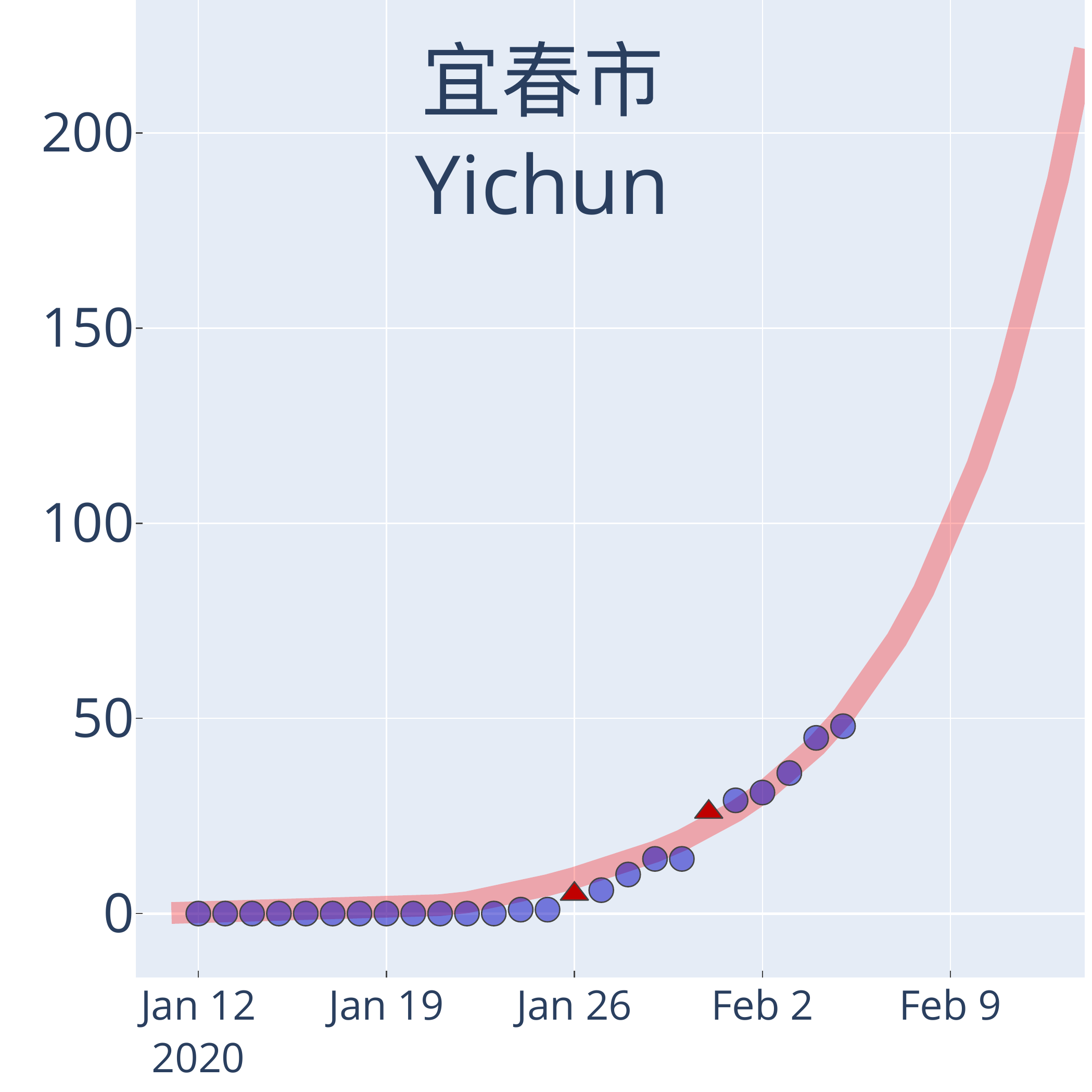}\hfill{}
\par\end{centering}
\begin{centering}
\hfill{}\includegraphics[width=0.24\textwidth]{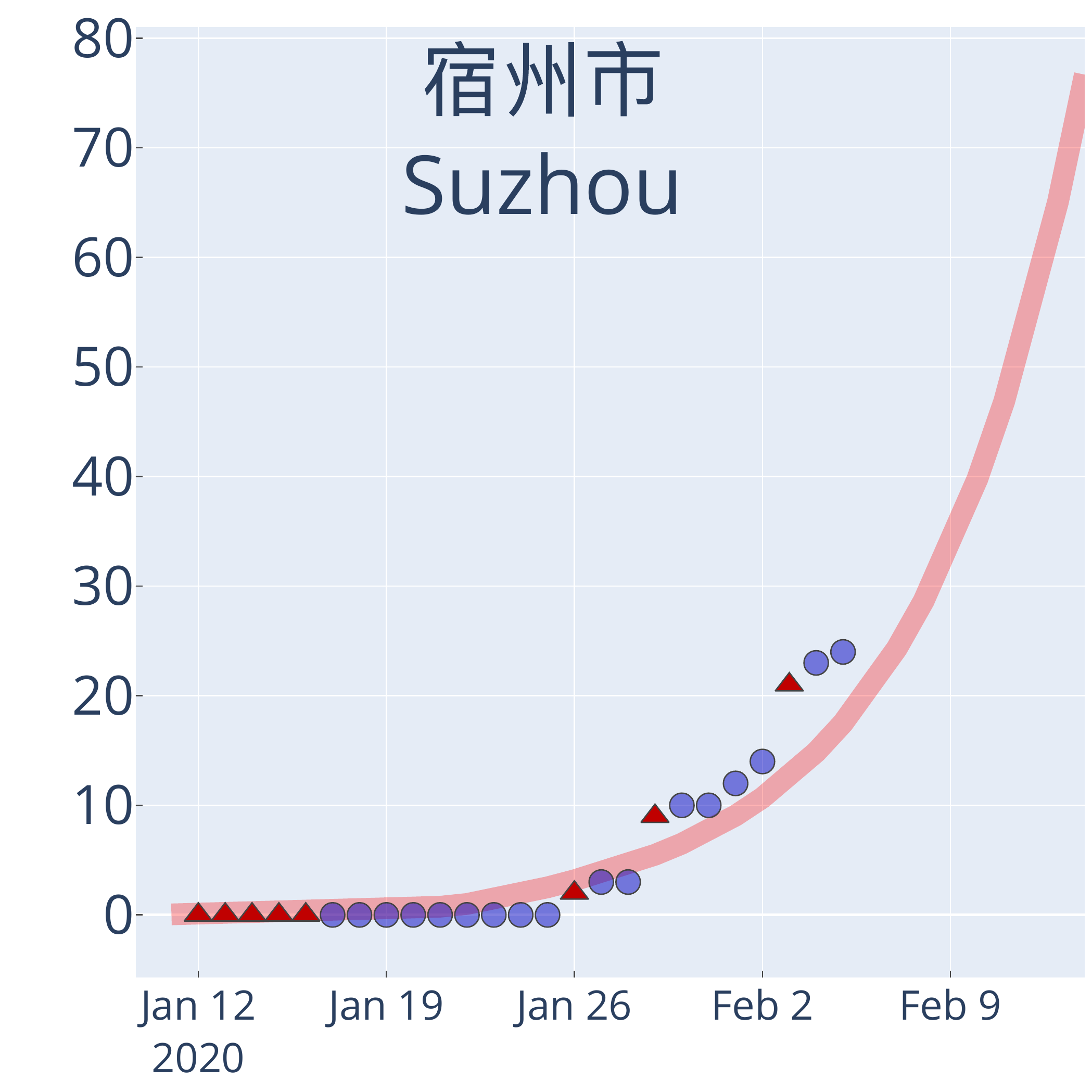}\hfill{}\includegraphics[width=0.24\textwidth]{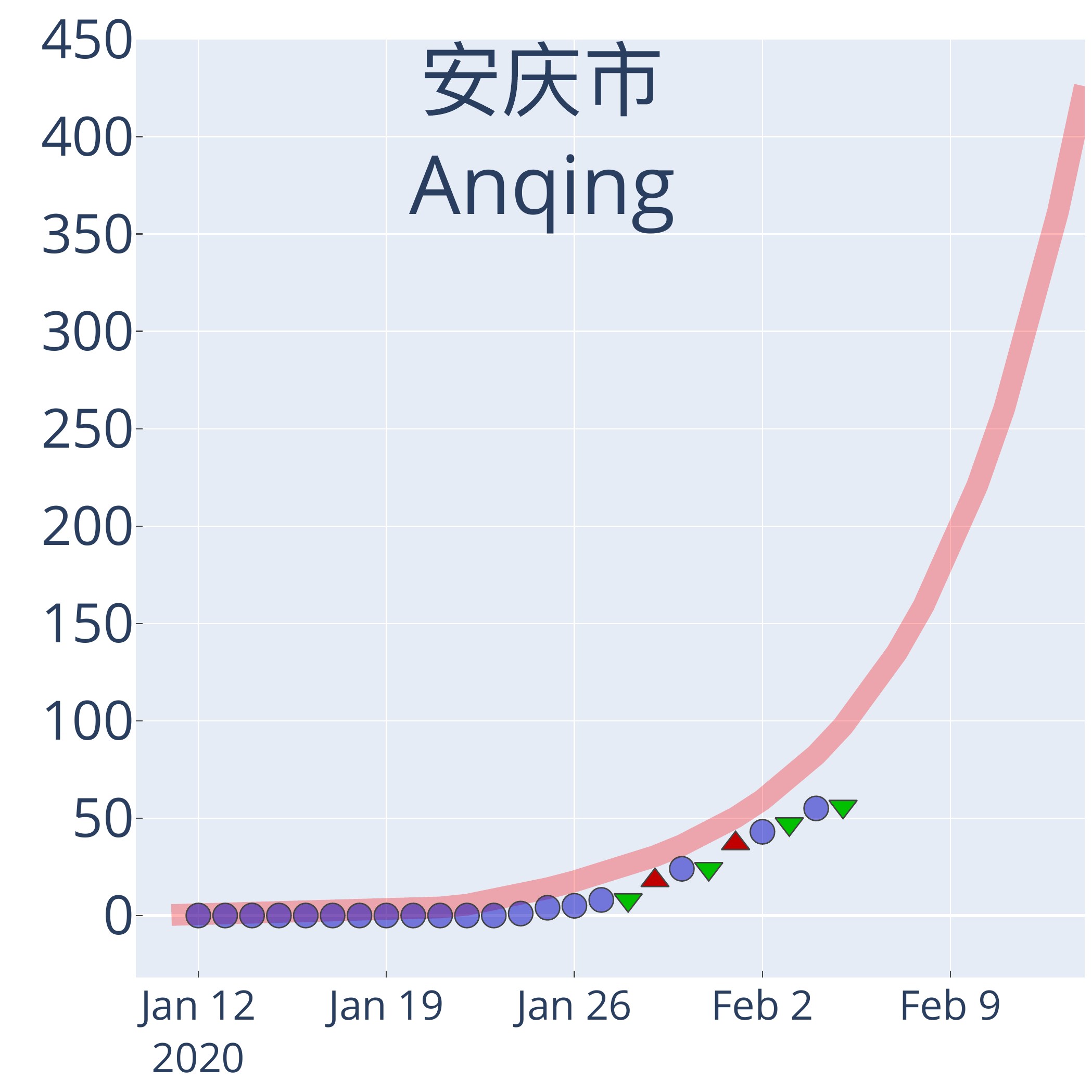}\hfill{}\includegraphics[width=0.24\textwidth]{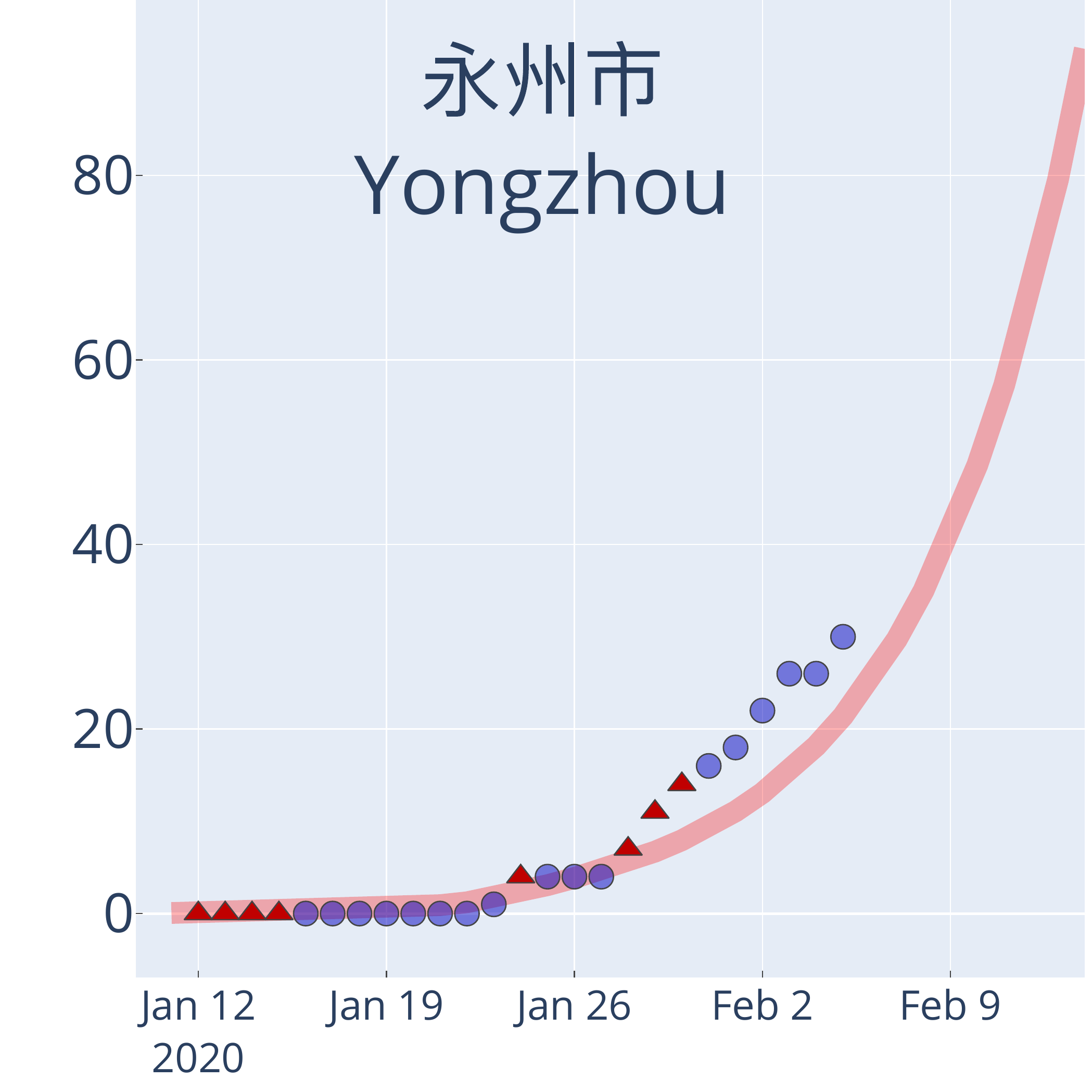}\hfill{}\includegraphics[width=0.24\textwidth]{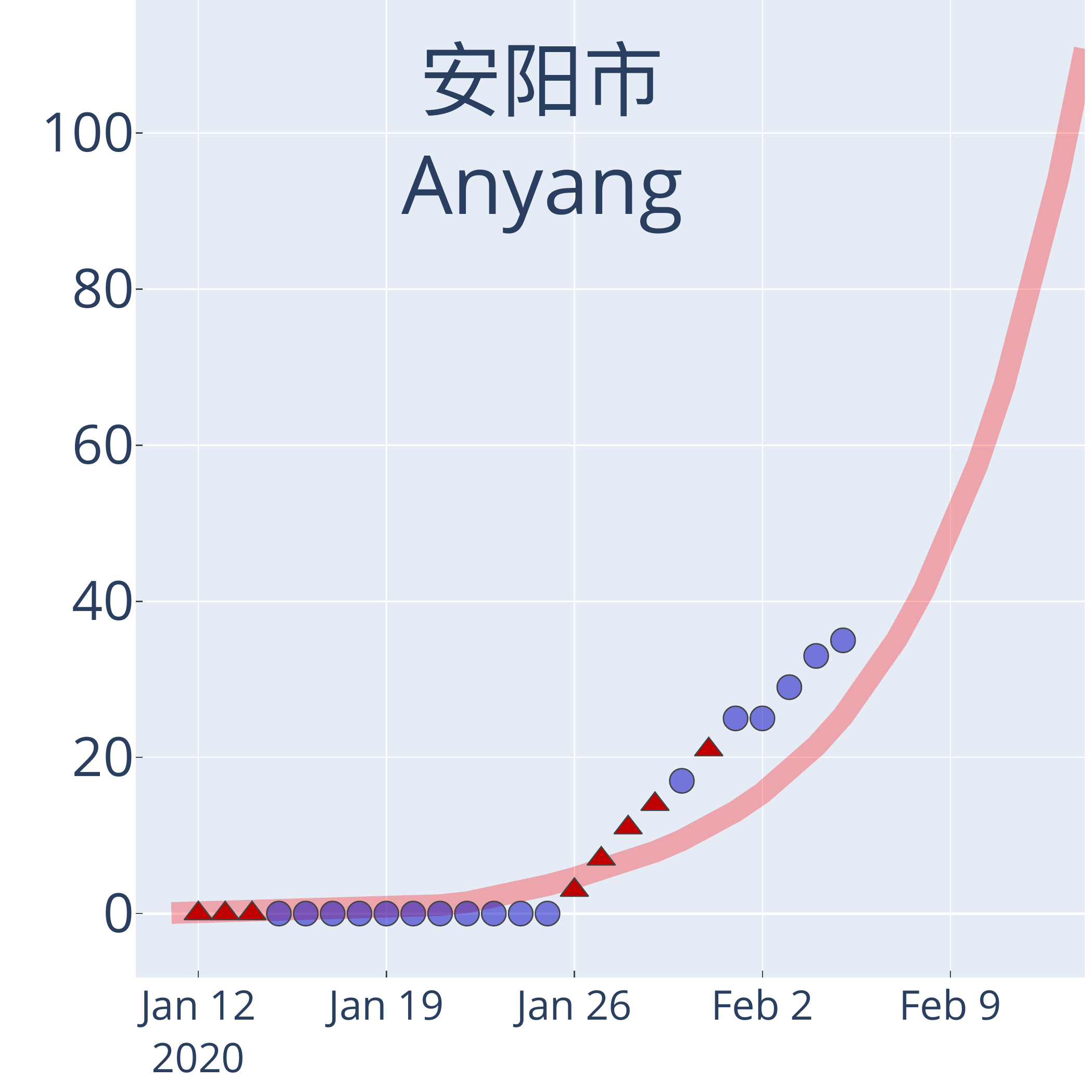}\hfill{}
\par\end{centering}
\caption{Simulation and forecasting of infections in major China cities and
comparison to accumulated cases. See Figure \ref{fig:sim-sample}
for detailed interpretation of the marks and legends used in the plots.}
\end{figure*}

\begin{figure*}[t]
\begin{centering}
\hfill{}\includegraphics[width=0.24\textwidth]{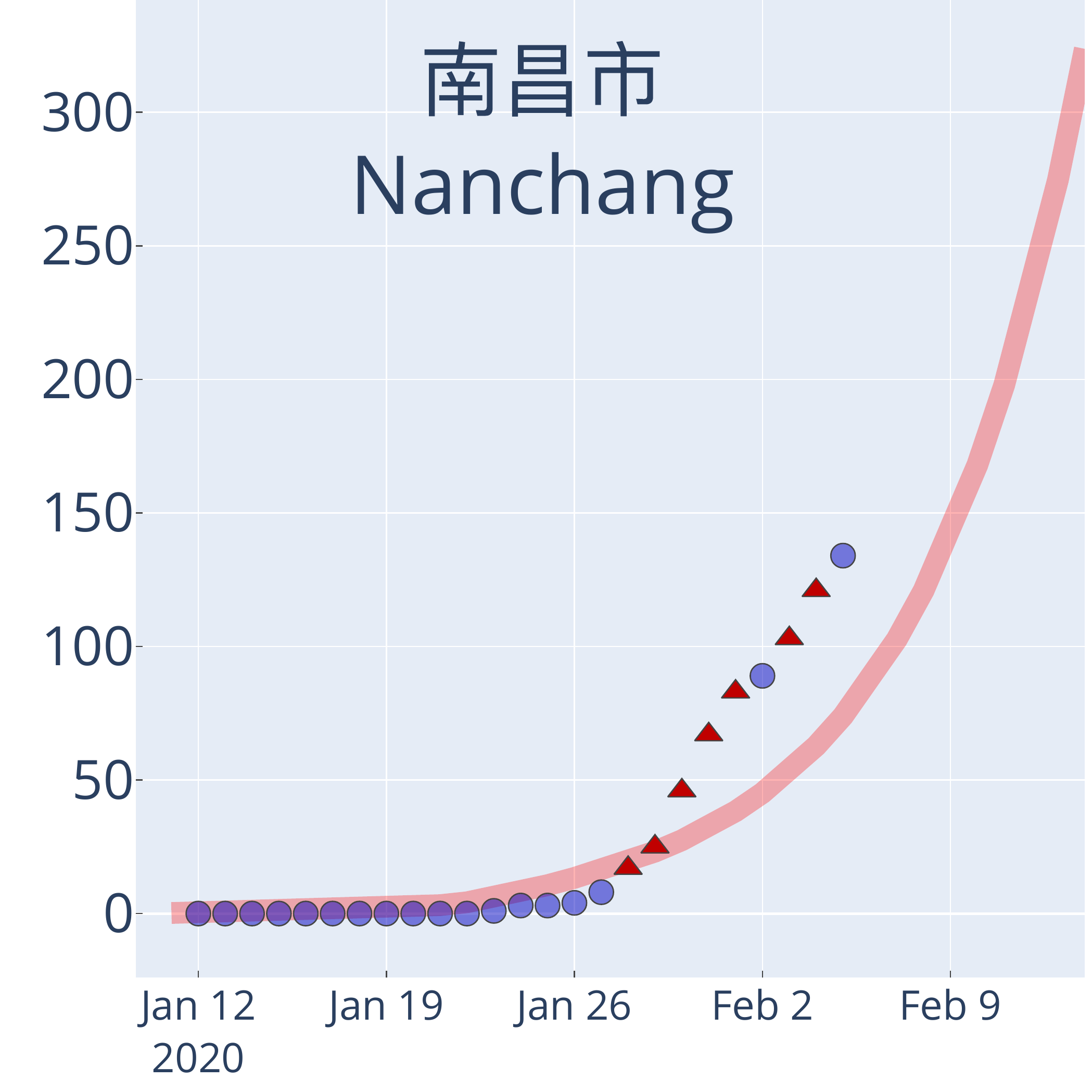}\hfill{}\includegraphics[width=0.24\textwidth]{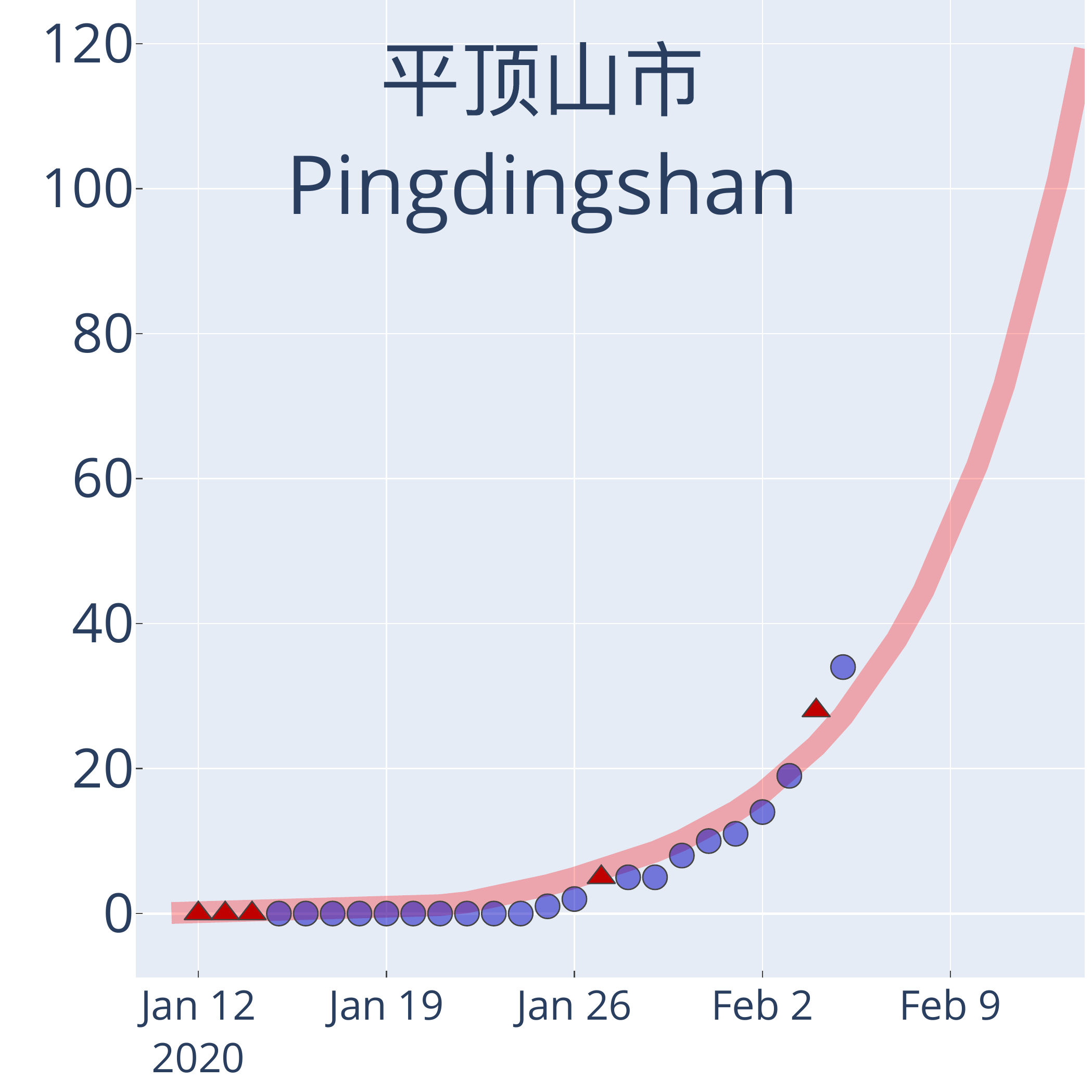}\hfill{}\includegraphics[width=0.24\textwidth]{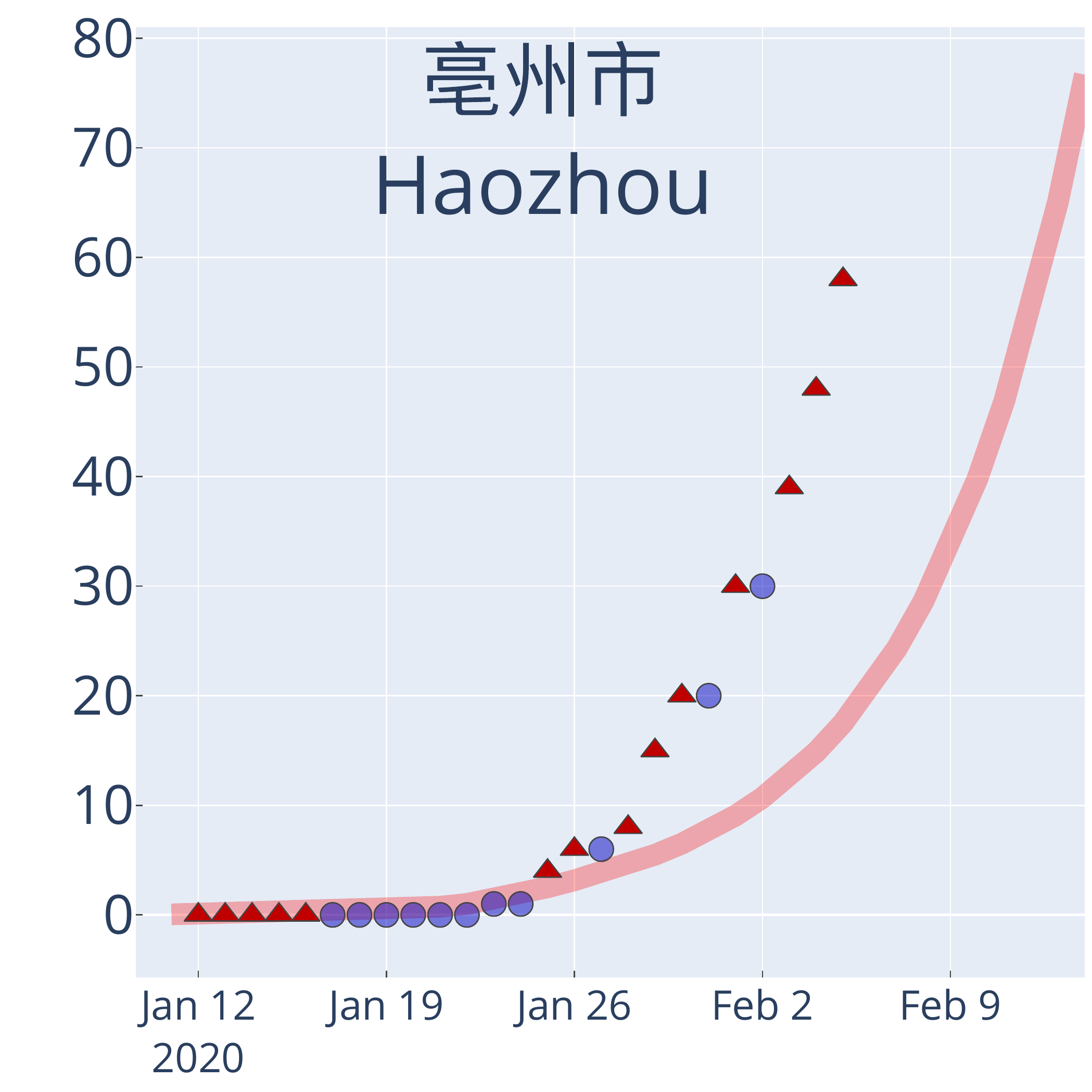}\hfill{}\includegraphics[width=0.24\textwidth]{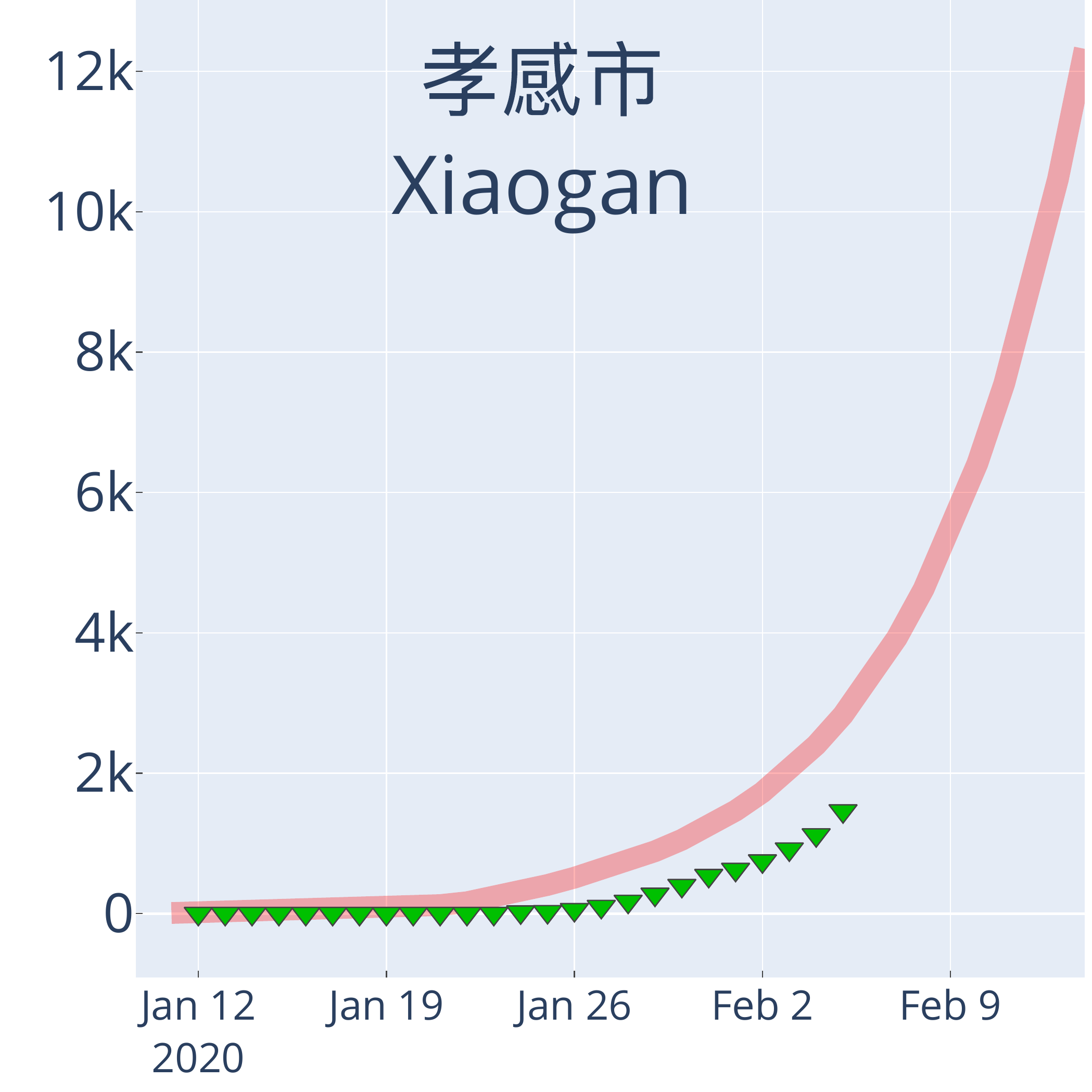}\hfill{}
\par\end{centering}
\begin{centering}
\hfill{}\includegraphics[width=0.24\textwidth]{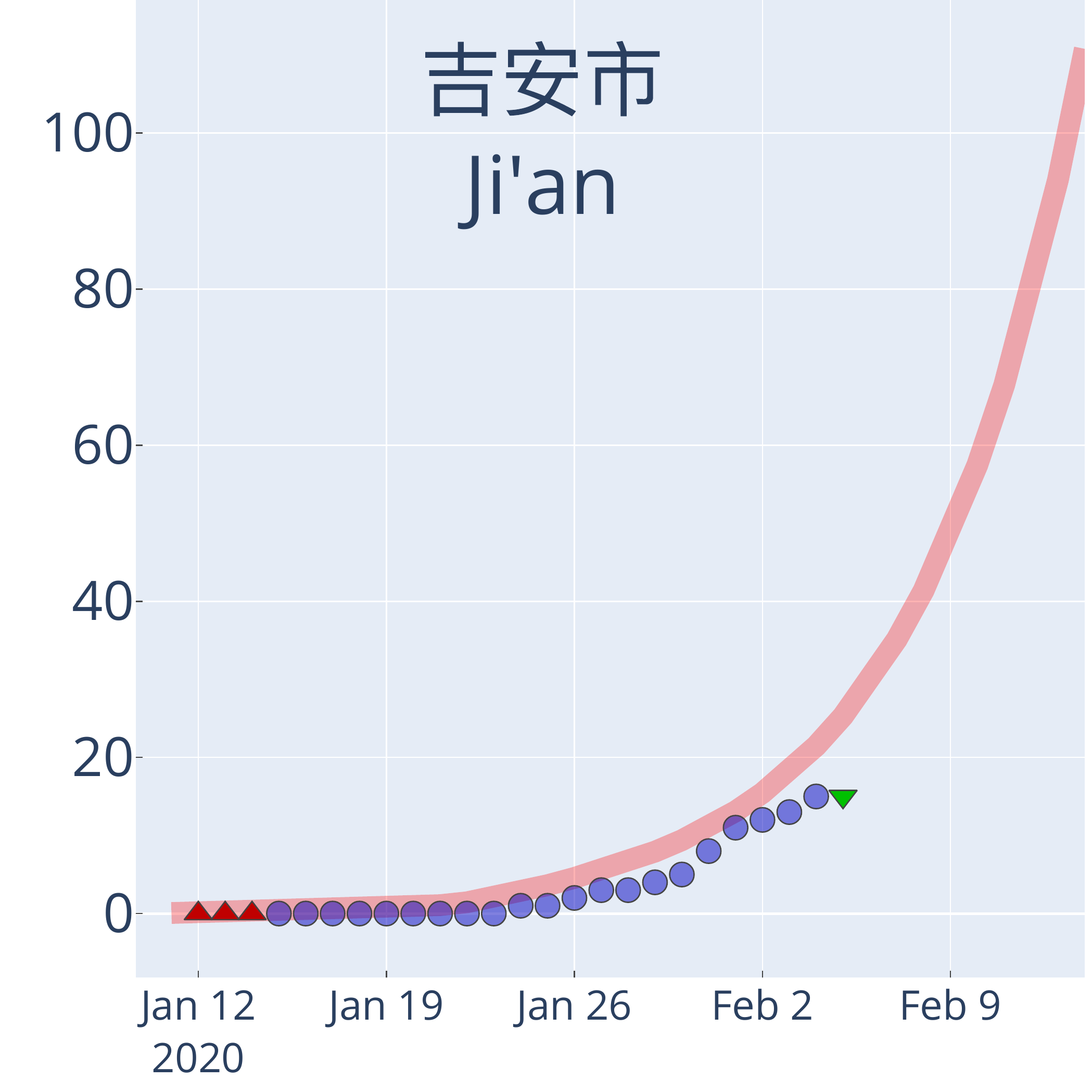}\hfill{}\includegraphics[width=0.24\textwidth]{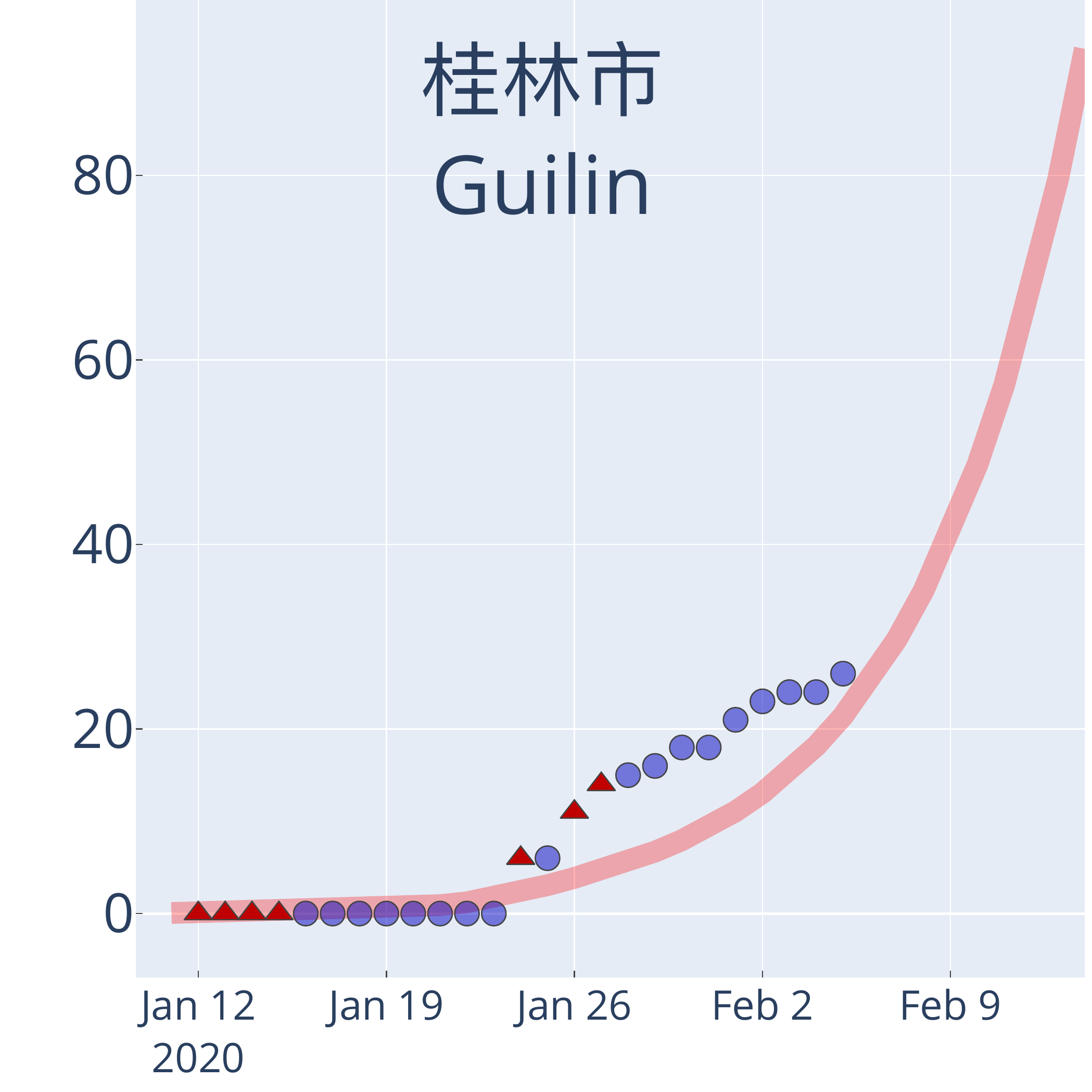}\hfill{}\includegraphics[width=0.24\textwidth]{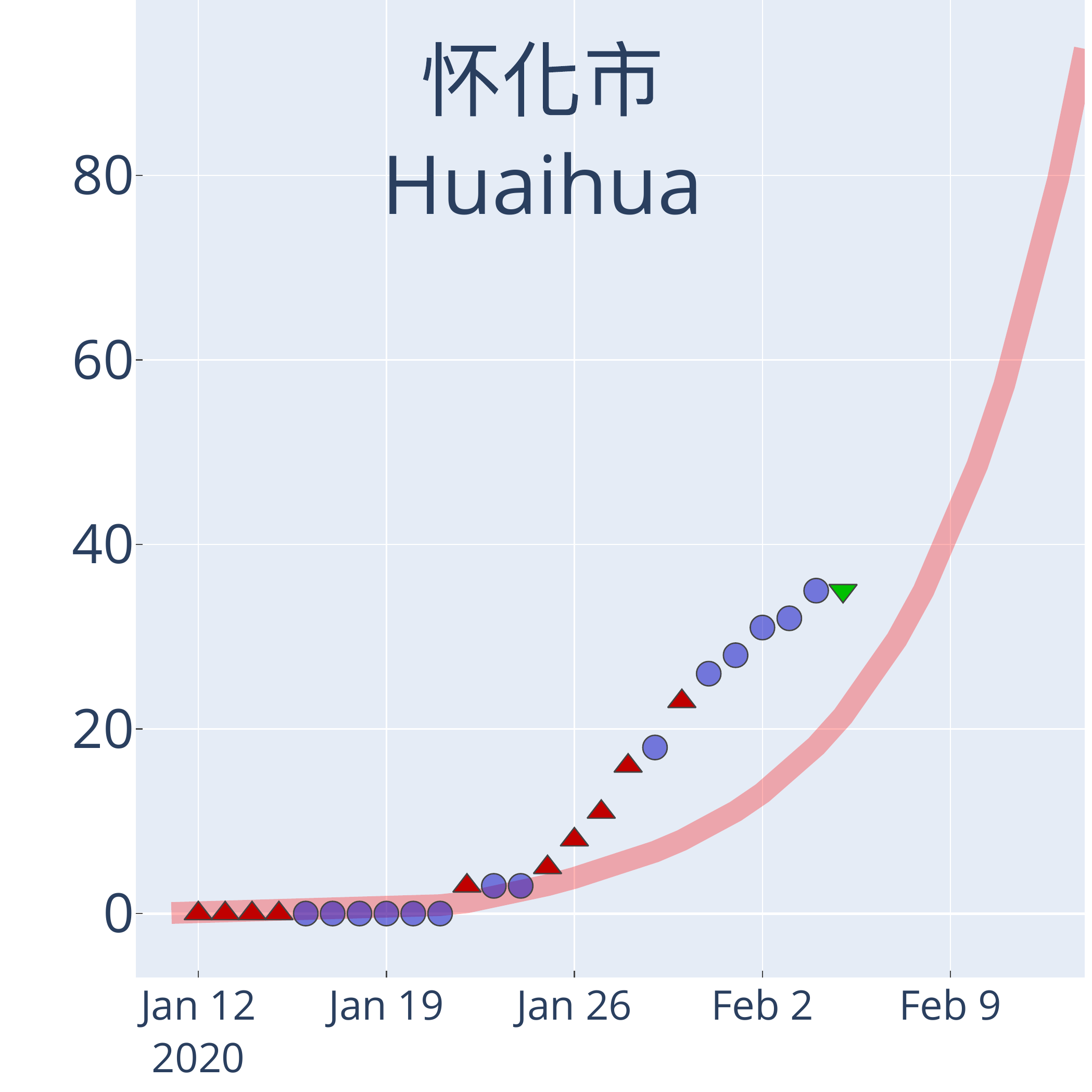}\hfill{}\includegraphics[width=0.24\textwidth]{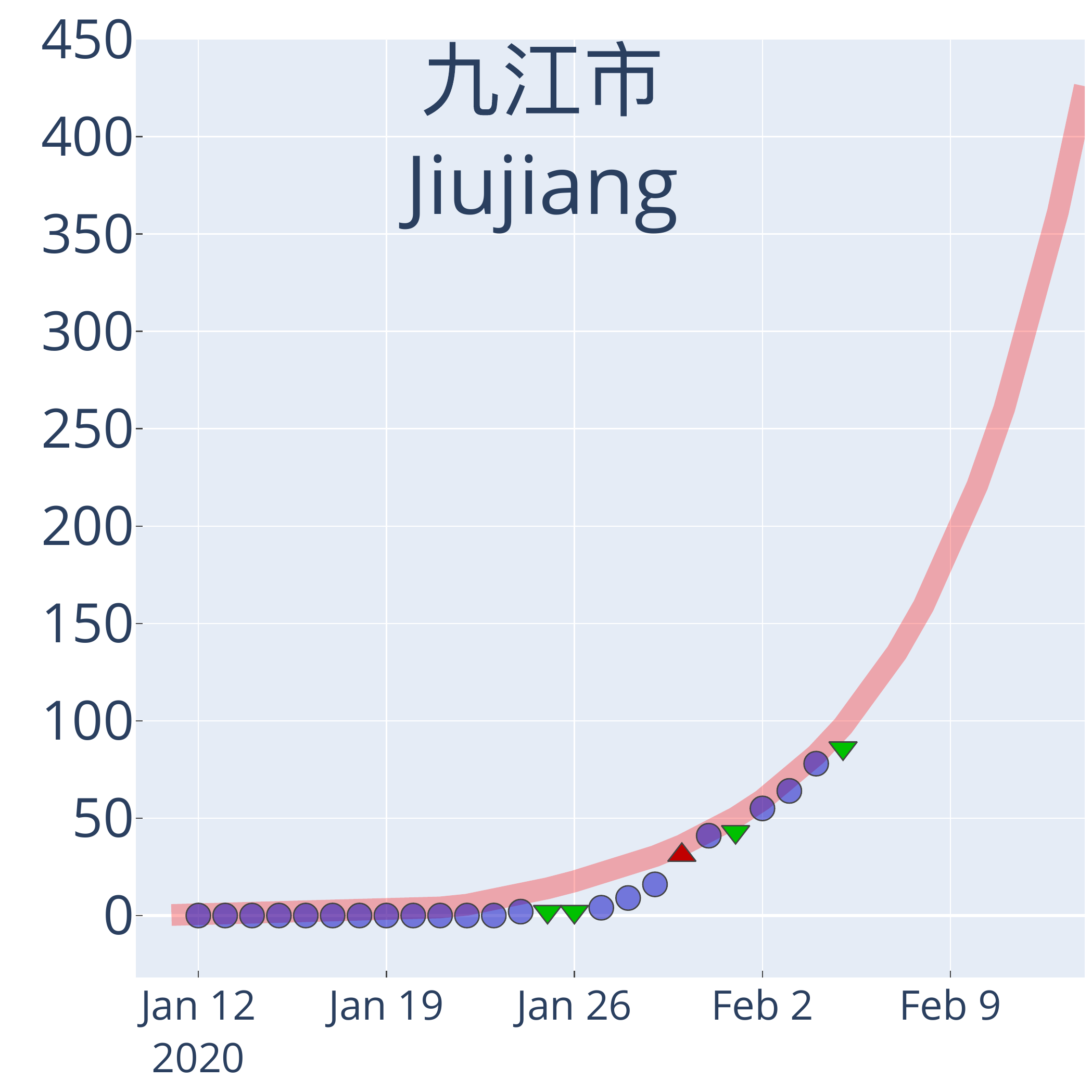}\hfill{}
\par\end{centering}
\begin{centering}
\hfill{}\includegraphics[width=0.24\textwidth]{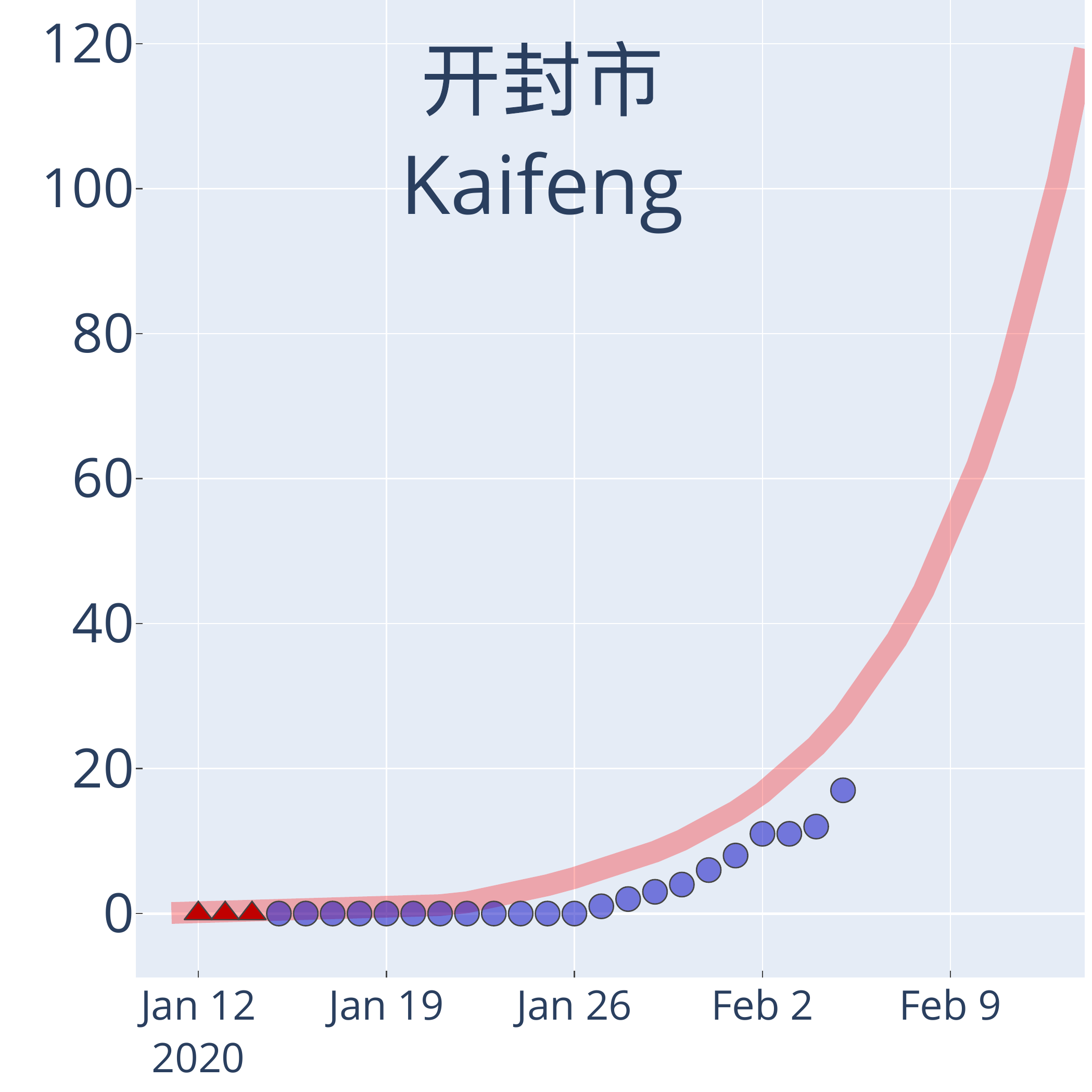}\hfill{}\includegraphics[width=0.24\textwidth]{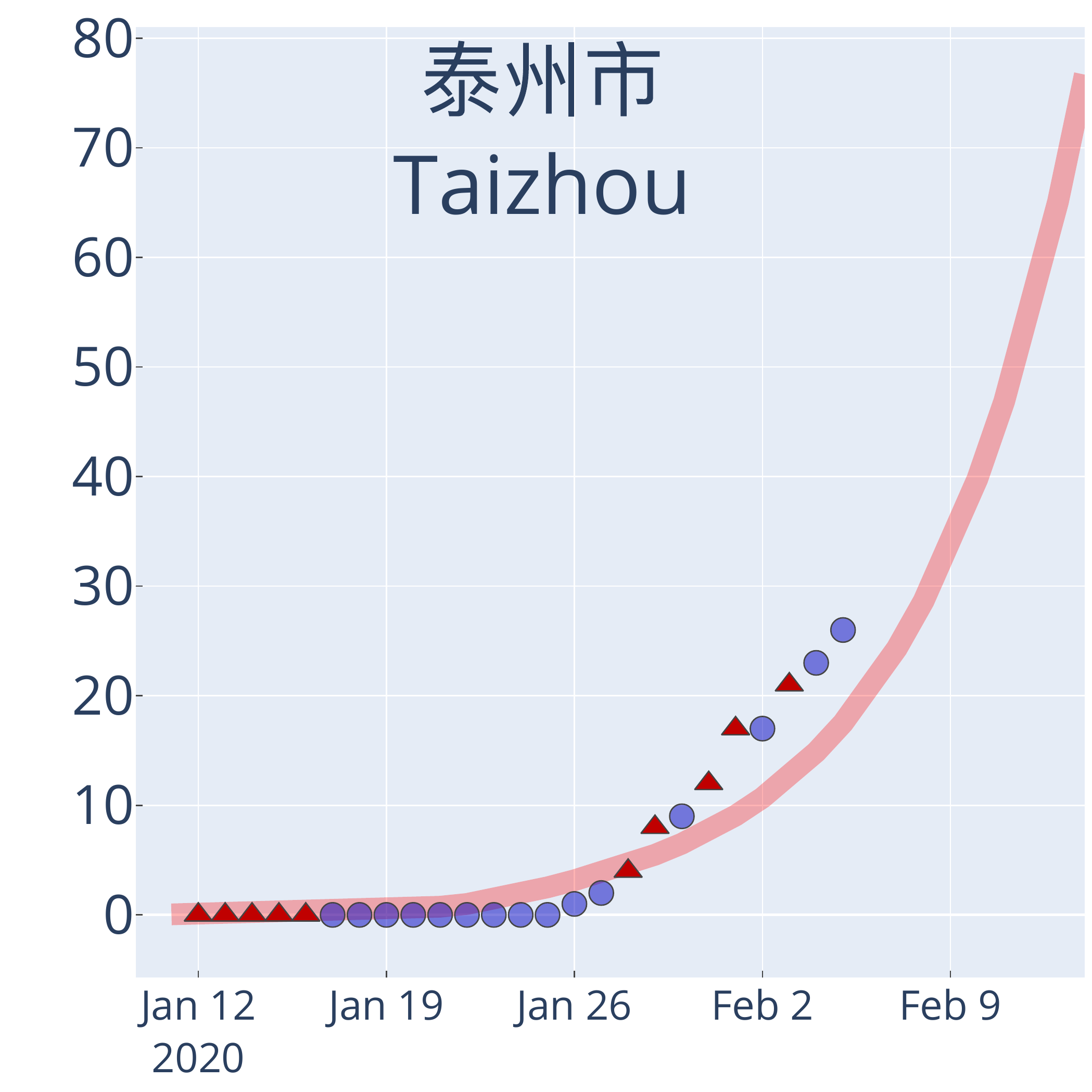}\hfill{}\includegraphics[width=0.24\textwidth]{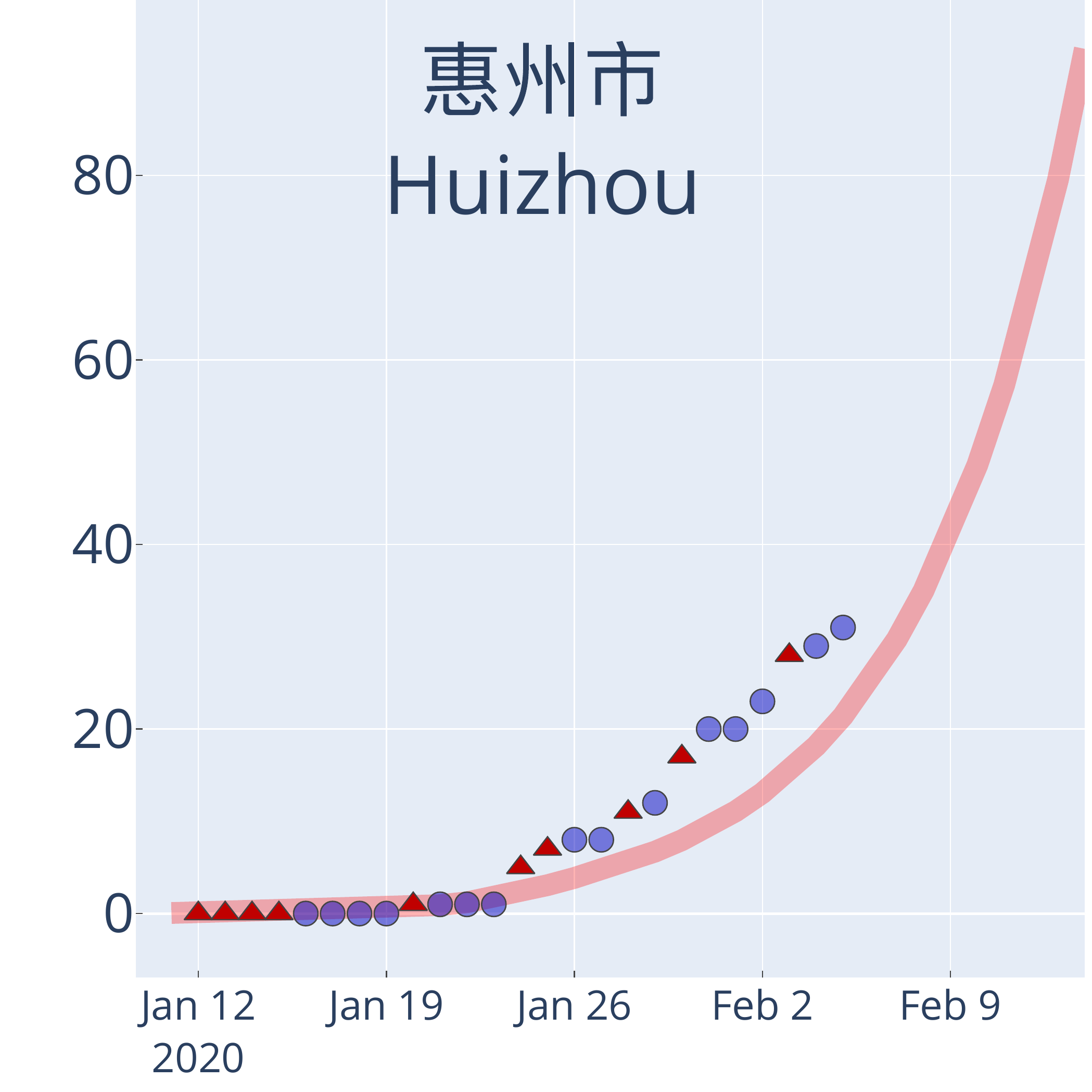}\hfill{}\includegraphics[width=0.24\textwidth]{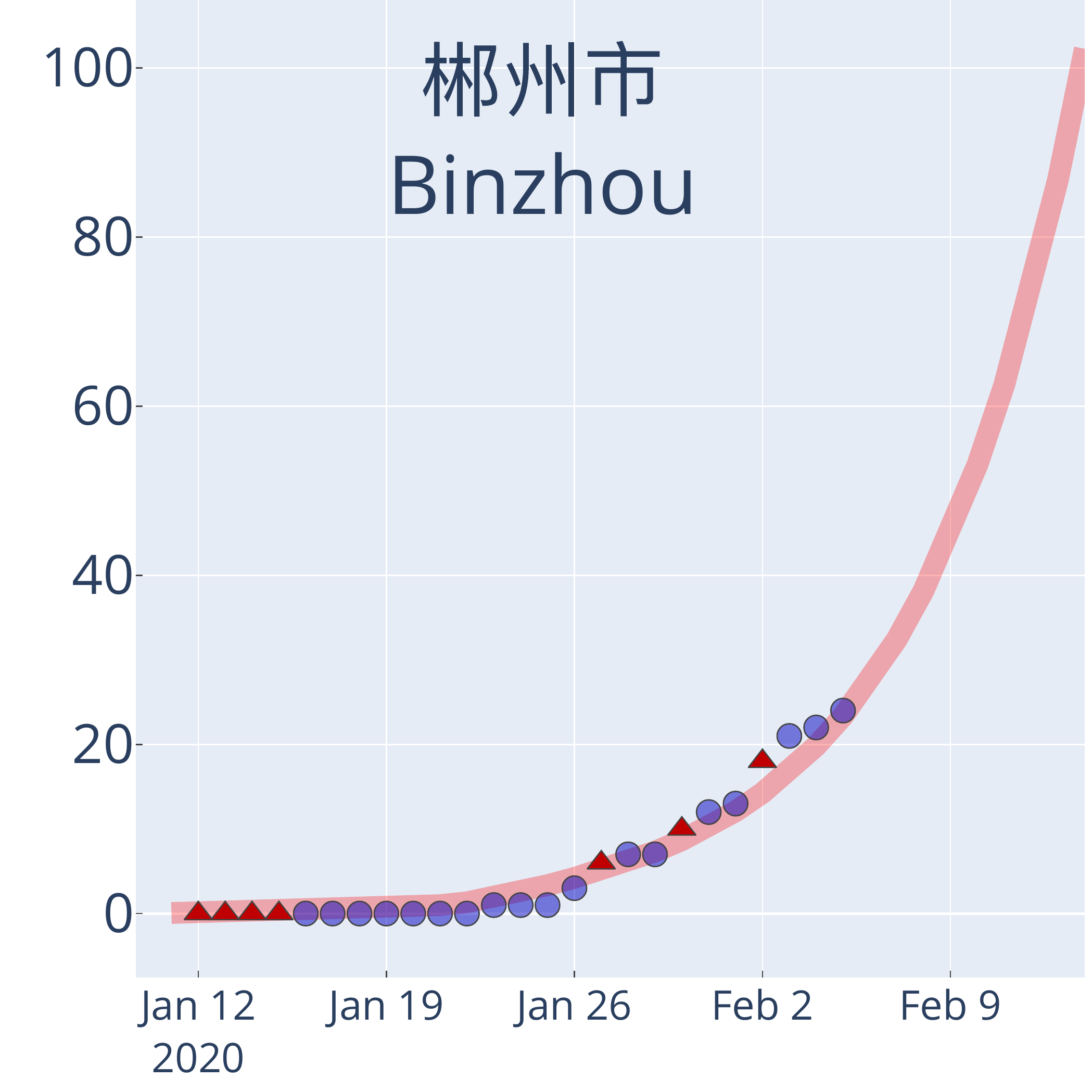}\hfill{}
\par\end{centering}
\begin{centering}
\hfill{}\includegraphics[width=0.24\textwidth]{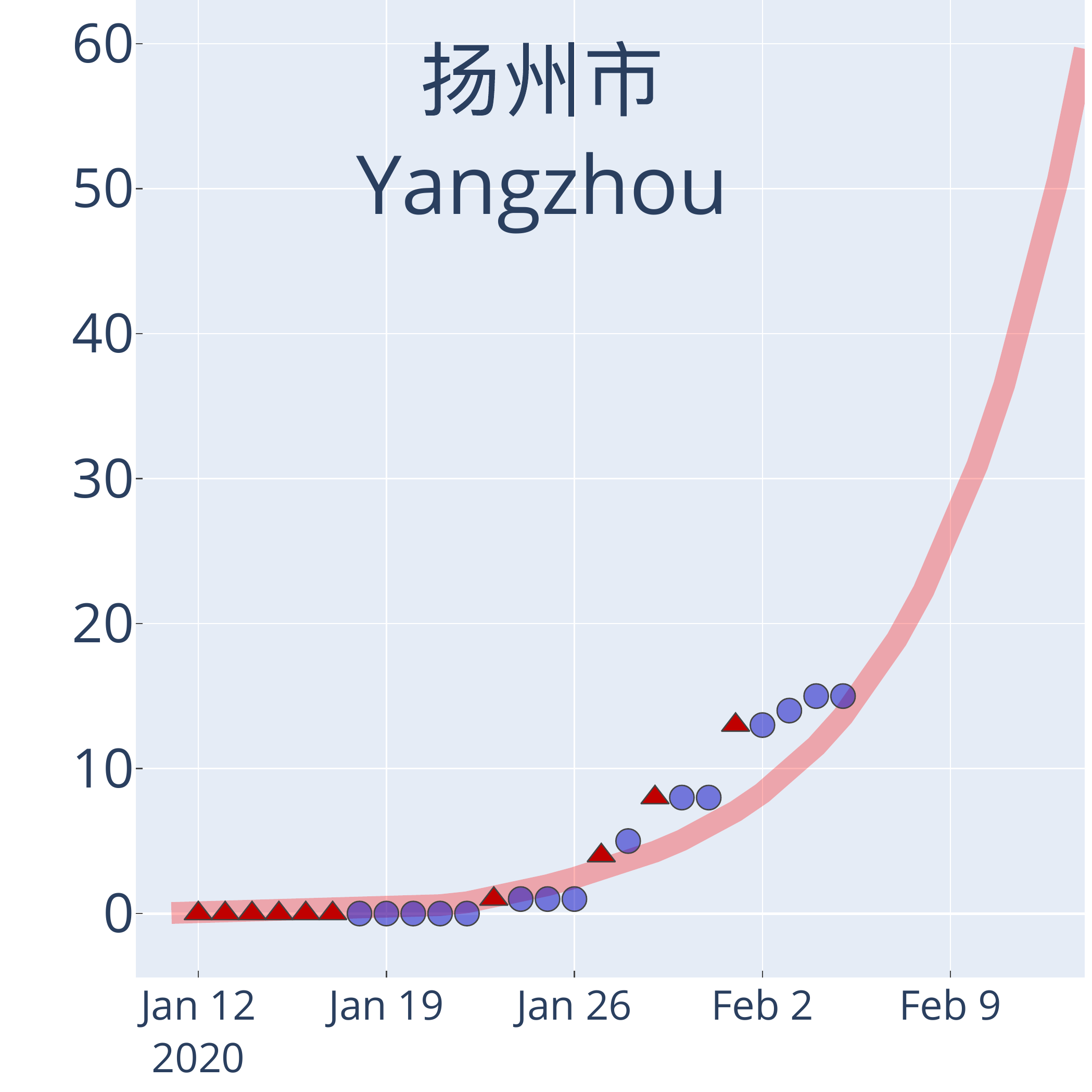}\hfill{}\includegraphics[width=0.24\textwidth]{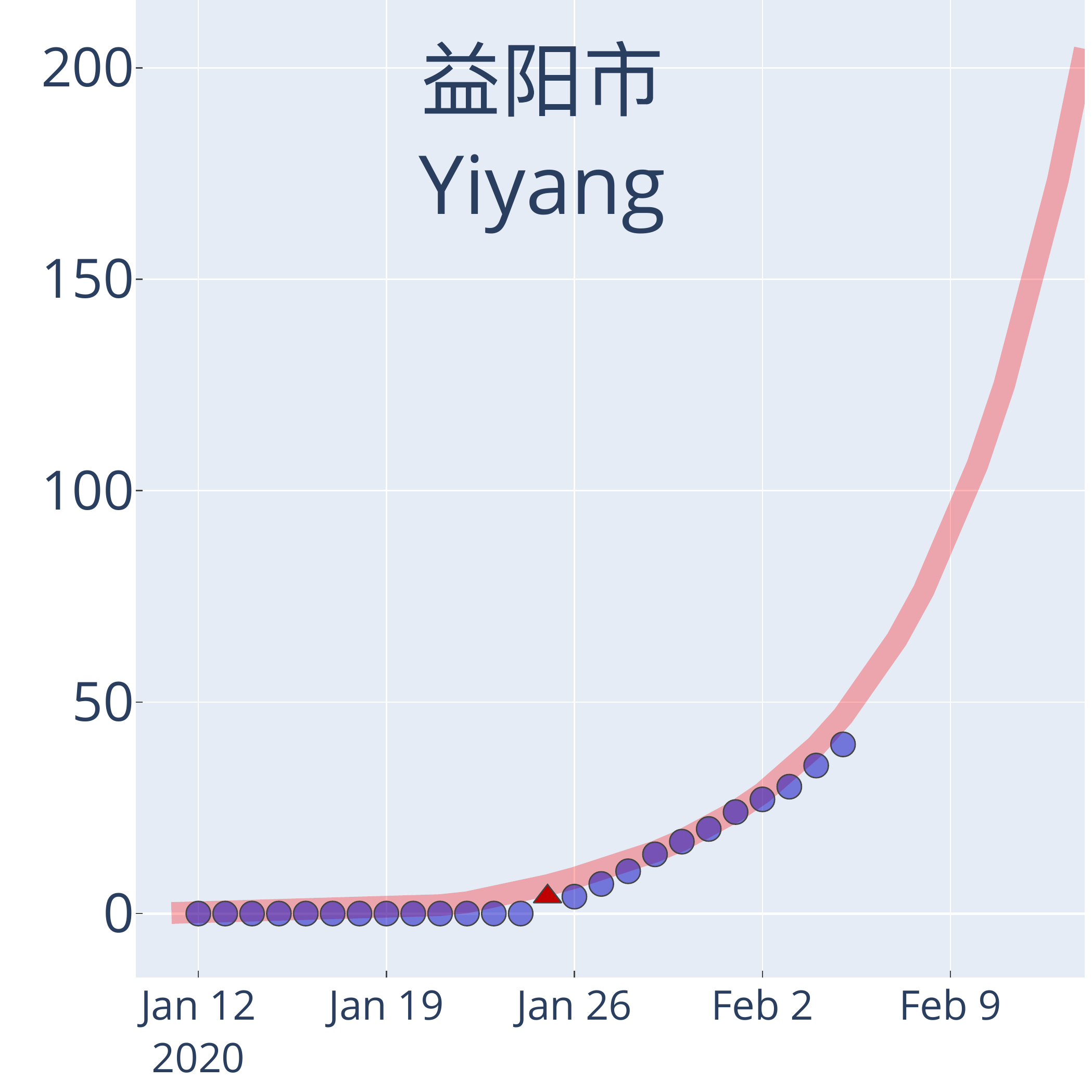}\hfill{}\includegraphics[width=0.24\textwidth]{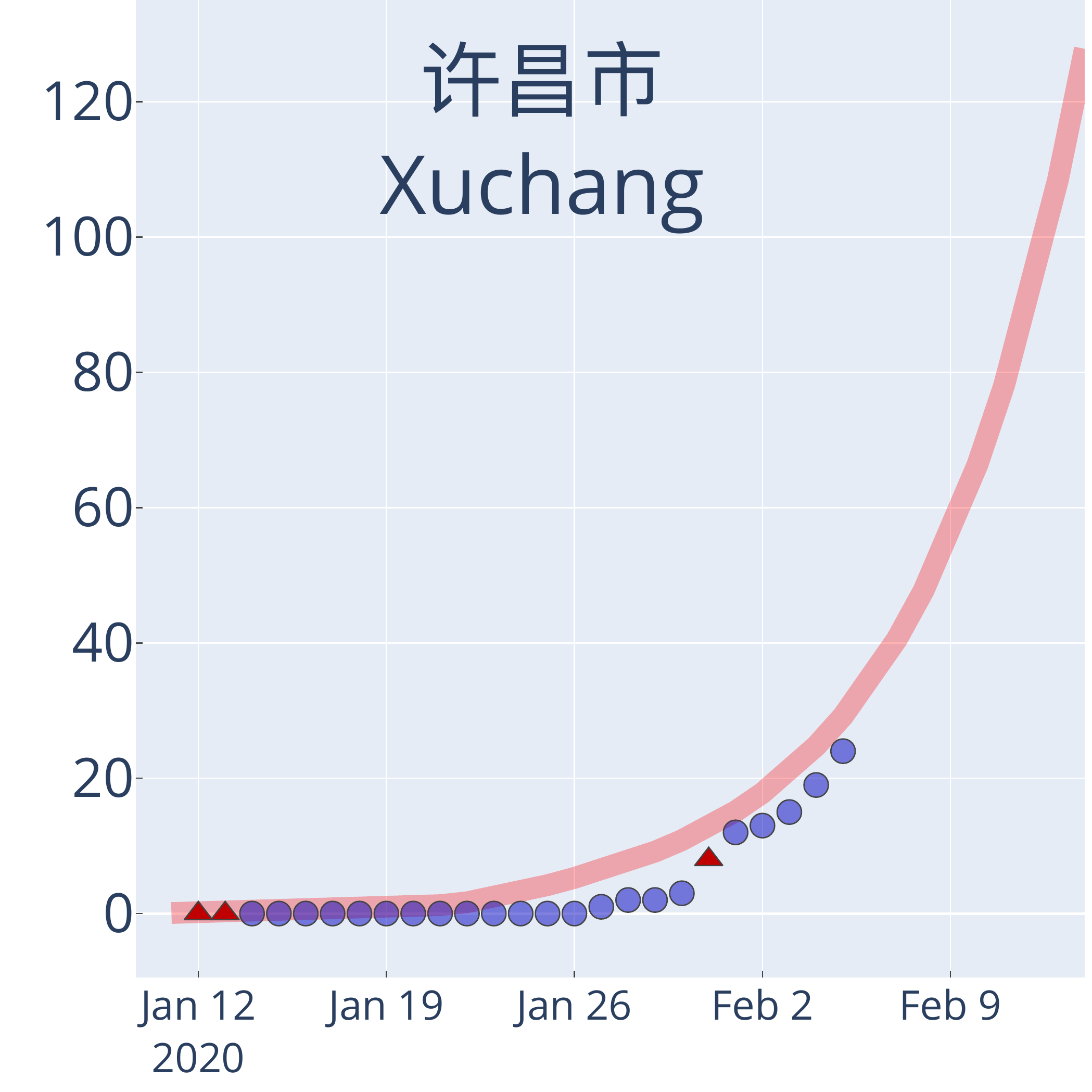}\hfill{}\includegraphics[width=0.24\textwidth]{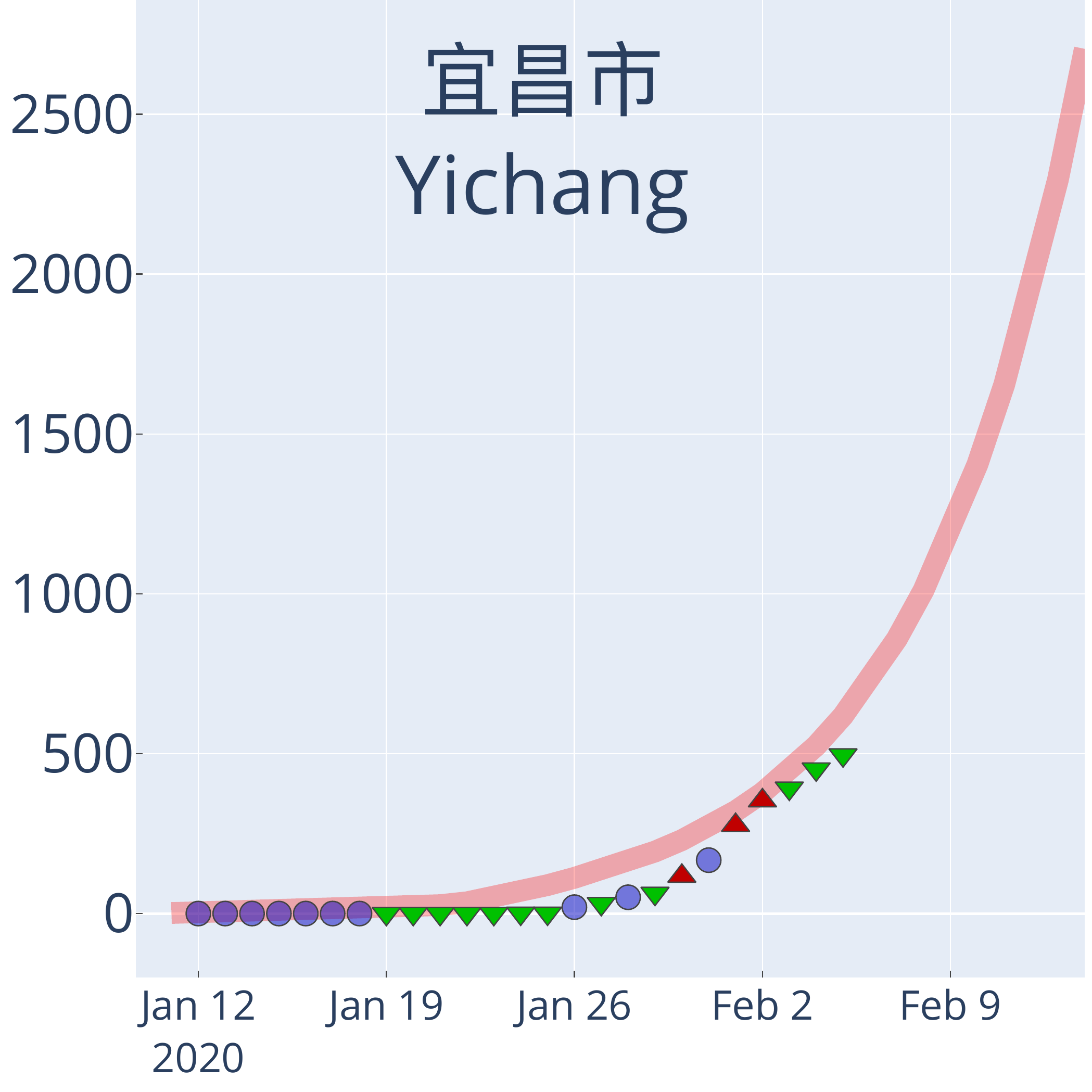}\hfill{}
\par\end{centering}
\begin{centering}
\hfill{}\includegraphics[width=0.24\textwidth]{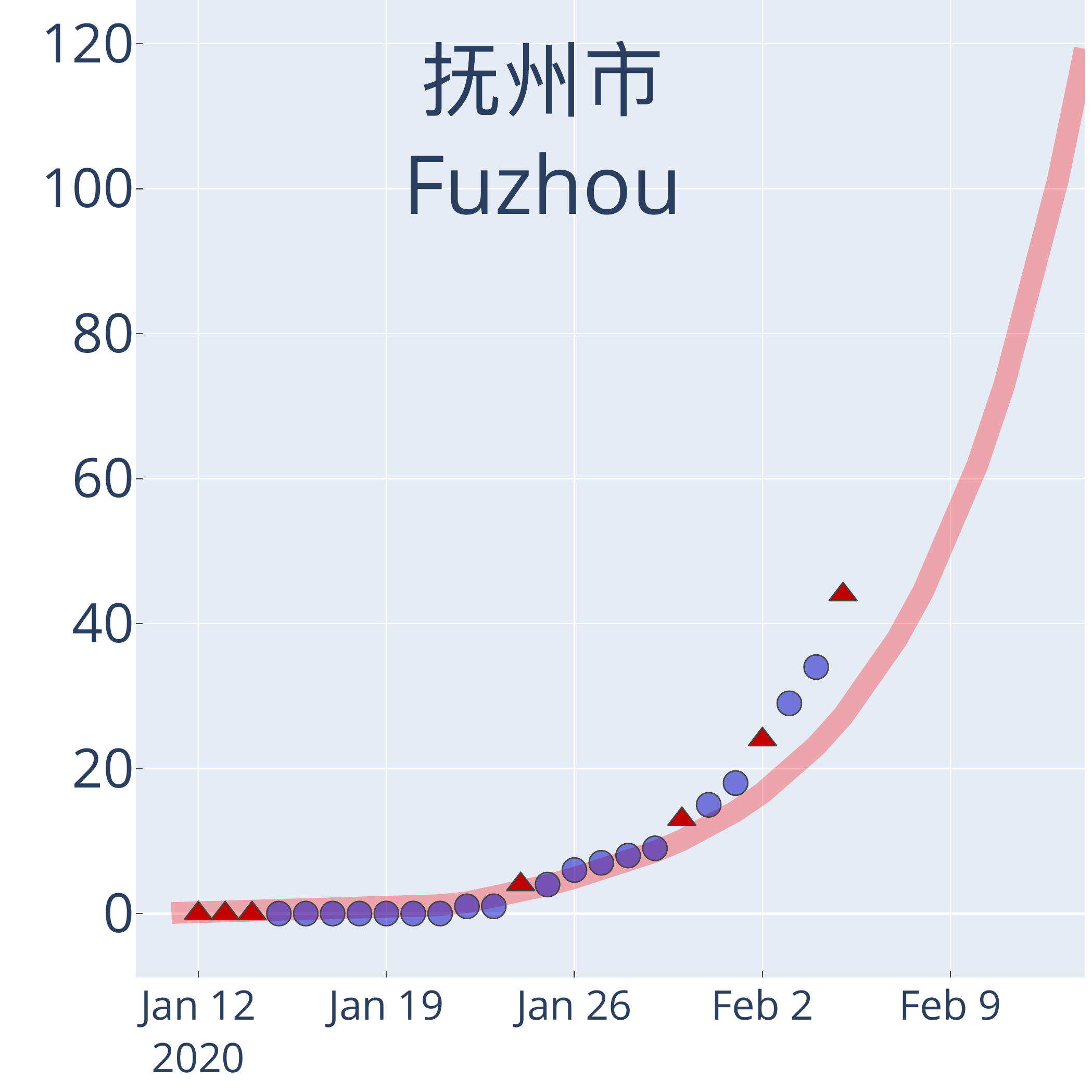}\hfill{}\includegraphics[width=0.24\textwidth]{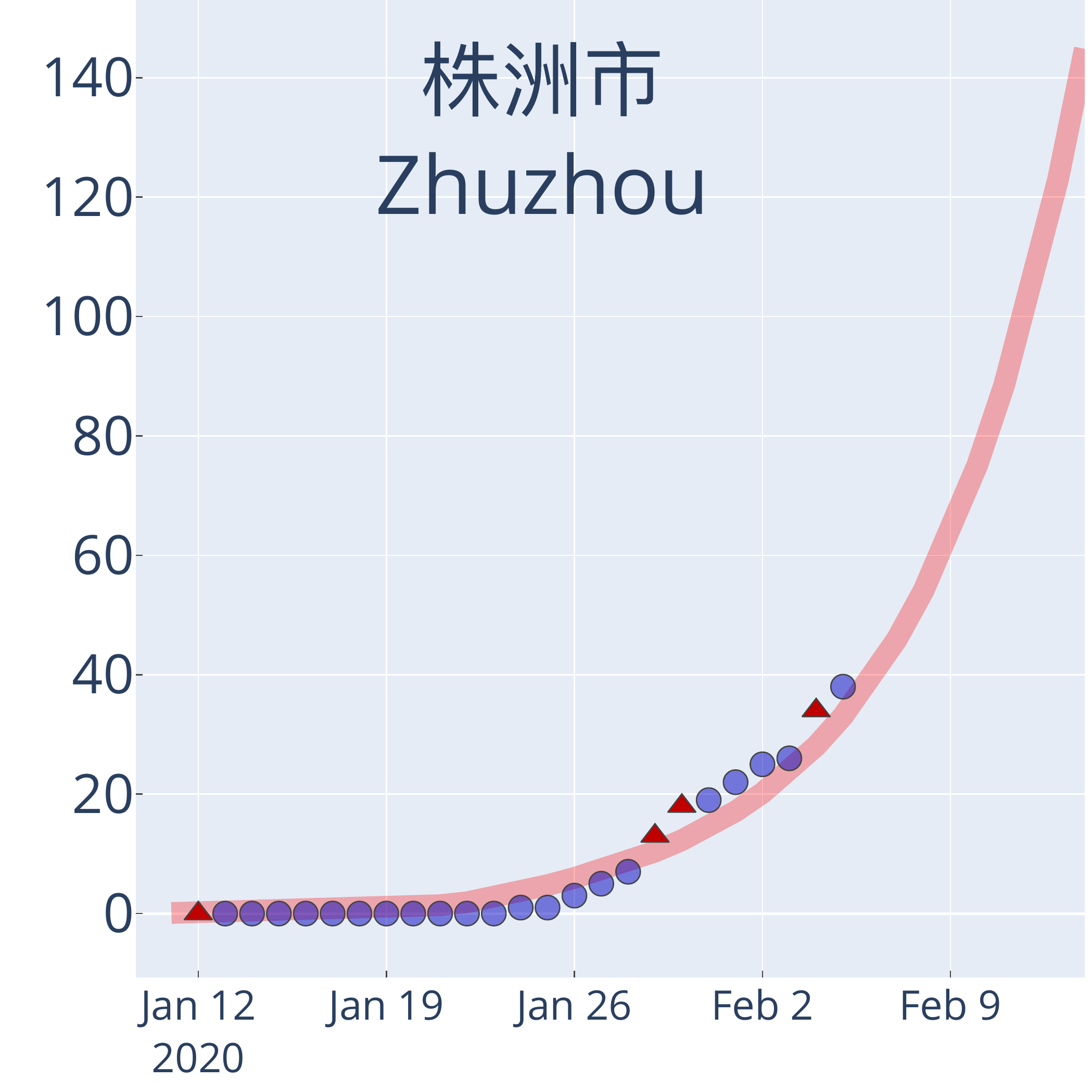}\hfill{}\includegraphics[width=0.24\textwidth]{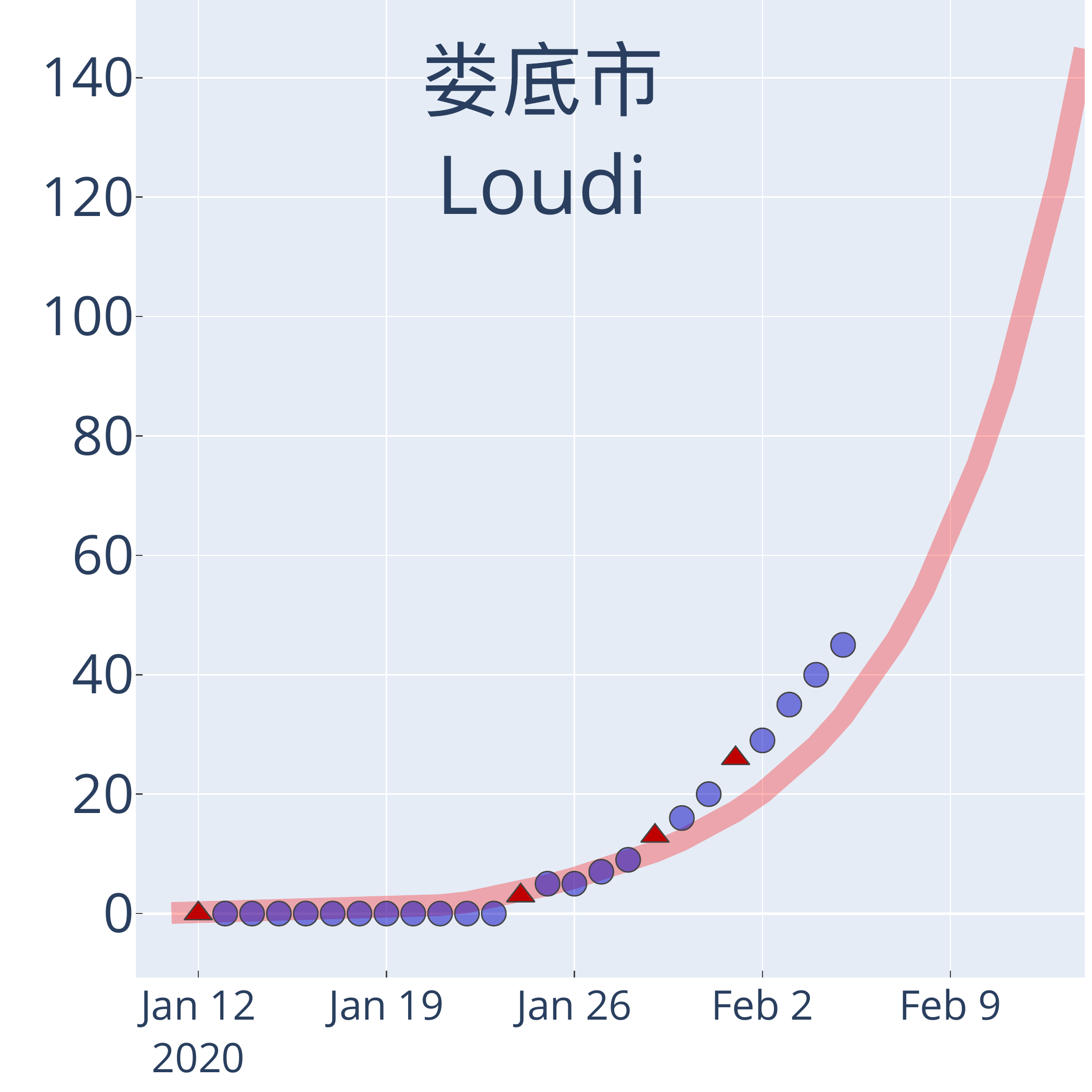}\hfill{}\includegraphics[width=0.24\textwidth]{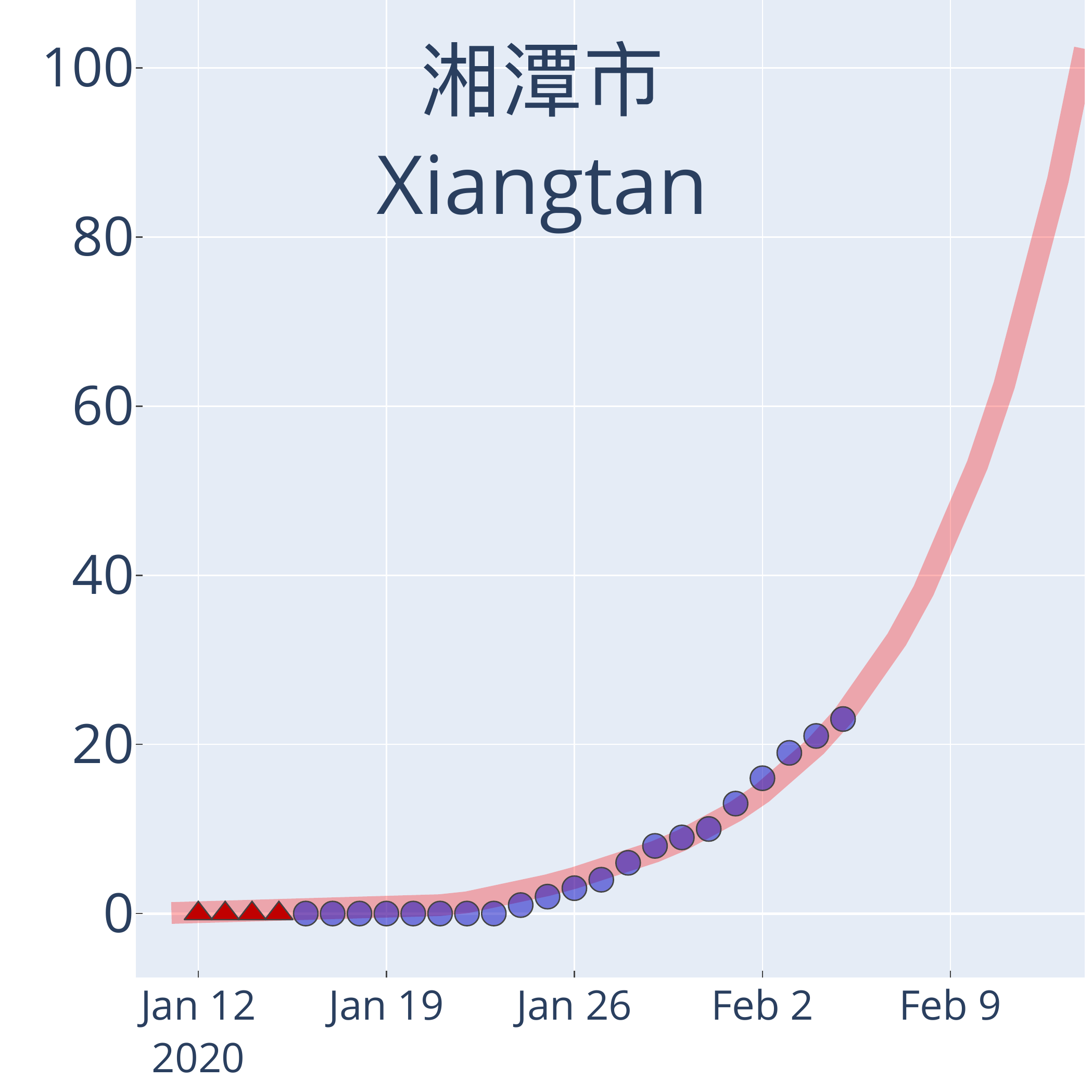}\hfill{}
\par\end{centering}
\caption{Simulation and forecasting of infections in major China cities and
comparison to accumulated cases. See Figure \ref{fig:sim-sample}
for detailed interpretation of the marks and legends used in the plots.}
\end{figure*}

\begin{figure*}[t]
\begin{centering}
\hfill{}\includegraphics[width=0.24\textwidth]{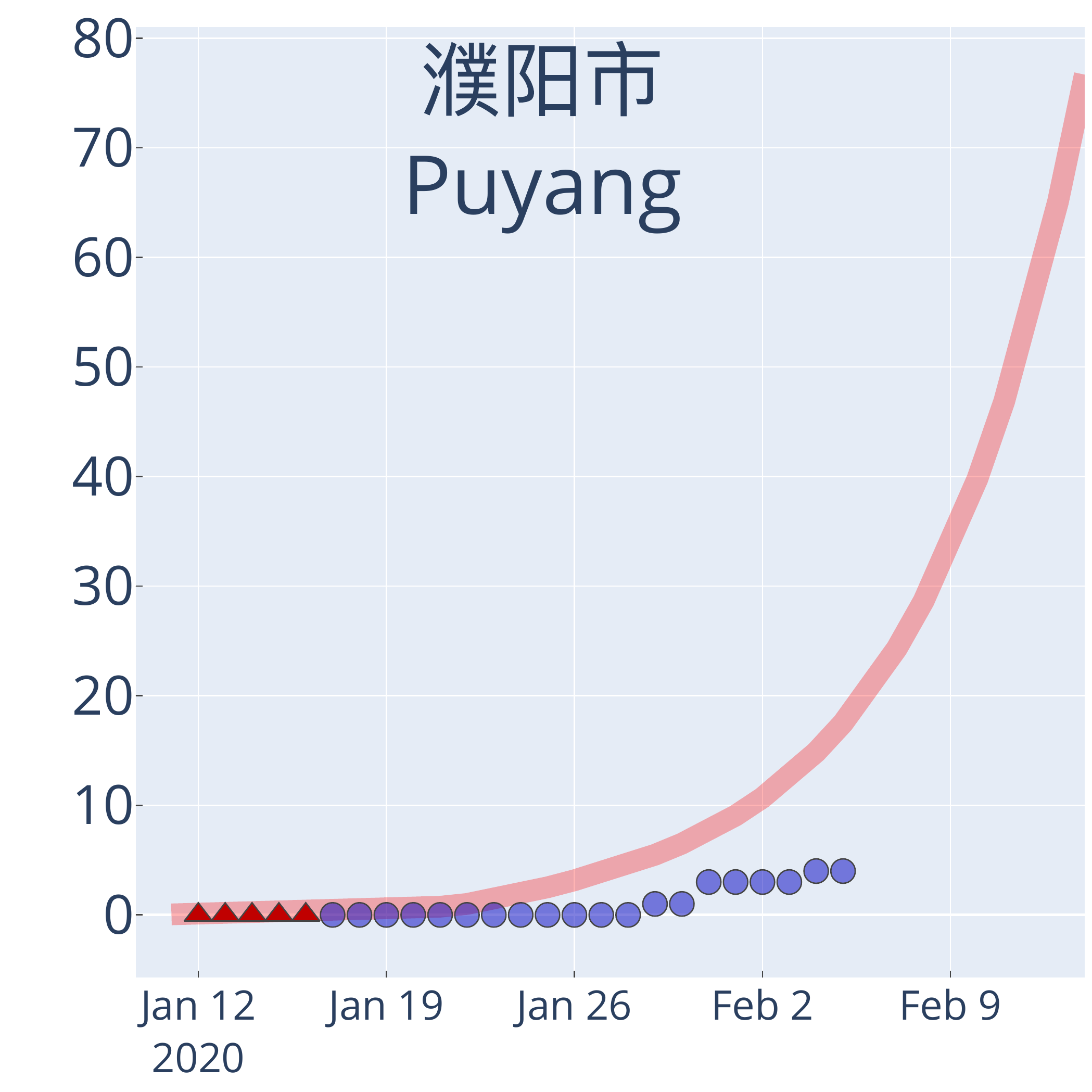}\hfill{}\includegraphics[width=0.24\textwidth]{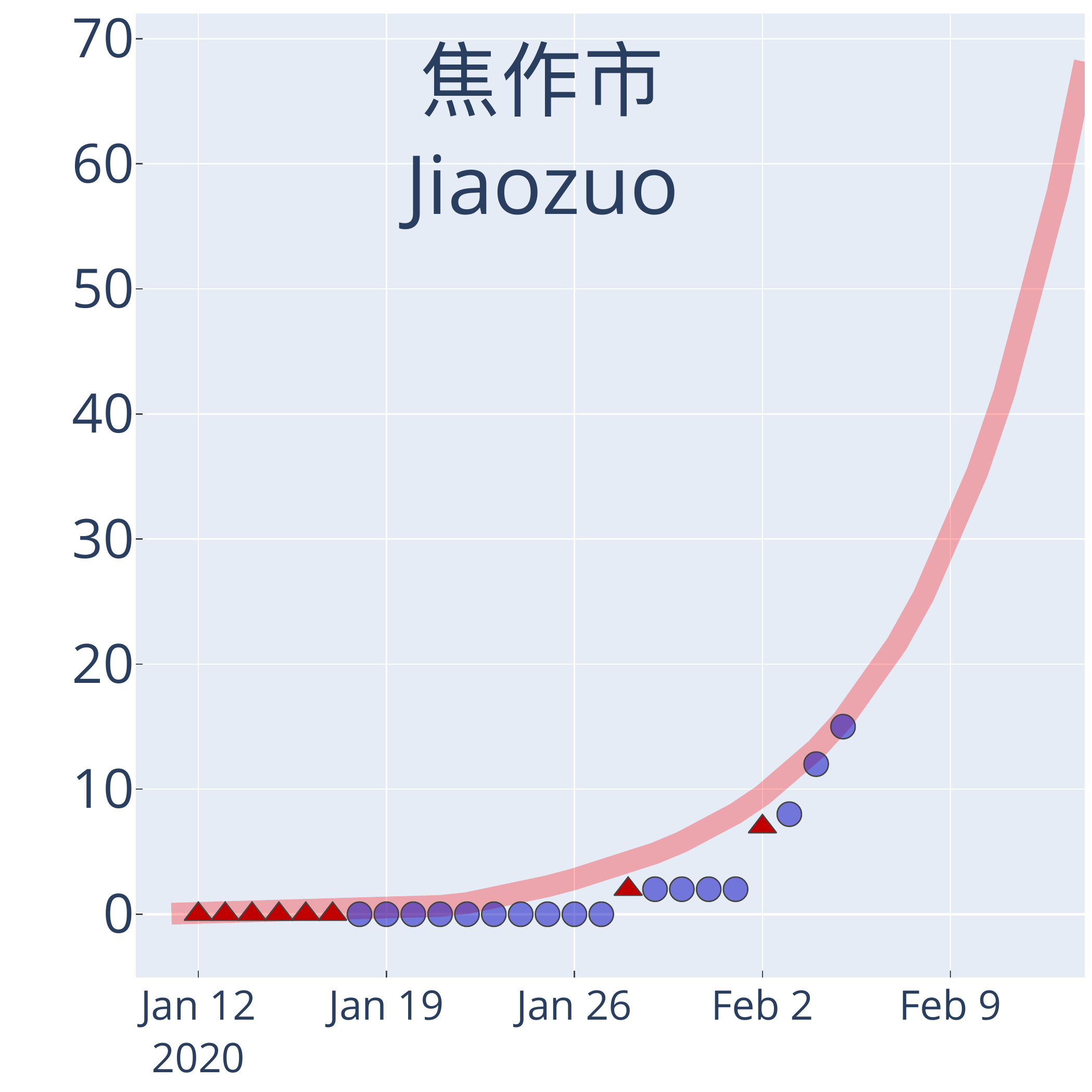}\hfill{}\includegraphics[width=0.24\textwidth]{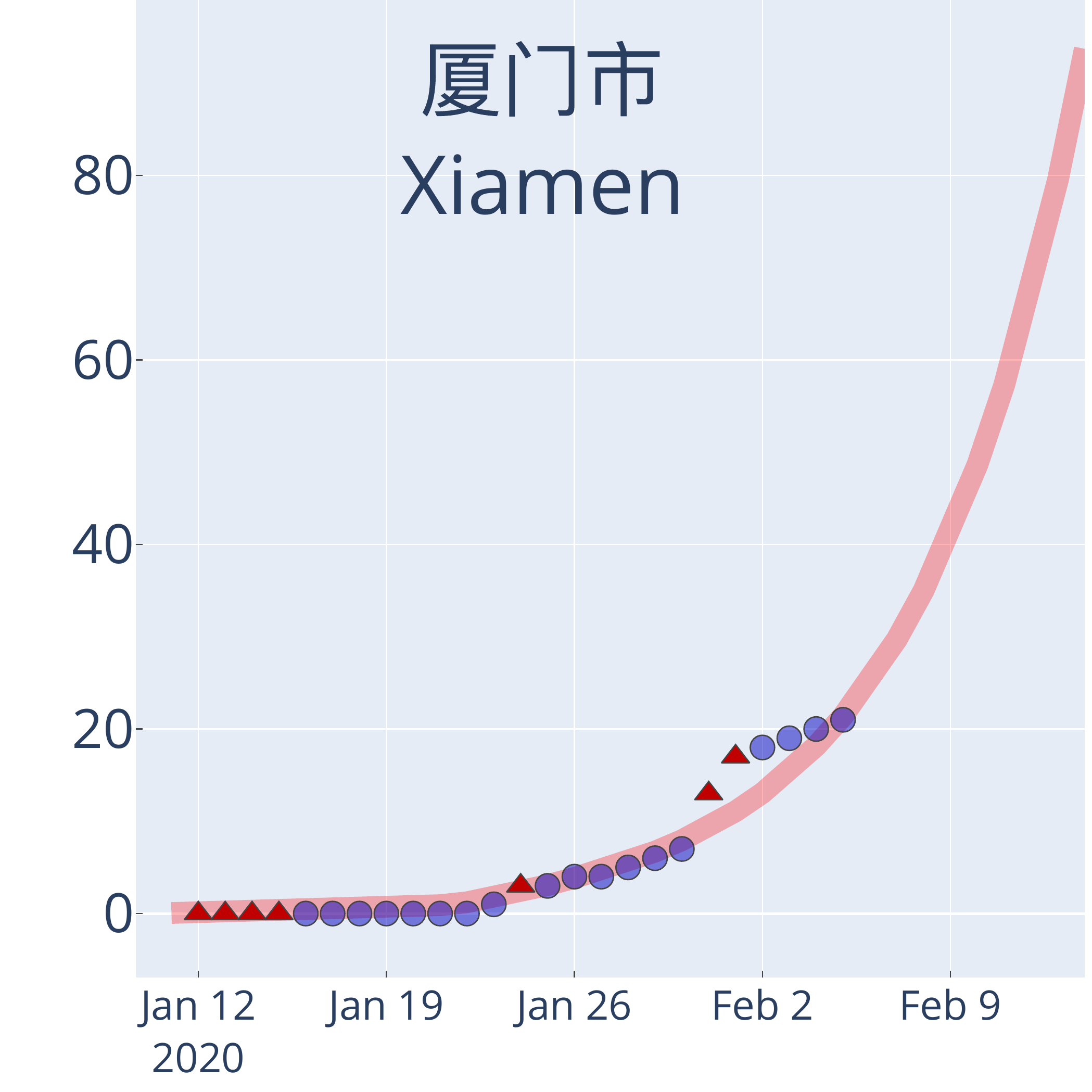}\hfill{}\includegraphics[width=0.24\textwidth]{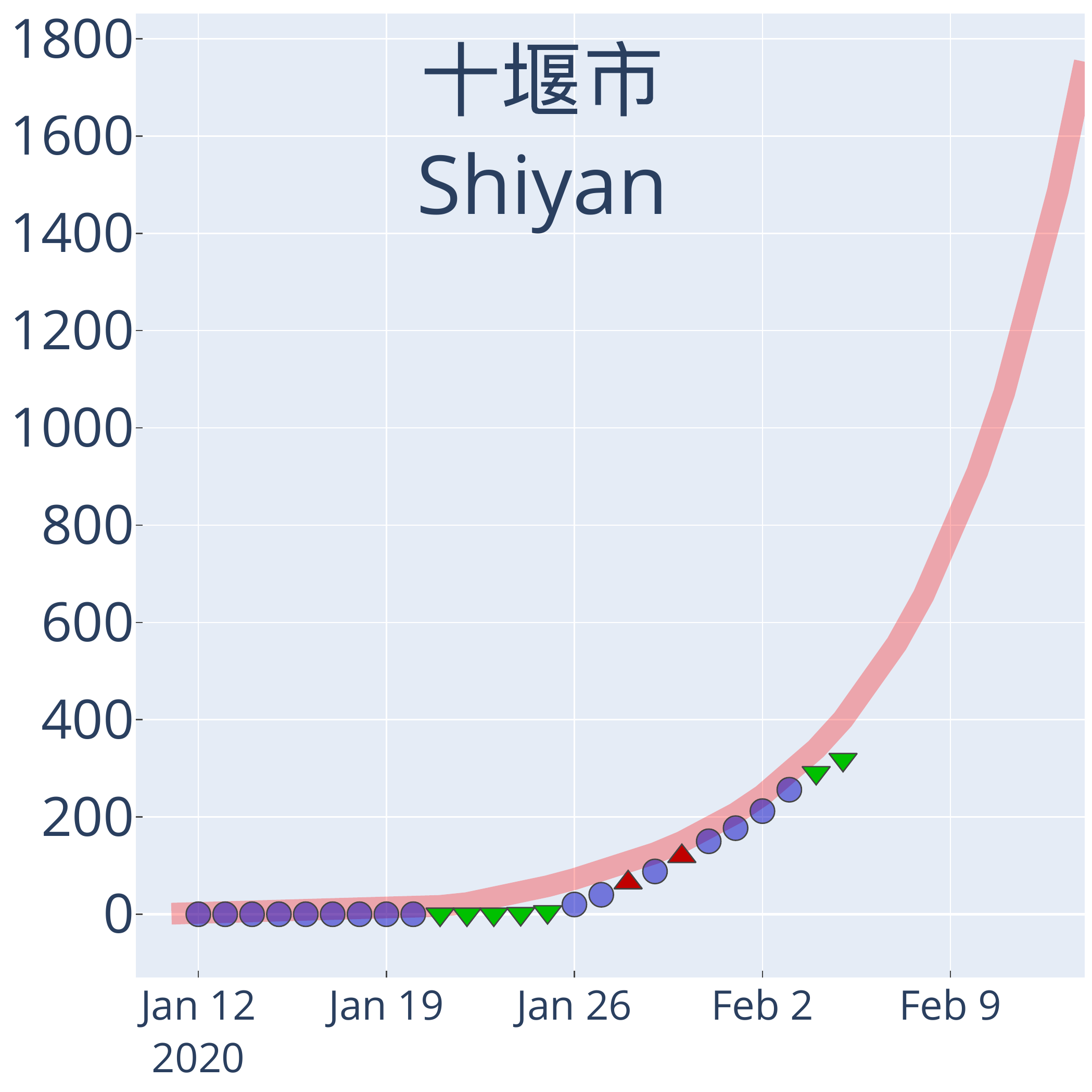}\hfill{}
\par\end{centering}
\begin{centering}
\hfill{}\includegraphics[width=0.24\textwidth]{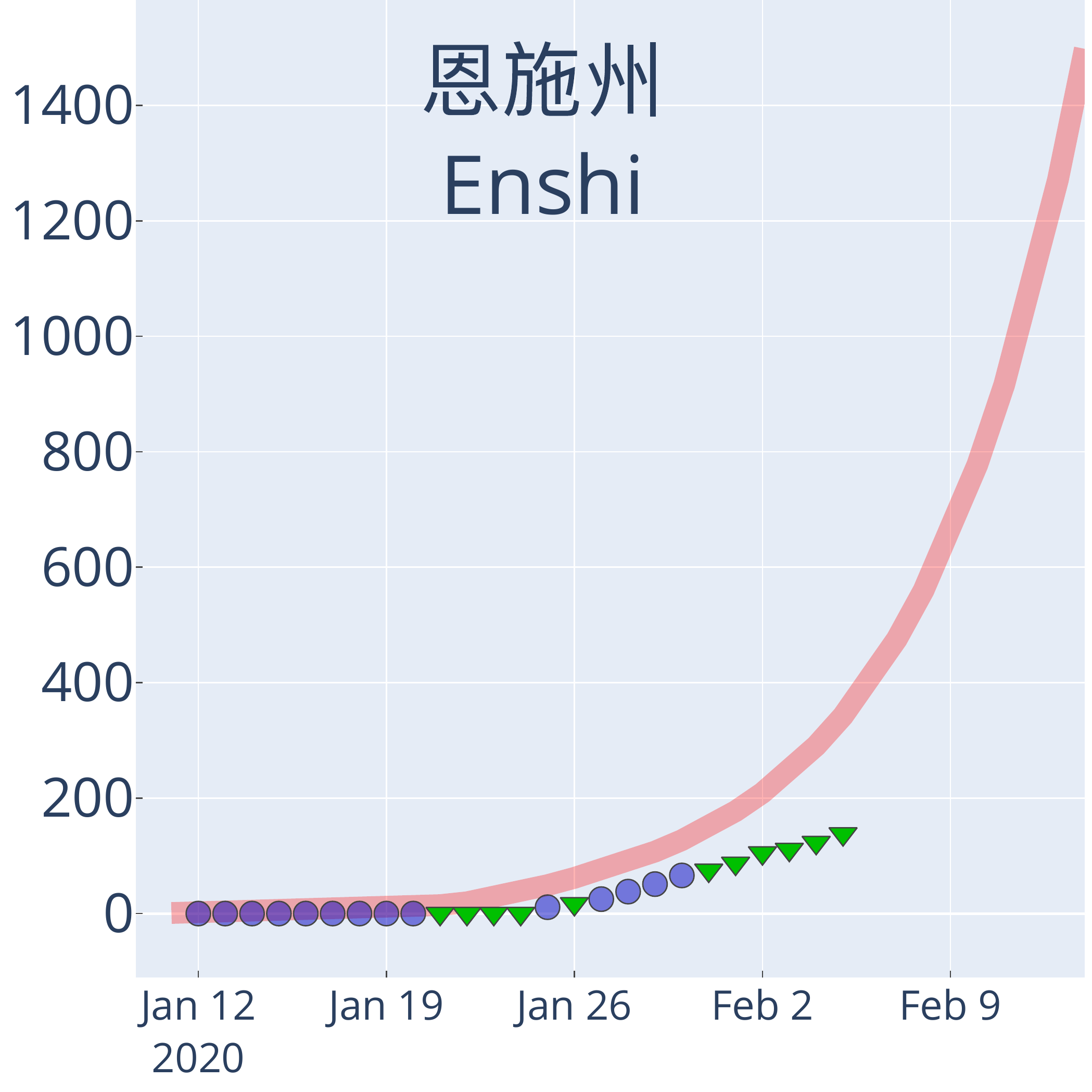}\hfill{}
\par\end{centering}
\caption{\label{fig:sim-1}Simulation and forecasting of infections in major
China cities and comparison to accumulated cases. See Figure \ref{fig:sim-sample}
for detailed interpretation of the marks and legends used in the plots.}
\end{figure*}

\bibliographystyle{plain}
\bibliography{biblo}

\end{document}